\RequirePackage{luatex85}
\documentclass{aa}  

\usepackage{graphicx}
\usepackage{txfonts}
\usepackage{amsmath}
\usepackage{xcolor}
\usepackage{float}
\usepackage{ragged2e}
\usepackage{graphicx}
\usepackage{multirow} 
\usepackage{hyperref}
\usepackage{subcaption}

\graphicspath{{./}{figures/}}

\begin{document}

\title{
The Outskirts of Abell\,1795:\\Probing Gas Clumping in the Intra-Cluster Medium
}

\author{
Orsolya E. Kov\'acs\inst{1},
Zhenlin Zhu\inst{2,3},
Norbert Werner\inst{1},
Aurora Simionescu\inst{2,3,4},
\and \'Akos Bogd\'an \inst{5}
}

\institute{
Department of Theoretical Physics and Astrophysics, Faculty of Science, Masaryk University, Kotl\'a\v{r}sk\'a 2, Brno, 611 37, Czech Republic, \email{o.e.kovacs@gmail.com}
\and
SRON Netherlands Institute for Space Research, Niels Bohrweg 4, 2333 CA Leiden, The Netherlands
\and
Leiden Observatory, Leiden University, Niels Bohrweg 2, 2300 RA Leiden, The Netherlands
\and
Kavli Institute for the Physics and Mathematics of the Universe, The University of Tokyo, Kashiwa, Chiba 277-8583, Japan
\and
Center for Astrophysics\,\textbar\,Harvard \& Smithsonian, 60 Garden Street, Cambridge, MA 02138, USA
}

\date{Received Month XX, 2023; accepted Month YY, 2023}

\abstract
{The outskirts of galaxy clusters host complex interactions between the intra-cluster and circumcluster media.
During the evolution of clusters, ram-pressure stripped gas clumps from infalling substructures break the uniformity of the gas distribution, which may lead to observational biases at large radii.
Assessing the contribution of gas clumping, however, poses observational challenges, and requires robust X-ray measurements in the background-dominated regime of cluster outskirts.
}
{The main objectives of this study are isolating faint gas clumps from field sources and from the diffuse emission in the Abell\,1795 galaxy cluster, then probing their impact on the observed surface brightness and thermodynamic profiles.}
{We performed imaging analysis on deep \textit{Chandra} ACIS-I observations of the Abell\,1795 cluster outskirts, extending out to $\sim\!1.5r_{200}$ with full azimuthal coverage.
We built the $0.7-2.0$\,keV surface brightness distribution from the adaptively binned image of the diffuse emission, and looked for clumps in the form of $>\!+2\sigma$ surface brightness outliers.
Classification of the clump candidates was based primarily on \textit{Chandra} and \textit{SDSS} data.
Benefiting from the point source list resolved by \textit{Chandra}, we extracted the thermodynamic profiles of the intra-cluster medium from the associated \textit{Suzaku} XIS data out to $r_{200}$ using multiple point source and clump candidate removal approaches.}
{We identified 24 clump candidates in the Abell\,1795 field, most of which are likely associated with background objects, including AGN, galaxies, and clusters or groups of galaxies, as opposed to intrinsic gas clumps. 
These sources had minimal impact on the surface brightness and thermodynamic profiles of the cluster emission.
{Nevertheless, the measured entropy profile deviates from a pure gravitational collapse, suggesting complex physics at play in the outskirts, including potential electron-ion non-equilibrium and non-thermal pressure support.}}
{}


\titlerunning {The outskirts of Abell\,1795}
\authorrunning {Kov\'acs et al.}

\maketitle

\section{Introduction} \label{sec:introduction}

\begin{table*}
\caption{Particulars of the \textit{Chandra} ACIS-I observations of Abell\,1795. $t_{\rm exp}$ refers to the clean (i.e. flare-filtered) exposure time.}
\label{tab:observations}
\centering
\addtolength{\tabcolsep}{-.8pt} 
\begin{tabular}{ccccc|ccccc} 
\hline\hline        
Obs. ID & Obs. date & R.A. & Decl. & $t_{\rm exp} \rm (ks)$  & Obs. ID & Obs. date & R.A. & Decl. & $t_{\rm exp} \rm (ks)$ \\ 
\hline               
17632 & 2015-04-29 & 13 46 36.85 & +26 28 20.38 & 45.03 & 17615 & 2016-12-08 & 13 50 24.41 & +26 40 15.23 & 23.78 \\
17608 & 2015-03-07 & 13 46 52.90 & +26 15 14.10 & 40.94 & 17616 & 2016-11-08 & 13 46 43.44 & +26 53 58.97 & 25.51 \\
17478 & 2015-02-27 & 13 46 52.90 & +26 15 14.10 & 47.53 & 17617 & 2016-07-01 & 13 50 25.84 & +26 09 43.41 & 19.04 \\
17479 & 2015-04-28 & 13 46 57.83 & +26 35 50.18 & 44.80 & 17618 & 2016-09-01 & 13 47 30.83 & +26 26 15.97 & 56.95 \\
17221 & 2016-12-20 & 13 48 52.79 & +27 03 28.99 & 26.93 & 17619 & 2016-03-21 & 13 47 10.42 & +26 59 50.35 & 59.32 \\
17222 & 2016-04-03 & 13 50 32.37 & +26 16 16.71 & 56.22 & 17620 & 2016-04-01 & 13 47 35.12 & +27 03 00.21 & 55.96 \\
17223 & 2015-03-19 & 13 49 54.22 & +26 05 01.15 & 52.42 & 17621 & 2016-03-27 & 13 50 53.44 & +26 16 39.28 & 58.83 \\
17224 & 2016-11-18 & 13 46 28.32 & +26 44 58.65 & 29.89 & 17622 & 2016-04-25 & 13 47 57.87 & +26 14 59.75 & 36.40 \\
17225 & 2015-04-05 & 13 50 21.16 & +26 53 40.65 & 52.60 & 17623 & 2015-03-08 & 13 49 40.25 & +26 12 22.30 & 53.91 \\
17226 & 2015-08-25 & 13 50 37.84 & +26 29 07.89 & 53.34 & 17624 & 2015-03-29 & 13 50 49.96 & +26 51 49.10 & 52.71 \\
17227 & 2015-11-20 & 13 48 33.58 & +26 09 29.65 & 52.38 & 17625 & 2015-05-01 & 13 48 15.77 & +27 05 54.34 & 49.75 \\
17228 & 2017-05-07 & 13 48 52.30 & +26 35 36.20 & 59.28 & 17626 & 2015-05-15 & 13 49 50.67 & +27 05 06.51 & 32.93 \\
17229 & 2016-04-13 & 13 51 12.66 & +26 40 07.20 & 23.85 & 17627 & 2015-08-28 & 13 51 11.17 & +26 25 02.04 & 50.90 \\
17230 & 2015-04-26 & 13 48 37.00 & +27 00 16.08 & 52.38 & 17628 & 2015-11-18 & 13 48 59.89 & +26 11 51.36 & 50.39 \\
17231 & 2015-05-14 & 13 49 35.31 & +26 59 52.02 & 54.18 & 17629 & 2016-11-10 & 13 47 15.91 & +26 46 04.06 & 27.62 \\
17232 & 2016-04-10 & 13 47 22.07 & +26 23 22.48 & 55.70 & 17656 & 2015-05-16 & 13 49 50.67 & +27 05 06.51 & 20.75 \\
17597 & 2015-02-08 & 13 50 58.67 & +26 35 46.34 & 20.68 & 18822 & 2016-04-14 & 13 51 12.66 & +26 40 07.20 & 25.85 \\
17598 & 2015-02-11 & 13 48 53.20 & +26 07 23.79 & 47.07 & 18835 & 2016-04-27 & 13 47 57.87 & +26 14 59.75	& 18.50 \\
17599 & 2015-02-15 & 13 46 48.66 & +26 34 53.72 & 50.26 & 18877 & 2016-06-28 & 13 50 25.84 & +26 09 43.41 & 37.82 \\
17609 & 2015-03-20 & 13 49 29.66 & +26 02 00.67 & 51.06 & 19939 & 2016-11-11 & 13 47 15.91 & +26 46 04.06 & 25.89 \\
17610 & 2015-04-01 & 13 50 23.05 & +27 02 20.10 & 53.40 & 19944 & 2017-05-02 & 13 46 28.32 & +26 44 58.65 & 25.43 \\
17611 & 2015-05-03 & 13 48 47.60 & +27 07 29.04 & 50.92 & 19946 & 2016-11-30 & 13 46 43.44 & +26 53 58.97 & 30.83 \\
17612 & 2016-12-03 & 13 50 02.02 & +26 24 41.67 & 55.27 & 19964 & 2016-12-25 & 13 48 52.79 & +27 03 28.99 & 23.80 \\
17613 & 2015-08-30 & 13 51 17.12 & +26 29 05.69 & 52.34 & 19972 & 2017-04-08 & 13 47 56.31 & +26 51 32.07 & 32.13 \\ 
17614 & 2015-11-25 & 13 47 52.08 & +26 07 38.81 & 55.20 &  &  &  & &  \\
\hline 
\end{tabular}
\end{table*}

\begin{table}
\caption{Particulars of the \textit{Suzaku} XIS observations of Abell\,1795}
\label{tab:suzaku_observations}
\centering
\addtolength{\tabcolsep}{-2.1pt} 
\begin{tabular}{ccccc} 
\hline\hline        
Obs. ID & Obs. date & R.A. & Decl. & $t_{\rm exp} \rm (ks)$ \\    
\hline    
800012050 & 2005-12-12 & 13 48 53.45 & +26 12 00.4 & 40.12 \\
804082010 & 2009-06-28 & 13 49 55.27 & +26 52 45.5 & 40.03 \\
804083010 & 2009-06-29 & 13 47 52.15 & +26 18 51.5 & 38.36 \\
804084010 & 2009-06-26 & 13 47 17.76 & +26 41 42.0 & 35.48 \\
800012030 & 2005-12-11 & 13 48 53.52 & +26 59 58.2 & 30.64 \\
800012040 & 2005-12-11 & 13 48 53.47 & +26 24 02.5 & 26.07 \\
800012020 & 2005-12-10 & 13 48 53.30 & +26 47 57.5 & 24.41 \\
804085010 & 2009-06-27 & 13 47 59.40 & +26 35 24.7 & 24.27 \\
800012010 & 2005-12-10 & 13 48 53.78 & +26 36 03.6 & 13.10 \\
\hline
\end{tabular}
\end{table}

X-ray analysis of the intracluster medium (ICM) permeating galaxy clusters is an important diagnostic tool for cosmology, the history of cluster growth, and large-scale structure formation in general.
In the related studies, the hot, tenuous ICM gas is commonly treated as a homogeneous medium, but unresolved, small-scale density inhomogeneities, referred to as ICM clumps, may complicate this picture.
The presence of clumps is predicted by simulations \citep{2011ApJ...731L..10N,2013MNRAS.432.3030R}, and attributed primarily to internal gas motions in the clusters' virialization region ($r>r_{500}$\footnote{$r_{500}$ ($r_{200}$) is the radius within which the mean enclosed matter density is equal to $500$ ($200$) times the critical density of the Universe at the cluster's redshift.}) caused by accretion and mergers.
Accordingly, the degree of clumping is predicted to increase with the cluster radius and becomes significant at $r>r_{\rm 200}$.
Unresolved clumps may contribute to the ICM's density profile, which has the net effect of introducing a systematic bias in the measured ICM masses.

The direct detection of individual clumps, however, is yet to be realized in the observational data.
As these density fluctuations occupy only $\sim\!1\%$ of the cluster volume \citep{2013MNRAS.428.3274Z} and populate the background-dominated regime of the outskirts, their detection is complicated and requires high spatial resolution and high detector sensitivity.
Thanks to its low and well-constrained instrumental background, robust results were achieved with \textit{Suzaku} in the field of ICM physics of the cluster outskirts.
These data probe the cluster outskirts out to, and even beyond the $r_{200}$ virial radius \citep[e.g.][]{2009PASJ...61.1117B,2009MNRAS.395..657G,2009A&A...501..899R,2010ApJ...714..423K,2011Sci...331.1576S,2012PASJ...64...95S,2012MNRAS.427L..45W,2013ApJ...775....4S,2014MNRAS.437.3939U,2021ApJ...908...17Z} and many of them imply deviations from the hydrostatic equilibrium due to the presence of non-gravitational processes.
A main finding of these works was the flattening of the entropy profile at the outer cluster regions suggesting, as a promising explanation, an overestimated density due to unresolved clumping.
Profiles obtained from the stacked \textit{ROSAT} data of $31$ galaxy clusters also support this scenario \citep{2012A&A...541A..57E}.
Alternative explanations suggest, for example, non-equilibrium electrons \citep{2010PASJ...62..371H,2011PASJ...63S1019A} or non-thermal pressure support \citep{2012ApJ...758...74B,2018A&A...614A...7G} in the outskirts.

\textit{Chandra} observations of cluster outskirts, although they provide the much-needed angular resolution, are sparser.
This is because an analogous study requires ultra-deep \textit{Chandra} exposures of bright nearby clusters, especially their outskirts, to overcome the high instrumental background with a reasonable number of pointings, which limits the list of possible sources.
Up to now, the outskirts of only two nearby galaxy clusters, Abell\,1795 and Abell\,133, were explored by \textit{Chandra} in detail. 
These observations provide full azimuthal coverage of the clusters' virialization region extending beyond $r_{200}$.
Abell\,133 is the subject of a previous clumping study by \cite{2014MNRAS.437.1909M}, and a companion paper of Zhu et al. 2023 (submitted), while in this paper we concentrate on the outskirts of Abell\,1795.

Abell\,1795 is a bright galaxy cluster with an X-ray flux of $F_{0.1-2.4,\textrm{keV}} = 6.8 \times 10^{-11} \, \rm erg \, s^{-1} \, cm^{-2}$\,\citep{1998MNRAS.301..881E} and a temperature of $kT = 6.1$\,keV\,\citep{2006ApJ...640..691V}.
It is covered by \textit{Chandra} observations out to a projected radius of $\gtrsim$\,$40\,\arcmin$, while $r_{200}$ of Abell\,1795 is $\sim$\,$26\arcmin$ \citep{2009PASJ...61.1117B}.
In these \textit{Chandra} data, we resolve and localize potential ICM inhomogeneities, then investigate their origin individually.
With the set of available \textit{Suzaku} observations, we complement our study with spectral analysis.
Benefiting from the source list obtained from the high-resolution \textit{Chandra} data, we extract thermodynamic profiles and probe the impact of resolved and unresolved ICM inhomogeneities.
A surface brightness profile of diffuse gas and a log$N-$log$S$ plot of point sources are also built.
We investigate deviations in the surface brightness profile introduced by \textit{Chandra}'s and \textit{Suzaku}'s different angular resolutions and measure azimuthal variations in four directions.
The projected profiles obtained in this study extend beyond, while the deprojected profiles extend out to $r_{200}$.

This paper is organized as follows.
In Section\,\ref{sec:dataanalysis}, we describe the data reduction and data analysis for both \textit{Chandra} and \textit{Suzaku} observations; in Section\,\ref{sec:results}, we report the methods and results; in Section\,\ref{sec:discussion}, we assess the clumping bias and elaborate further on its influence, and discuss surface brightness profile features.

Throughout the paper, we adopt a standard $\Lambda$CDM cosmology with $H_{0}\,=\,70\, \rm km^{-1} \, s^{-1} \,Mpc^{-1} $, $\rm\Omega_M = 0.3$, and $\rm\Omega_{\Lambda} = 0.7$. 
The redshift of Abell\,1795 is $z=0.0622$\,\citep{2006ApJ...640..691V}, where the 
projected scale of $1'$ corresponds to $71.9$\,kpc.
The hydrogen column density towards Abell\,1795 is $N_{\textrm{H}} = 1.0\times 10^{20}\,\rm cm^{-2}$ \citep{2013MNRAS.431..394W}.
Errors are quoted at the $1$$\sigma$ level unless otherwise stated.

\section{Observations and data reduction} \label{sec:dataanalysis}

\begin{figure*}
\centering
{\includegraphics[width=1\textwidth]{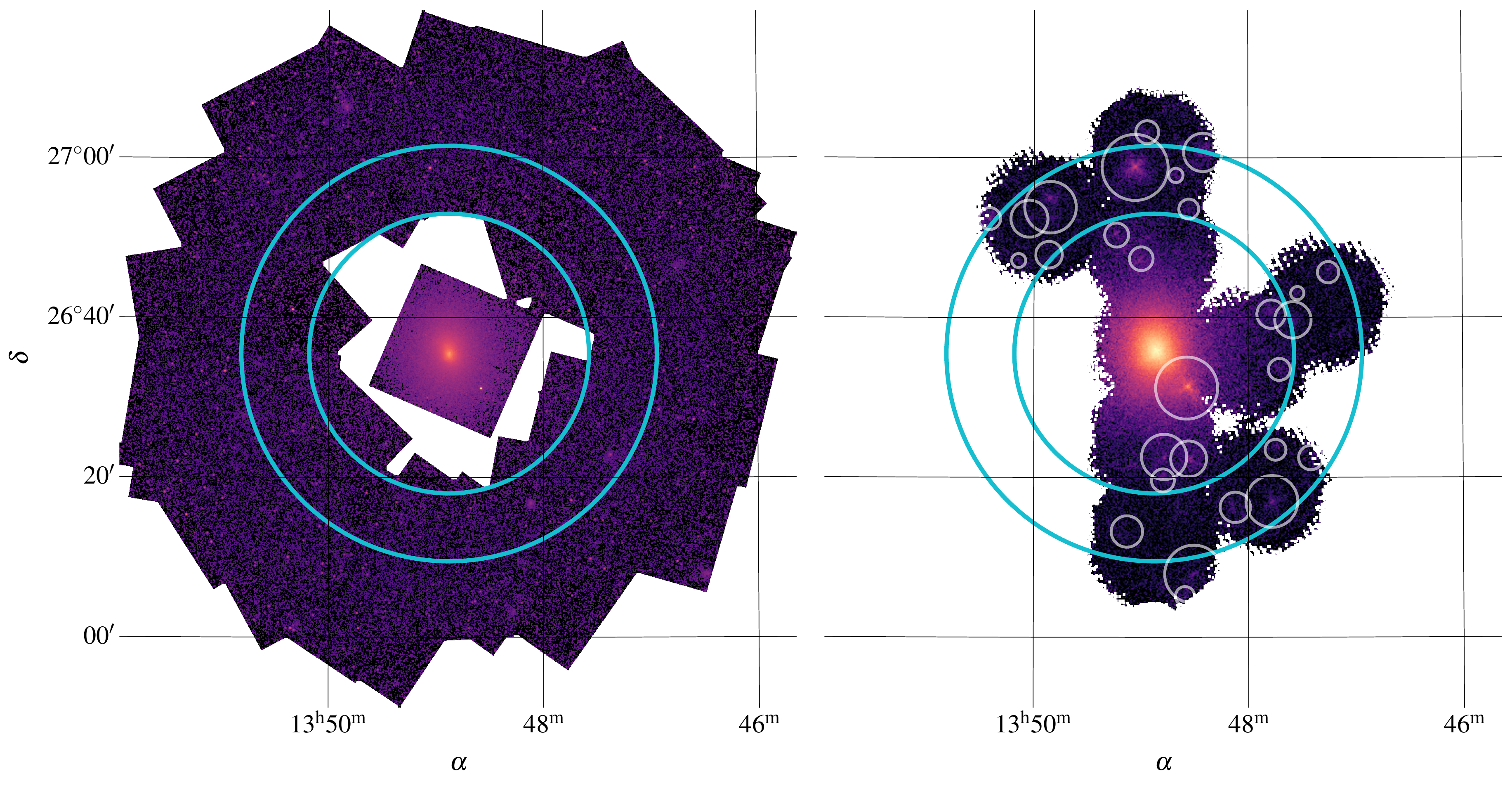}}
\caption{FOV comparison between the exposure and background corrected $0.7-2$\,keV \textit{Chandra} ACIS-I mosaic (left panel) and the exposure and vignetting corrected $0.7-7$\,keV \textit{Suzaku} XIS mosaic (right panel) of Abell\,1795. Images are to scale, but the smoothness and the color scales are not identical due to the different image sensitivities.
The location of $r_{500}$ and $r_{200}$ is marked with the inner and outer circles, respectively, and on the right panel we also marked the \textit{Suzaku} resolved point source regions with white.
}
\label{fig:mosaic}
\end{figure*}

\subsection{\textit{Chandra}}
\label{sec:chandra}
In the imaging analysis, we utilized primarily \textit{Chandra} observations of the cluster outskirts around the galaxy cluster Abell\,1795.
For this, 49 observations were downloaded from the public archive.
Out of the $49$ observations, $1$ observation (observation ID $17228$) points towards the cluster center, while the remaining $48$ observations cover the periphery of the cluster
within $\sim$\,$40\,\arcmin$, in an annular region (Figure\,\ref{fig:mosaic}, left panel).
The observations were taken with the ACIS-I {(Advanced CCD Imaging Spectrometer)} array in very faint (VFAINT) telemetry mode with a total exposure time of $2.26\,\rm Ms$ (Table\,\ref{tab:observations}).
Although Abell\,1795 is a frequently observed calibration source, calibration observations are not used in the {main} analysis, as they map the cluster center primarily at low/high \texttt{CHIPY} pixels {leading to PSFs larger than the optimal.} 
Data reduction was carried out using standard CIAO tools (version 4.13) and the CALDB version 4.9.2.1.

First, we reprocessed the data with \texttt{chandra\_repro} to apply the same calibration correction for each observation and to clean the ACIS particle background for VFAINT observations. 
Then, we removed observation periods with high background count rates caused by flares. 
Flare filtering was carried out using the \texttt{deflare} script with the \texttt{lc\_clean} routine.
We note that ACIS-I observations are not particularly sensitive to high background periods, and thus, the exposure times are not affected significantly.
On average, $<\!8\%$ of the {exposure time was} removed from the data.

Next, we constructed two sets of exposure maps for each observation to correct for vignetting and variations of quantum efficiency across the CCD.
For tasks involving point sources, we assumed an absorbed power-law spectrum with a slope of $\Gamma = 1.6$, typical for AGN \citep[e.g.][]{2000MNRAS.316..234R,2005A&A...432...15P}.
For the analysis of ICM emission, we assumed a \textit{phabs(apec)} spectrum {(in \texttt{Xspec}\,\citep{1996ASPC..101...17A} terminology)} with a temperature of $kT = 6.1\,\rm keV$ and an abundance of $0.2$ solar.
Assigning these spectral models analogously, we built spectrally weighted point spread function (PSF) maps for each observation with $90\%$ encircled counts fraction (ECF).

Finally, individual observations were co-added to obtain a mosaic image of the cluster (Figure\,\ref{fig:mosaic}, left panel).
To this end, we used the \texttt{merge\_obs} routine with the default `psfmerge' setting, which takes the minimum ECF at each pixel.
In addition, to improve the cluster--background flux ratio of the observations, we applied an energy filter of $0.7-2$\,keV.
Except for point source detection (Section\,\ref{sec:sourcedetection}), this filter was used in the analysis.

\subsubsection{Background subtraction}  \label{sec:backgroundsubtraction}

Background subtraction in point source analysis was carried out using local background regions.
A local background, however, cannot be determined when probing ICM inhomogeneities, due to the ICM's extended nature.
It is a common practice to model the ICM background using the ACIS blank-sky dataset.
These observations, however, do not take into account the anisotropy of the cosmic X-ray background (CXB), and resolve different CXB fractions due to different exposure times.
Therefore, their use is not optimal for background-critical objects, such as clumps.
When probing the ICM clumping, we, therefore, relied on the ACIS stowed background dataset instead.

ACIS stowed background observations were collected with ACIS in the ``stowed'' position, i.e., with HRC {(High Resolution Camera)} in the focal plane.
Thus, it only includes the instrumental background produced by the solar wind \citep{2014A&A...566A..25B}, which has a constant spectral shape with varying normalization.
Stowed background observations were taken between $2002$ and $2012$, while the outskirts of Abell\,1795 were observed from $2015$.
We, therefore, utilized only the latest stowed data available in CALDB obtained between $2009$ and $2011$ with a total exposure time of $\sim$\,$240.35$\,ks.

Stowed data reduction was carried out using the following steps.
First, we applied VFAINT filtering then we reprocessed the data with \texttt{acis\_process\_events}.
To account for the correct normalization between the stowed and cluster data, we followed the steps discussed in \cite{2006ApJ...645...95H}.
Specifically, we calculated normalization factors from the count rate ratios between the cluster observation and the corresponding stowed data in the $9.5-12$\,keV band, where the effective area of ACIS is negligible.
Thus, with no astrophysical sources producing counts on the detector, the instrumental background remains the only contributor.
In addition, we also corrected the readout artifacts of ACIS-I.
We produced a readout background for each observation using the \texttt{make\_readout\_bkg} routine\,\citep{2000ApJ...541..542M}.
The appropriate scaling was then applied in the same manner as for the stowed data.
Finally, the sum of the scaled stowed and readout background was subtracted from each observation.

The X-ray background, besides the instrumental component, includes a sky component.
It can be subdivided into the CXB produced by the unresolved population of AGN \citep{2001ApJ...551..624G,2003ApJ...583...70M,2006ApJ...645...95H} and the Galactic foreground produced by the Local Hot Bubble (LHB) and the Galactic halo and disk \citep{1995ApJ...454..643S,2000ApJ...543..195K}.
These components are constant in time, and spatial variations of the foreground can be neglected within the Abell\,1795 FOV.
Therefore, during imaging analyses, a constant sky background level is assumed.

\subsubsection{Point source detection} \label{sec:sourcedetection}

Point sources in the field of Abell\,1795 were identified with the CIAO \texttt{wavdetect} tool.
We searched for point sources in the broad band ($0.5-7$\,keV), on the total counts mosaic image.
For the wavelet function, we specified pixel scales employing the ``$\sqrt{2}$ sequence'' from $\sqrt{2}$ to $32\sqrt{2}$, which selects both point-like and extended features.
A signal threshold of $<$\,$10^{-8}$ was supplied for \texttt{wavdetect} to reduce the probability of spurious detections, and detection of faint ICM features, such as clumps.
We set the `ellsigma' parameter to $5$ to produce sufficiently large source cells.
In most cases, this ensures the inclusion of $\sim$\,$100\%$ of the source counts, while residual source counts are not expected to significantly contribute to the ICM's diffuse emission.
PSF information and effective area at each sky position were also provided for \texttt{wavdetect}.
Finally, 1516 sources were identified.
We used this source list for source exclusion when carrying out analyses described in Section\,\ref{sec:clumping} and \ref{sec:sb}.

To build the $\log{N}-\log{S}$ plot for sources in the field of Abell\,1795, point source detection was carried out separately in the soft band ($0.5-2$\,keV) with the \texttt{wavdetect} parameters left unchanged.
This run thus revealed $1137$ sources.
\begin{figure}
\centering
\includegraphics[width=.525\textwidth]{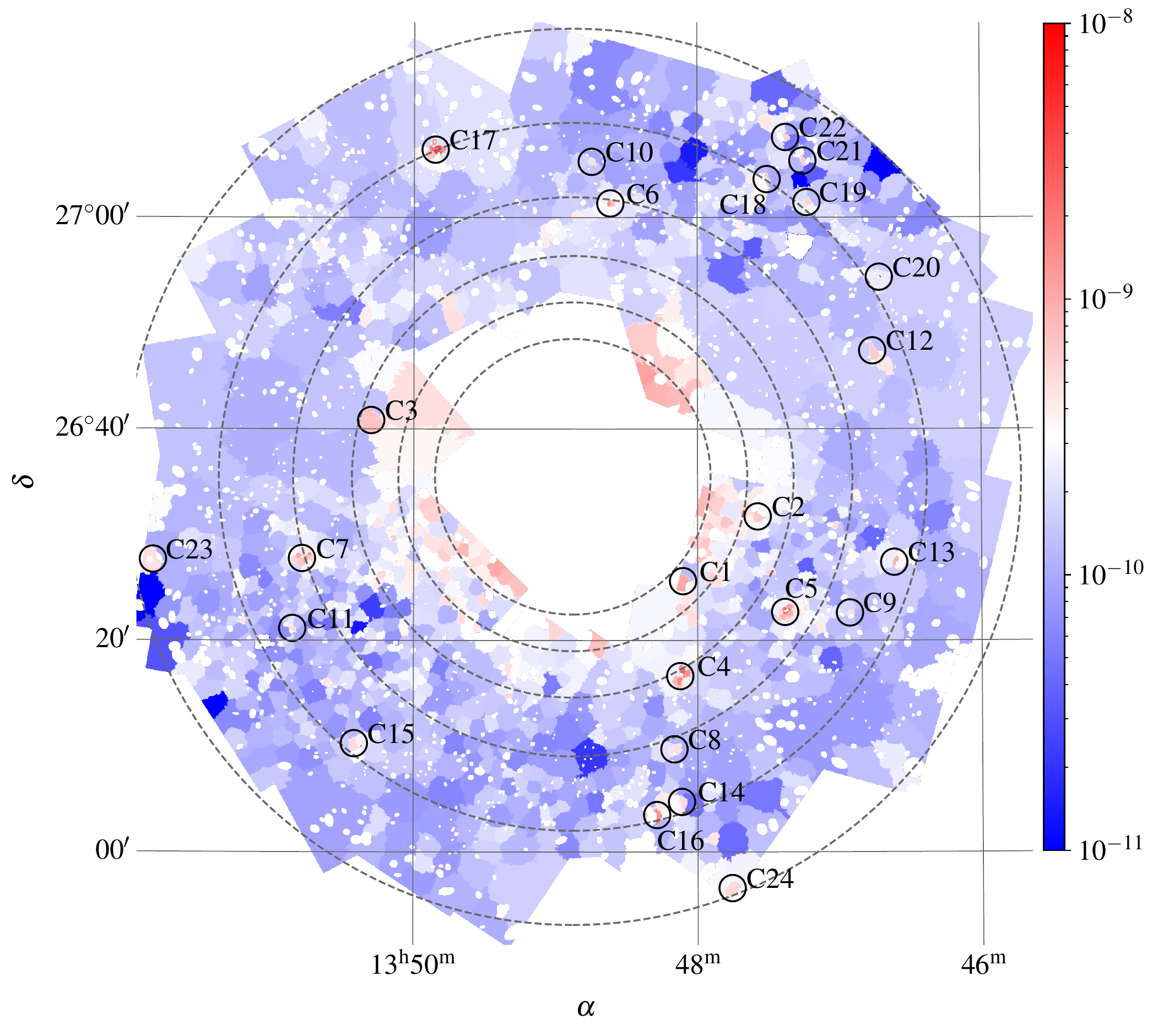}
\caption{Adaptively-binned image of Abell\,1795 outskirts in the $0.7-2$\,keV energy band; the image is instrumental background- and exposure-corrected. Each bin has a SNR of $\gtrsim 3$, and {a surface brightness} as indicated by the colorbar in photon\,$\rm / cm^{2} /\, s\, /\mathrm pixel$ units. 
Small white ellipses scattered throughout the field represent excluded point sources.
We used this image to measure the {surface brightness} distribution within each annulus {(dashed-line circles)} and to isolate outliers that may originate from ICM inhomogeneities. {Outlier positions are shown with small circles marked from C1 to C24; for the actual size and shape of outliers see Figure\,\ref{fig:stamps}.}}
\label{fig:contbin}
\end{figure}

\begin{table*}
\caption{Surface brightness distribution parameters from the adaptively binned image in $5$ annuli,  including the number of $>\!2\,\sigma$ outlier bins ($N_{\rm outliers}$), the Fisher-Pearson skewness coefficient of the distributions, and the {$b_X = \mathrm{SB_{mean}/SB_{median}}$} emissivity bias (Section\,\ref{sec:fluxbias}) {related to clumping.}}
\label{tab:histograms}
\centering
\addtolength{\tabcolsep}{-1pt} 
\begin{tabular}{cccccccccc} 
\hline\hline        
\multirow{2}{*}{Annulus} & 
\multirow{2}{*}{$r_{\rm in}$} & 
\multirow{2}{*}{$r_{\rm out}$} & 
\multirow{2}{*}{$N_{\rm bins}$} & 
\multirow{2}{*}{{$N_{\rm outliers}$}}  &
{$\mathrm{SB}_{\rm fit\,centroid,\,0.7-2\,\rm keV}$} &
$\mathrm{SB}_{\rm mean,\,0.7-2\,\rm keV}$ & 
$\mathrm{SB}_{\rm median,\,0.7-2\,\rm keV}$ & 
\multirow{2}{*}{Skewness} & 
\multirow{2}{*}{$b_X$} \\
 &  &  &  &  & \multicolumn{3}{c}{{($10^{-10}$\,photon\,$\rm cm^{-2} \, s^{-1} \, \rm arcsec^{-2}$)}}  & &  \\
\hline               
1 & $13.03\arcmin$ & $16.49\arcmin$ & $76$ & {$1$} & {$3.96$} & $4.11^{+0.22}_{-0.20}$ & $3.71^{+0.48}_{-0.36}$ & $0.82\pm0.34$ & $1.11 \pm 0.12$\\
2 & $16.49\arcmin$ & $20.88\arcmin$ & $144$ & {$2$} & {$2.53$} & $2.81^{+0.18}_{-0.12}$ & $2.40^{+0.11}_{-0.08}$ & $4.01\pm1.53$ & $1.17 \pm 0.07$\\
3 & $20.88\arcmin$ & $26.44\arcmin$ & $256$ & {$3$} & {$1.99$} & $3.64^{+0.89}_{-0.49}$ & $1.83^{+0.09}_{-0.08}$ & $9.38\pm2.08$ & $1.99 \pm 0.19$\\
4 & $26.44\arcmin$ & $33.47\arcmin$ & $368$ & {$11$} & {$1.54$} & $2.96^{+0.62}_{-0.39}$ & $1.50^{+0.05}_{-0.04}$ & $9.43\pm1.87$ & $1.99 \pm 0.17$\\
5 & $33.47\arcmin$ & $42.38\arcmin$ & $238$ & {$7$} & {$1.60$} & $2.73^{+0.94}_{-0.44}$ & $1.47^{+0.06}_{-0.06}$ & $11.47\pm2.34$ & $1.85 \pm 0.26$\\
\hline
\end{tabular}
\end{table*}

\subsection{Suzaku}
\label{sec:suzaku}
Due to its low and stable instrumental background (non-X-ray background, NXB), we relied on data from the \textit{Suzaku} X-ray Imaging Spectrometer (XIS) to probe the thermodynamic profiles of the cluster outskirts.
{The spectral analysis was refined by incorporating} the \textit{Chandra} point source list to bypass \textit{Suzaku}'s low angular resolution, which causes unresolved point sources to add to the uncertainty of the measured properties of the diffuse emission.
In addition, the surface brightness profile of Abell\,1795 was also extracted from the \textit{Suzaku} observations to compare with the Chandra profile. 

The analyzed \textit{Suzaku} observations of Abell\,1795 are listed in Table\,\ref{tab:suzaku_observations}, while the full band mosaic is presented in the right panel of Figure\,\ref{fig:mosaic}.
Note that the \textit{Suzaku} data does not provide full azimuthal coverage, and has a smaller radial extent ($\sim$\,$30 \arcmin$) than the Chandra observations.

The data reduction of XIS observations follows the procedure detailed in previous works \citep{2013ApJ...775....4S,2014MNRAS.437.3939U,2021A&A...652A.147Z}.
To summarize, we utilized the cleaned event files produced by the standard screening process available in the HEAsoft package (version 6.26)\,\footnote{\url{https://heasarc.gsfc.nasa.gov/docs/suzaku/analysis/abc/node9.html}} with all the recommended XIS screening criteria applied.
Observation periods with low geomagnetic cut-off rigidity (COR\,$\leq 6$\,GV) were removed.
The vignetting correction was carried out by utilizing ray-tracing simulations of a flat field image.
{Based on the solar proton flux curve measured by the WIND spacecraft's solar wind experiment instrument, we also ensured that no contamination from the solar wind charge exchange (SWCX) during the clean exposure occurred.}

For the imaging analysis, $0.7-2$\,keV images from XIS\,0, 1, and 3 detectors were extracted along with the corresponding NXB obtained from the night--Earth data.
To reduce the effect of systematic uncertainties caused by vignetting, a $30\arcsec$ width region around the detector edges, and pixels with an effective area less than half of the on-axis value were removed from the background-subtracted images (Figure\,\ref{fig:mosaic}, right panel).
We then masked the point sources as detailed in Section\,\ref{sec:suzaku-cxb} and \ref{sec:sbazimuthallyavg}.

Spectral extraction was carried out with \texttt{XSELECT} {for $5$ concentric annuli with widths of $5\arcmin$ centered around the cluster and measured from $5\arcmin$ to $30\arcmin$,} thus, excluding the cluster core.
The Redistribution Matrix Files (RMF) and the Auxiliary Response Files (ARF) were produced in parallel with the mission-specific FTOOLS \texttt{xisrmfgen} and \texttt{xisimarfgen} commands, respectively.
Finally, point sources were removed using three different approaches relying both on the \textit{Suzaku} and the \textit{Chandra} source list, which we explain in Section\,\ref{sec:suzaku-cxb}.
During the spectral fit of the ICM emission (Section\,\ref{sec:thermo_prof}), we accounted for the foreground, the CXB, and the NXB contribution, which we detail in the following subsections.

\subsubsection{Modeling the foreground emission}\label{sec:suzaku-foreground}

The contribution of the LHB and Galactic components (Section\,\ref{sec:backgroundsubtraction}) were modeled using the mission-specific \texttt{X-Ray Background Tool}\,\citep{2019ascl.soft04001S} on \textit{ROSAT} \textit{All-sky Survey} (RASS) diffuse background maps, which we downloaded for an annulus between $1.5 \rm r_{200}$ and $2.5 \rm r_{200}$ around Abell\,1795.
The extracted spectrum was fitted in \texttt{Xspec} with a $tbabs \times(apec + power$-$law) + apec$ model, where the unabsorbed and absorbed \textit{apec} describes the LHB and the Galactic emission, respectively, while the absorbed \textit{power-law} represents the fraction of the unresolved CXB in the RASS data.
With the CXB parameters fixed at values adopted from \cite{2000ApJ...543..195K}, we obtained plasma temperatures of $0.093\,\rm keV$ and $0.197\,\rm keV$, and \textit{apec} normalizations of $6.82 \times 10^{-4} \,\rm cm^{-5}$ and $1.39 \times 10^{-3} \,\rm cm^{-5}$ for the LHB and Galactic emission, respectively.

The $\sim$\,$0.6$\,keV hot foreground component originating from the Galactic Plane \citep{2009PASJ...61..805Y} is not important in the case of Abell\,1795 because of the cluster's relatively large galactic latitude ($b=+77.1553$).
{We confirmed this by adding a $0.6\,\rm keV$ \textit{apec} component to our model, and finding that its normalization could not be constrained, so the $\chi^2$ statistics did not improve.}

\subsubsection{Point source removal and modeling the CXB}\label{sec:suzaku-cxb}
The CXB contribution was estimated depending on the treatment of point sources.
In the first approach (hereafter referred to as the first round or R1), only \textit{Suzaku} resolved sources were removed from the data.
These sources were identified on the broad-band \textit{Suzaku} mosaic  by \texttt{wavdetect} in a manner similar to that described in Section\,\ref{sec:sourcedetection}.
The resulting source list collects $29$ objects, which are marked on the right panel of Figure\,\ref{fig:mosaic}.
The flux of the remaining, unresolved sources was estimated from the point source excluded spectrum of the outermost annulus, where the ICM temperature is expected to drop below $2.5$\,keV.
{We fit the spectra in \texttt{XSpec} with a power-law in the $4-7$\,keV band, where the CXB emission dominates, and obtained a best-fit photon index of $\Gamma = 1.5$ and a normalization of $1 \times 10^{-3}\, \rm photons/keV/cm^2/s$ at $1\,\rm keV$.}

In the second round (R2) of spectral extraction, we accounted for the limits on \textit{Suzaku}'s angular resolution by extending the R1 point source list with the \textit{Chandra} broad-band source catalog (Section\,\ref{sec:sourcedetection}) {down to a $2-8$\,keV flux threshold of $\sim$$1.2\times10^{-14} \, \rm erg \,cm^{-2} \, s^{-1} $.
Based on the hard band log$N-$log$S$ plot built in the same way as described in Section\,\ref{sec:logn-logsplot}, $95\!\%$ of the sources above this flux threshold are resolved in the current set of \textit{Chandra} data.}
To account for the broader PSF of \textit{Suzaku}, the radius of each \textit{Chandra} source was increased to $1\arcmin$ as a first approximation.
Note that the spatial resolution of XIS detectors is $\sim\!1.8\,\arcmin$.
For completeness, we also added sources that were identified separately on those calibration observations that cover the data gap in \textit{Chandra} data (Figure\,\ref{fig:mosaic}, left panel).
This step was needed because of the partial overlap between the \textit{Suzaku} pointings and the \textit{Chandra} data gap.
{Processing and merging of the calibration observations were carried out following the standard \texttt{CIAO} procedure, while point source detection was identical to that described in Section\,\ref{sec:sourcedetection}.}
Considering the extended source list, the reduced CXB flux of R2 was estimated from the cluster's $\mathrm{log}\,N - \mathrm{log}\,S$ plot (Section \ref{sec:logn-logsplot}) using \texttt{CXBTools} \citep{2015A&A...575A..37M,Jelle2017}.
This yielded a CXB flux of $1.14 \times 10^{-11} \, \rm erg \, cm^{-2} \, s^{-1} \, deg^{-2}$ in the $2-8$\,keV band.
To this flux, we added the $50\%$ of the total $2-8$\,keV flux, $1.55 \times 10^{-13} \, \rm erg \, cm^{-2} \, s^{-1} \, deg^{-2}$, of the removed \textit{Chandra} sources to account for excess point source emission beyond source radii of $1\,\arcmin$ {which is below XIS's angular resolution.}
This approach was optimized to secure a FOV area that is sufficient for spectral extraction.
From the estimated CXB flux, a power-law photon index of $\Gamma = 1.4$ and normalization of $8.87 \times 10^{-4}\, \rm photons/keV/cm^2/s$ at $1$\,keV was obtained.

In the third round (R3) of spectral extraction, we probed the influence of ICM clumping, so we further extended the R2 source list with the clump candidate regions C1, C2, C6, C8, C10, and C12. The rest of the candidates are either not covered by \textit{Suzaku} or were already removed in R2 due to the overlap with a point source (Figure\,\ref{fig:contbin}).
Since the clump candidates' contribution to the total resolved source flux is negligible, {and these sources are essentially not part of the CXB,} we used the R2 CXB parameters here.

\begin{table*}
\caption{Clump candidate properties with $N_{\rm bins}$ showing the number of adjacent \texttt{contbin} bins forming a single source {and with $\sigma$ representing the candidate's significance measured in units of the log-normal fit's standard deviation in the host annulus. The possible source of the candidates' emission is also reported following the notation used in NED. For details, see Section\,\ref{sec:individualclumps}.}}
\label{tab:clumpcandidates}
\centering
{\begin{tabular}{ccccccccc}
\hline\hline
\multirow{2}{*}{ID} &
\multirow{2}{*}{RA} &
\multirow{2}{*}{Dec.} &
\multirow{2}{*}{$N_{\rm bins}$} &
\multirow{2}{*}{$\sigma$} &
$F_{0.7-2\,\rm keV}$ &
Area &
\multirow{2}{*}{Hardness ratio} &
Possible\\
& & & & & (photon\,$\rm cm^{-2} \, s^{-1}$) & (arcsec$^{2}$) & &  source\\
\hline
C1 & 13:48:06.18 & +26:25:36.96 & 1 & 2.03 & $(5.30\pm0.78)\times10^{-6}$ & 4992 & $0.68_{-0.11}^{+0.21}$ & $-$\\
C2 & 13:47:34.57 & +26:31:43.49 & 1 & 2.22 & $(1.88\pm0.21)\times10^{-6}$ & 2573 & $0.26_{-0.12}^{+0.52}$ & QSO\\
C3 & 13:50:18.07 & +26:40:50.04 & 1 & 2.23 & $(1.12\pm0.11)\times10^{-5}$ & 13694 & $0.36_{-0.11}^{+0.51}$ & QSO \\
C4 & 13:48:07.42 & +26:16:38.06 & 5 & 3.75 & $(9.02\pm0.58)\times10^{-6}$ & 5766 & $-0.12_{-0.12}^{+0.16}$ & GClstr\\
C5 & 13:47:23.09 & +26:22:41.29 & 9 & 3.65 & $(7.42\pm0.36)\times10^{-6}$ & 5009 & $-0.20_{-0.09}^{+0.11}$ & G (A1795 member)\\
C6 & 13:48:36.82 & +27:01:20.64 & 2 & 3.27 & $(1.67\pm0.17)\times10^{-6}$ & 1388 & $-0.53_{-0.15}^{+0.19}$ & G\\
C7 & 13:50:47.17 & +26:27:46.04 & 6 & 2.82 & $(4.90\pm0.26)\times10^{-6}$ & 6323 & $-$ & GGroup\\
C8 & 13:48:09.90 & +26:09:42.93 & 1 & 2.06 & $(1.86\pm0.17)\times10^{-6}$ & 3559 & $-0.02_{-0.33}^{+0.59}$ & QSO\\
C9 & 13:46:55.73 & +26:22:37.15 & 1 & 2.10 & $(5.31\pm0.75)\times10^{-7}$ & 1027 & $-$ & QSO\\
C10 & 13:48:44.71 & +27:05:16.77 & 1 & 2.09 & $(7.93\pm0.95)\times10^{-7}$ & 1482 & $-0.10_{-0.28}^{+0.54}$ & QSO\\
C11 & 13:50:51.15 & +26:21:09.22 & 1 & 2.09 & $(5.95\pm0.74)\times10^{-7}$ & 1189 & $0.58_{-0.09}^{+0.27}$ & QSO(?)\\
C12 & 13:46:45.91 & +26:47:24.09 & 1 & 2.55 & $(2.26\pm0.19)\times10^{-6}$ & 3401 & $-$ & GClstr\\
C13 & 13:46:36.96 & +26:27:24.04 & 1 & 3.18 & $(1.32\pm0.20)\times10^{-6}$ & 1386 & $0.61_{-0.17}^{+0.17}$ & QSO\\
C14 & 13:48:06.73 & +26:04:44.26 & 1 & 2.20 & $(1.33\pm0.20)\times10^{-6}$ & 2430 & $0.47_{-0.11}^{+0.40}$ & QSO/GClstr\\
C15 & 13:50:25.01 & +26:10:16.26 & 2 & 2.84 & $(2.58\pm0.21)\times10^{-6}$ & 3300 & $0.02_{-0.17}^{+0.31}$ & QSO\\
C16 & 13:48:17.21 & +26:03:29.99 & 2 & 4.17 & $(3.73\pm0.34)\times10^{-6}$ & 2232 & $-0.59_{-0.16}^{+0.30}$ & G/QSO\\
C17 & 13:49:51.01 & +27:06:25.29 & 11 & 4.60 & $(1.36\pm0.06)\times10^{-5}$ & 6352 & $-0.17_{-0.07}^{+0.08}$ & GClstr\\
C18 & 13:47:30.44 & +27:03:37.55 & 1 & 2.85 & $(1.19\pm0.18)\times10^{-6}$ & 1190 & $0.38_{-0.18}^{+0.21}$ & QSO\\
C19 & 13:47:13.59 & +27:01:26.47 & 1 & 2.38 & $(8.84\pm1.33)\times10^{-7}$ & 1559 & $-$ & clump(?)/G \\
C20 & 13:46:42.98 & +26:54:21.56 & 1 & 8.18 & $(1.21\pm0.23)\times10^{-6}$ & 82 & $0.08_{-0.16}^{+0.18}$ & QSO\\
C21 & 13:47:15.25 & +27:05:22.13 & 1 & 2.61 & $(8.82\pm1.47)\times10^{-7}$ & 1326 & $-0.17_{-0.29}^{+0.41}$ & QSO\\
C22 & 13:47:22.56 & +27:07:36.14 & 1 & 2.45 & $(8.11\pm1.28)\times10^{-7}$ & 1429 & $-$ & GClstr(?)\\
C23 & 13:51:50.11 & +26:27:39.34 & 1 & 2.11 & $(6.21\pm0.63)\times10^{-6}$ & 12427 & $-$ & $-$\\
C24 & 13:47:45.51 & +25:56:34.98 & 1 & 2.25 & $(3.13\pm0.45)\times10^{-6}$ & 6222 & $0.68_{-0.13}^{+0.11}$ & $-$\\
\hline
\end{tabular}}
\end{table*}
\subsubsection{Modeling the particle background}\label{sec:suzaku-particle-bg}

The NXB is produced by charged particles and $\gamma$-ray photons hitting the detector.
Its spectrum was extracted with \texttt{xisnxbgen} from the weighted, integrated night--Earth observations taken in a time window of $\pm 150$ days of the Abell\,1795 observations.
Instead of directly subtracting it from the cluster data, we tailored them to our observations by modeling their spectra.
For more details, see \cite{2021A&A...652A.147Z}.

Because of their different design, front-illuminated (FI) CCDs (XIS\,0 and XIS\,3) and the back-illuminated (BI) CCD (XIS\,1) have different sensitivities to the NXB\,\citep{2007SPIE.6686E..0JH,2008PASJ...60S..11T}, resulting in different spectral shapes.
Accordingly, we modeled the FI NXB spectra with a power-law representing the continuum, to which we added $9$ Gaussian emission lines representing the instrumental lines between $1$\,keV and $12$\,keV. In the BI XIS spectral model, we added another broad Gaussian component to account for the continuum bump above $7$\,keV.

We fitted the FI and BI NXB models to the convolution of the NXB and a diagonal RMF for each observation.
During the fit, we fixed the line centroid parameter, adopted from \cite{2008PASJ...60S..11T}, of those instrumental lines, which lie above $5$\,keV, where the contribution of ICM emission of the cluster outskirts is negligible.

{Finally, the NXB normalization factor was constrained during spectral modeling from the $7-12$\,keV source spectra based on similar principles as discussed in Section\,\ref{sec:backgroundsubtraction}.}

\begin{figure}
\includegraphics[trim={.2cm 1.7cm 0 0},clip, width=.46\textwidth]{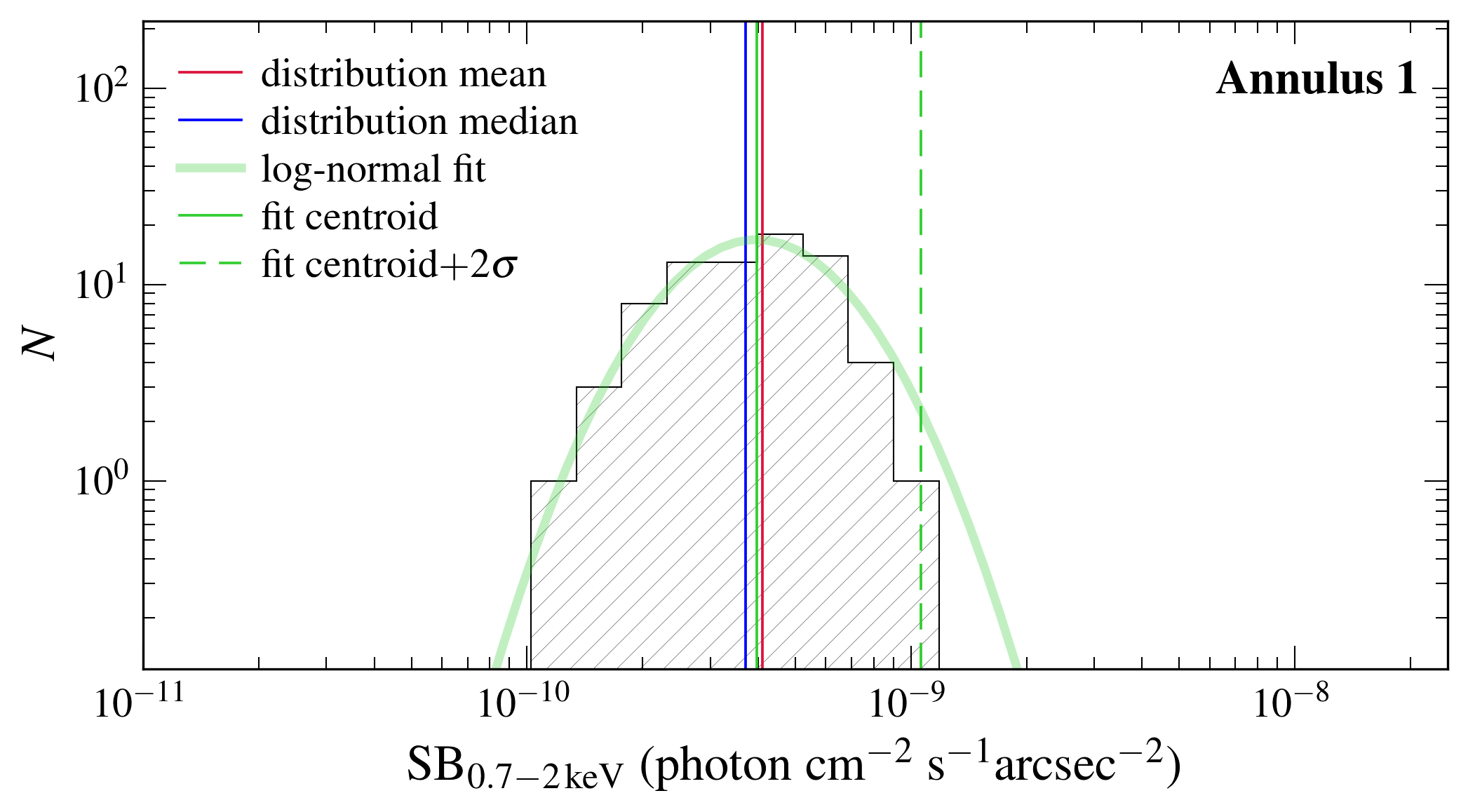}
\includegraphics[trim={.2cm 1.7cm 0 0},clip, width=.46\textwidth]{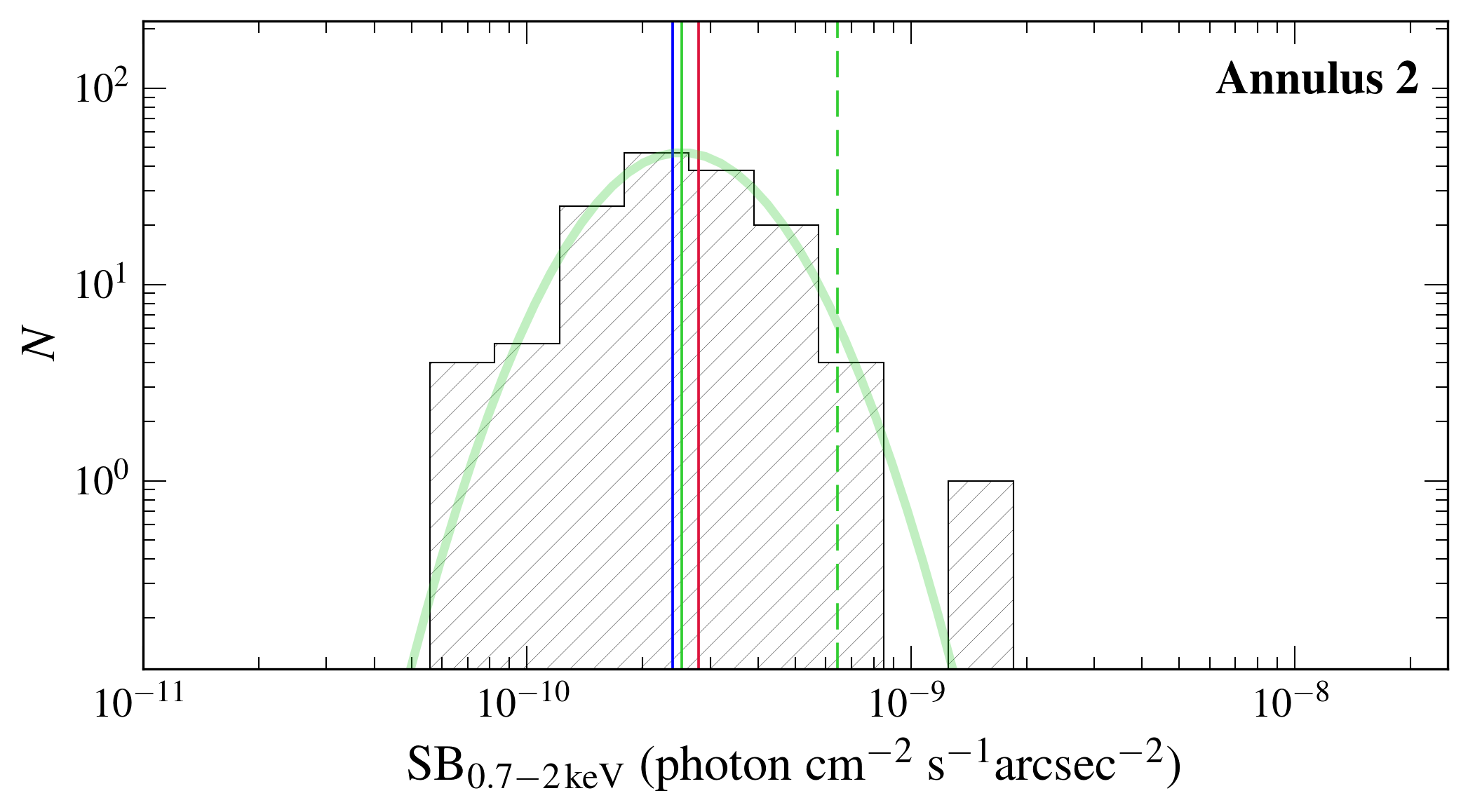}
\includegraphics[trim={.2cm 1.7cm 0 0},clip, width=.46\textwidth]{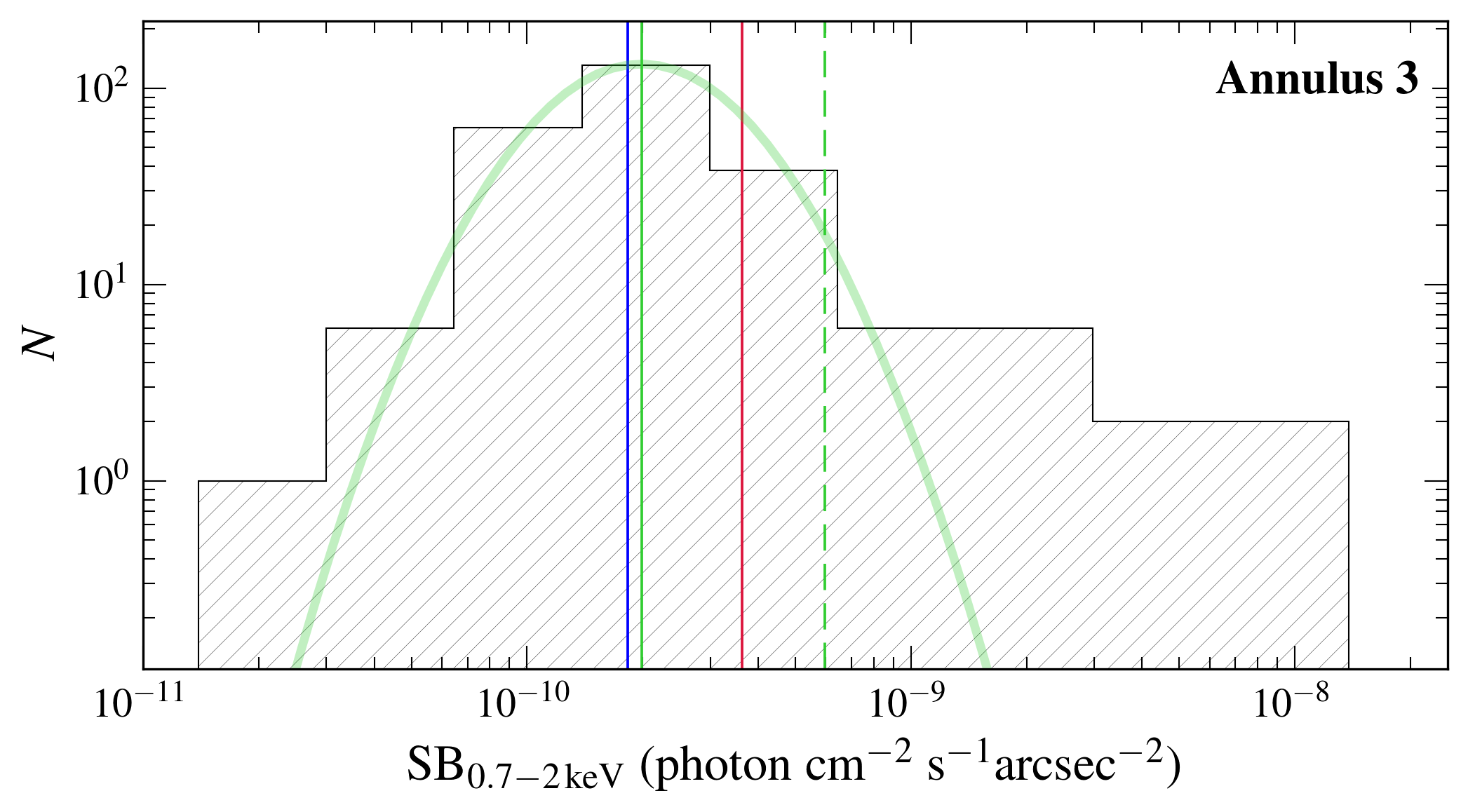}
\includegraphics[trim={.2cm 1.7cm 0 0},clip, width=.46\textwidth]{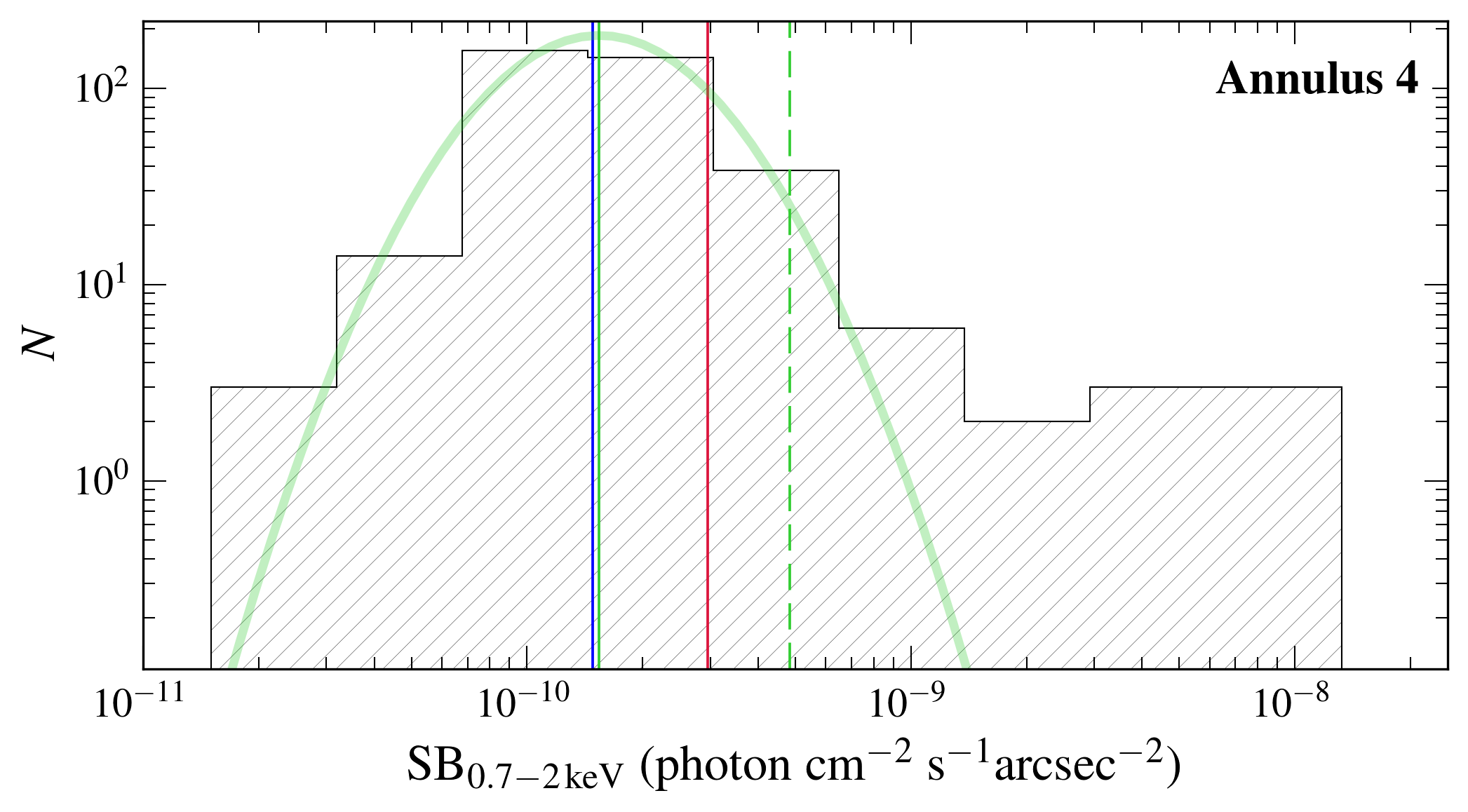}
\includegraphics[trim={.2cm 0 0 0},clip, width=.46\textwidth]{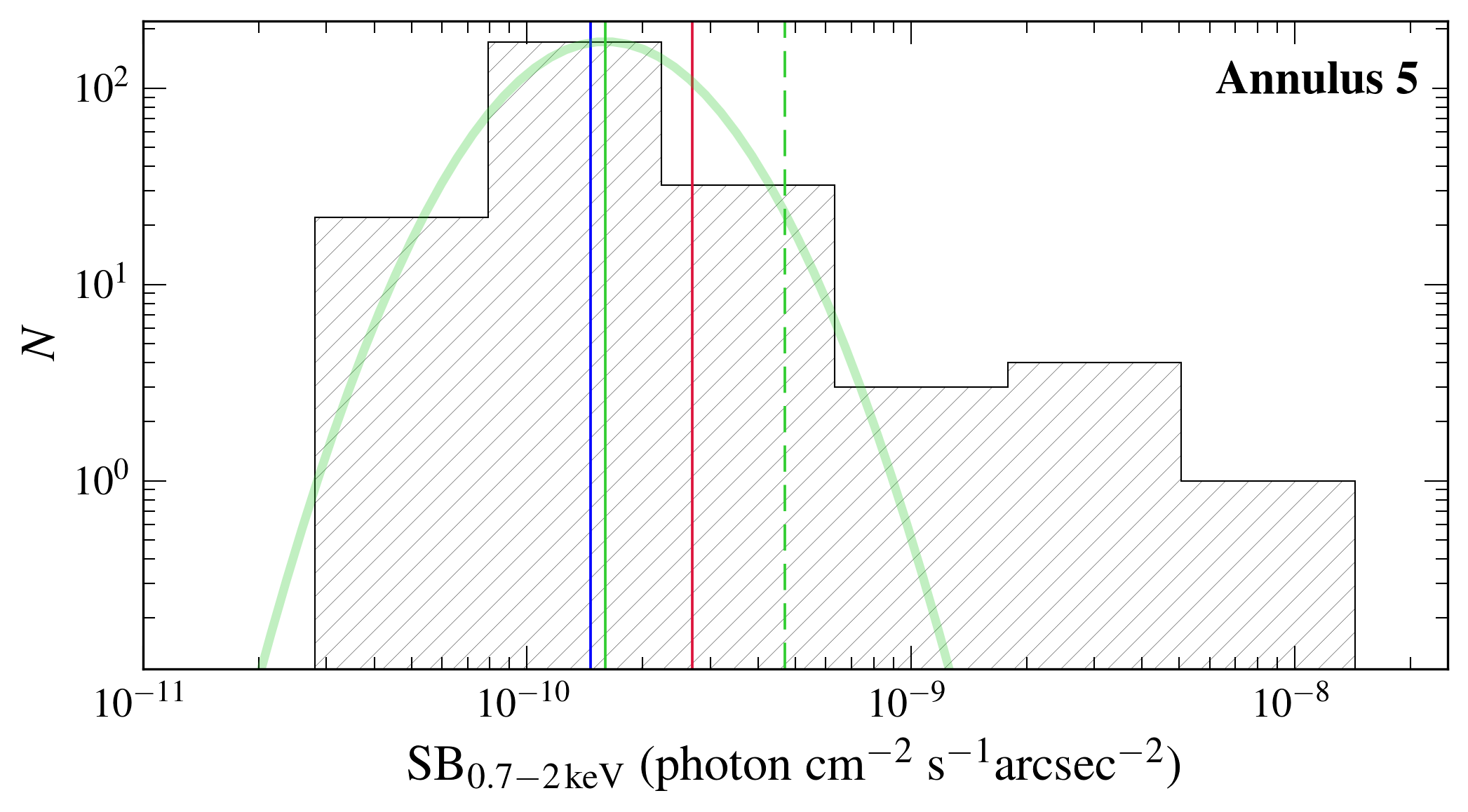}
\caption{$(0.7-2)\,\rm keV$ surface brightness distribution of the ICM in the outskirts of Abell\,1795 obtained in $5$ annuli, which extend out to $\sim$\,$42.5\arcmin$ as listed in Table\,\ref{tab:histograms}.
{Overall, the median {surface brightness} coincides with the peak of the distribution, but the mean, especially at larger radii, is shifted towards high values, which suggests the presence of unresolved gas inhomogeneities \citep{2013MNRAS.428.3274Z}.
We selected the $>\!+2\,\sigma$ outliers of the log-normal fit within each annulus, which resulted in a total of $24$ sources to form our list of clump candidates.}
\label{fig:histograms}}
\end{figure}

\subsubsection{{Systematic uncertainties}}\label{sec:suzaku-syst-uncert}

{Spatial variations across the Suzaku FOV might be present in the foreground X-ray emission, as well as in the CXB (called the cosmic variance), and our spectroscopic measurements are the subject of the corresponding systematic uncertainties.
A coarse measure for the uncertainty is provided here for both background components and demonstrated on the projected temperature profile in Appendix\,\ref{appendix:systematic}}.

{The spatial variation of the diffuse galactic X-ray emission was assessed from RASS spectra extracted in four directions around Abell\,1795 beyond $1.5\,r_{200}$ within $2^{\circ}$ diameter circles.
We fitted each spectra with a $tbabs \times apec$ model and found a $25\!\%$ variance in the best-fit normalizations}.

{The $\sigma^2$ cosmic variance in the expected $B = \int_{0}^{S_{\mathrm{excl}}} \frac{dN}{dS} SdS$ brightness of unresolved X-ray point sources with flux $S < S_{\mathrm{excl}}$ over a given $\Omega$ solid angle can be expressed, following \cite{2009PASJ...61.1117B}, as}
\begin{equation}
\sigma^{2} = \frac{1}{\Omega} \int_{0}^{S_{\rm excl}} \frac{dN}{dS} \times  S^{2} dS,
\end{equation}
{where ${dN}/{dS}$ is the differential number of AGN per unit solid angle and flux, which we approximated with a broken power-law with parameters adopted from \cite{2012ApJ...752...46L}.
For $S_{\mathrm{excl}}$ we adopted the limiting flux of R1 ($\sim\!2.6\times10^{-14} \, \rm erg \,cm^{-2} \, s^{-1}$) and R2 (R3) ($1.2\times10^{-14} \, \rm erg \,cm^{-2} \, s^{-1}$, Section\,\ref{sec:suzaku-cxb}).
Over the $5\,\arcmin-30\,\arcmin$ radial range, we found an average $\sigma / B$ relative cosmic variance of $\sim\!6.5\%$ and $\sim\!5.1\%$ in R1 and R2 (R3), respectively.
As expected, excluding more of the brightest point sources, the primary sources of spatial variations, scales down the cosmic variance.
However, it caused only a modest improvement from R1 to R2 (R3).
It is the result of the relatively large limiting flux applied for R$2$ (Section\,\ref{sec:suzaku-cxb}), which was optimized so that spectral extraction from the outermost annulus remains feasible given the limited coverage and spatial resolution of \textit{Suzaku}, which, in turn, yielded only modest difference in the limiting fluxes between R1 and R2.
}

\section{Results} \label{sec:results}

\begin{figure*}[h!]
\flushleft
\includegraphics[trim=2cm 2cm 2cm 1cm, clip, width=.245\textwidth]{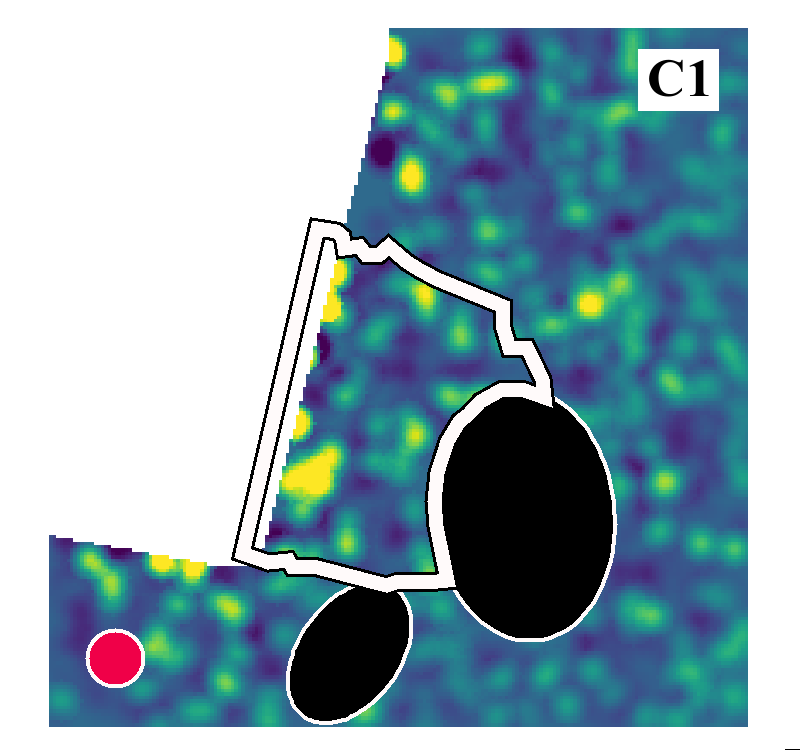}
\hspace{-.1cm}
\includegraphics[trim=2cm 2cm 2cm 1cm, clip, width=.245\textwidth]{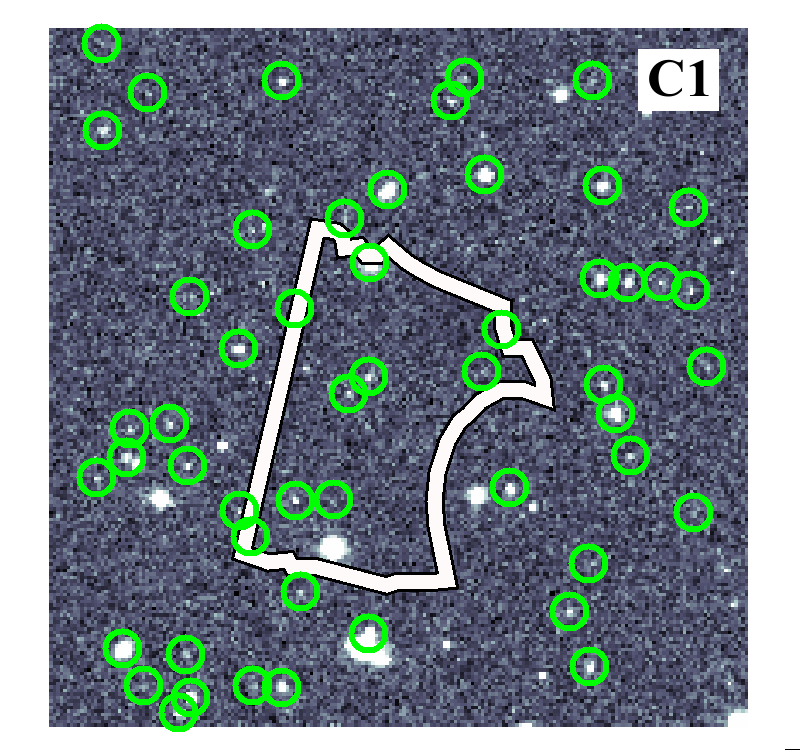}
\hspace{.05cm}
\includegraphics[trim=2cm 2cm 2cm 1cm, clip, width=.245\textwidth]{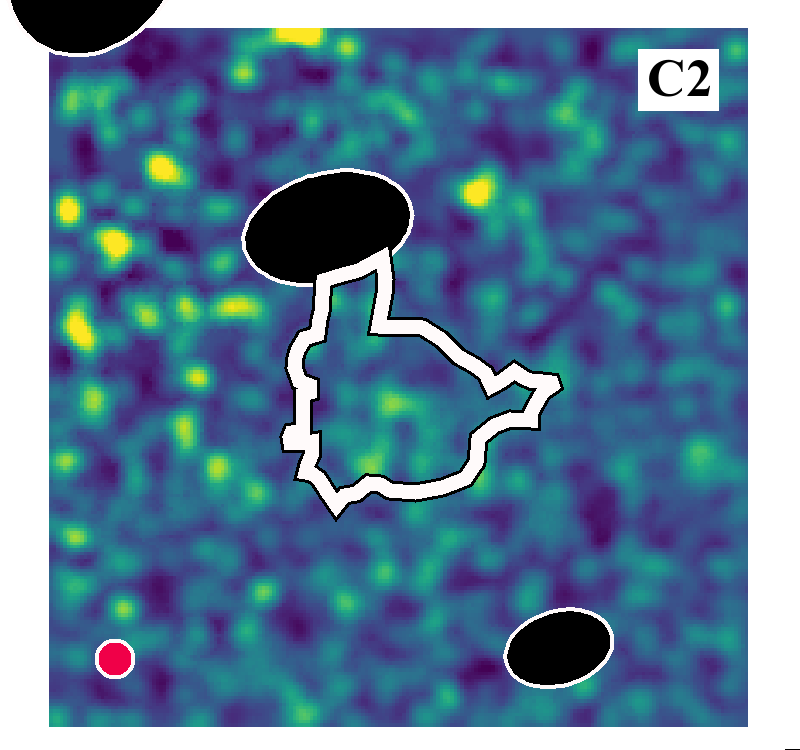}
\hspace{-.1cm}
\includegraphics[trim=2cm 2cm 2cm 1cm, clip, width=.245\textwidth]{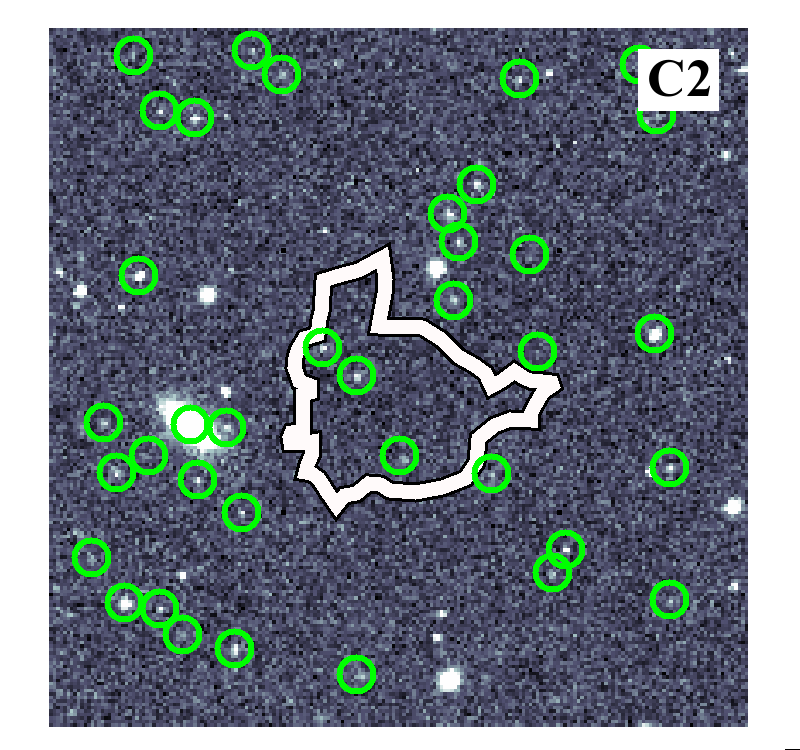} \\
\vspace{.02cm}
\includegraphics[trim=2cm 2cm 2cm 1cm, clip, width=.245\textwidth]{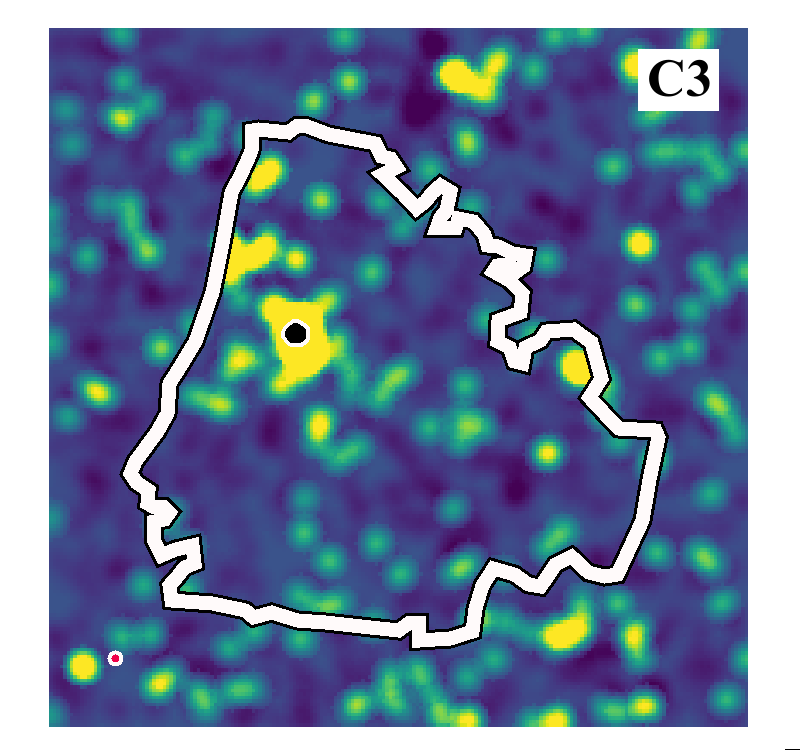}
\hspace{-.1cm}
\includegraphics[trim=2cm 2cm 2cm 1cm, clip, width=.245\textwidth]{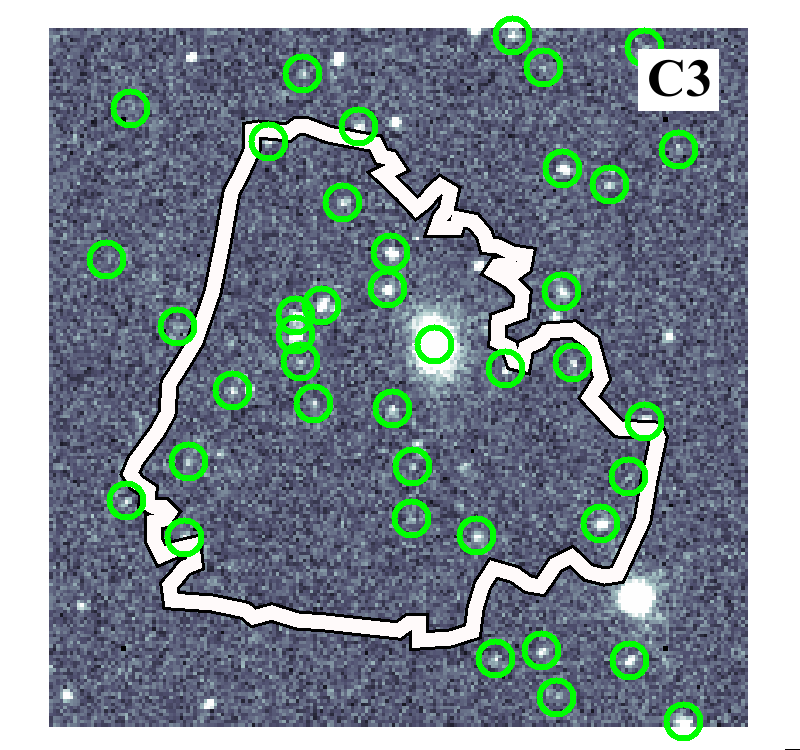}
\hspace{.05cm}
\includegraphics[trim=2cm 2cm 2cm 1cm, clip, width=.245\textwidth]{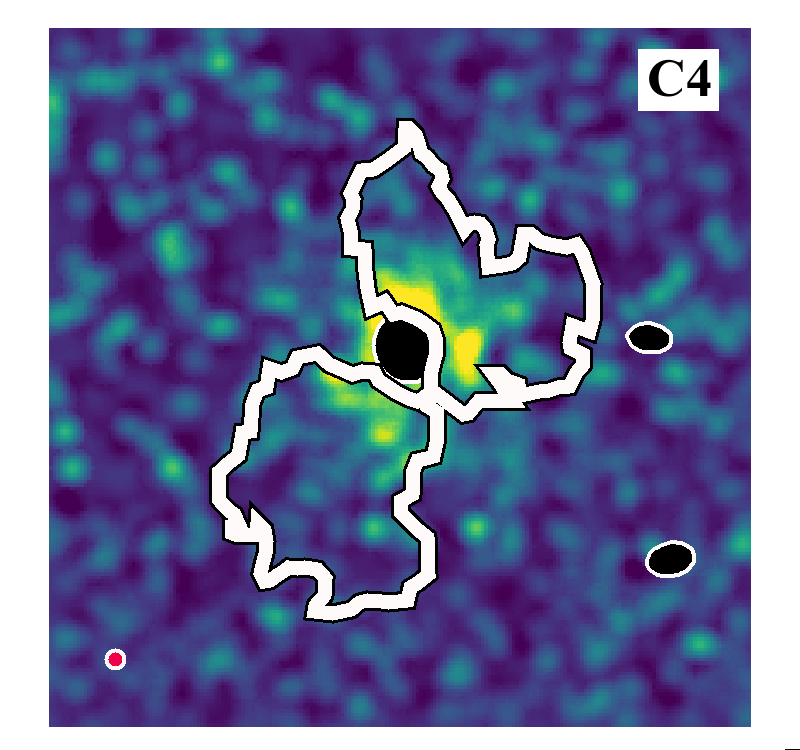} 
\hspace{-.1cm}
\includegraphics[trim=2cm 2cm 2cm 1cm, clip, width=.245\textwidth]{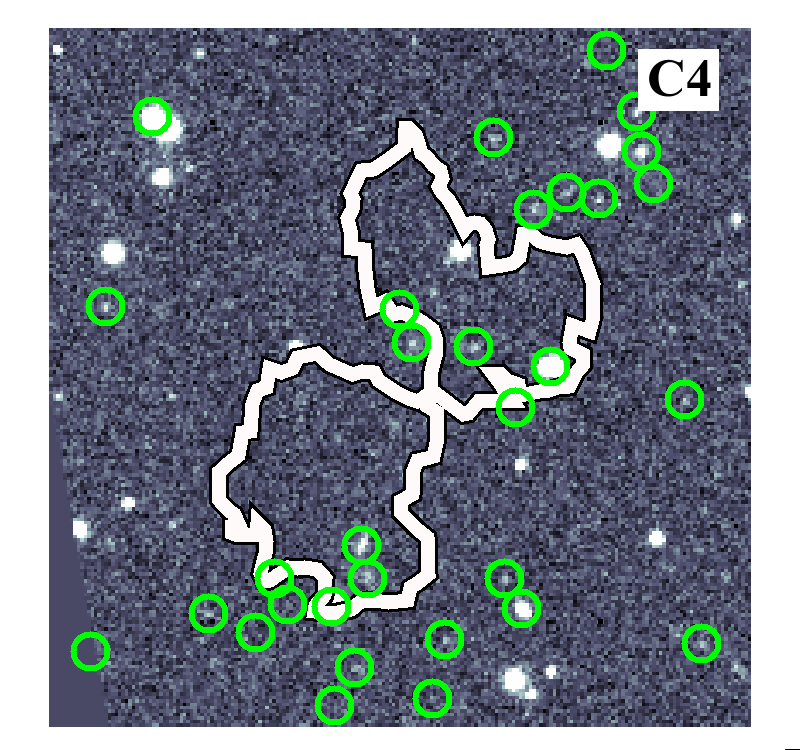} \\
\vspace{.02cm}
\includegraphics[trim=2cm 2cm 2cm 1cm, clip, width=.245\textwidth]{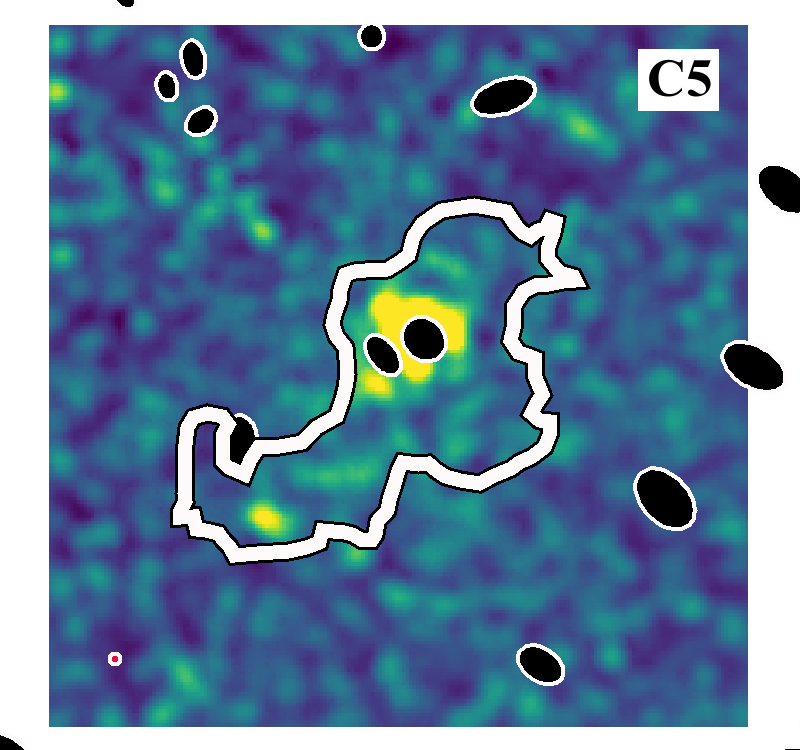}
\hspace{-.1cm}
\includegraphics[trim=2cm 2cm 2cm 1cm, clip, width=.245\textwidth]{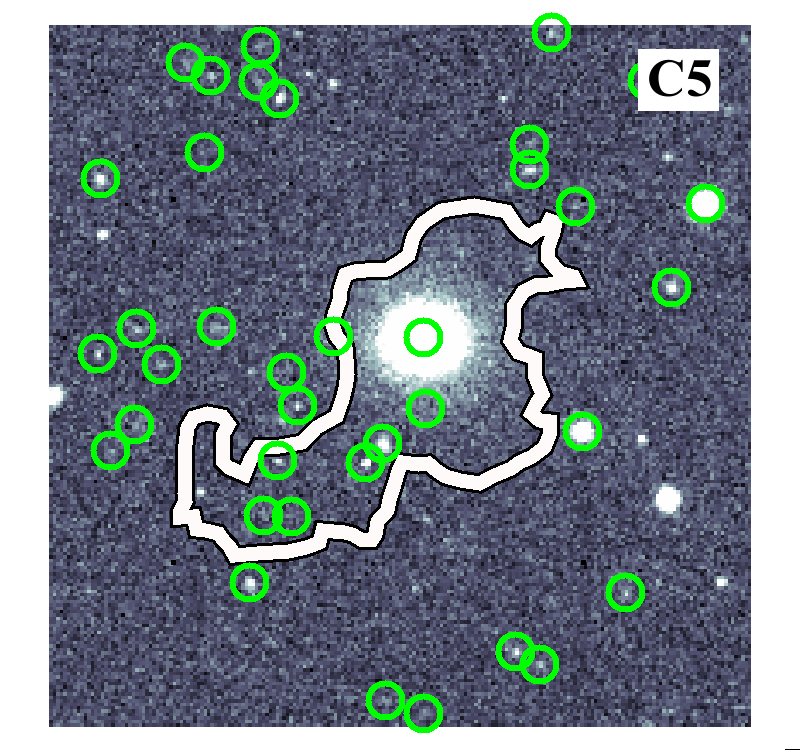}
\hspace{.05cm}
\includegraphics[trim=2cm 2cm 2cm 1cm, clip, width=.245\textwidth]{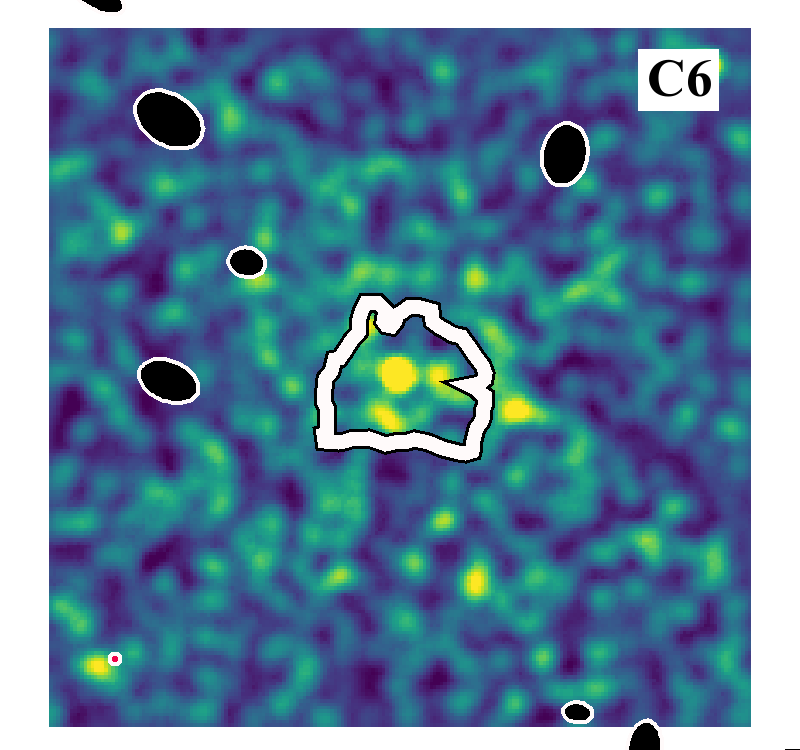}
\hspace{-.1cm}
\includegraphics[trim=2cm 2cm 2cm 1cm, clip, width=.245\textwidth]{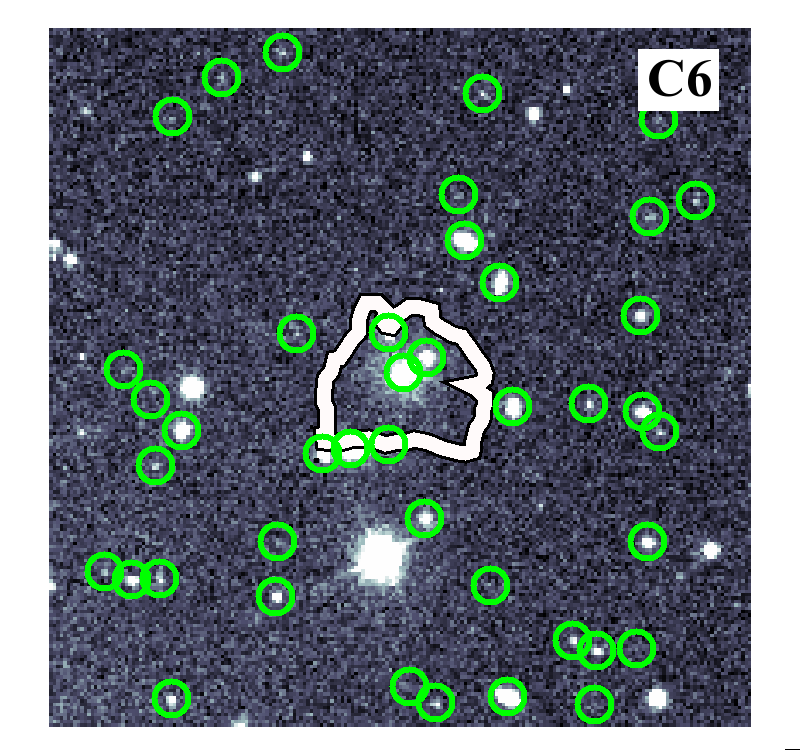} \\
\vspace{.02cm}
\includegraphics[trim=2cm 2cm 2cm 1cm, clip, width=.245\textwidth]{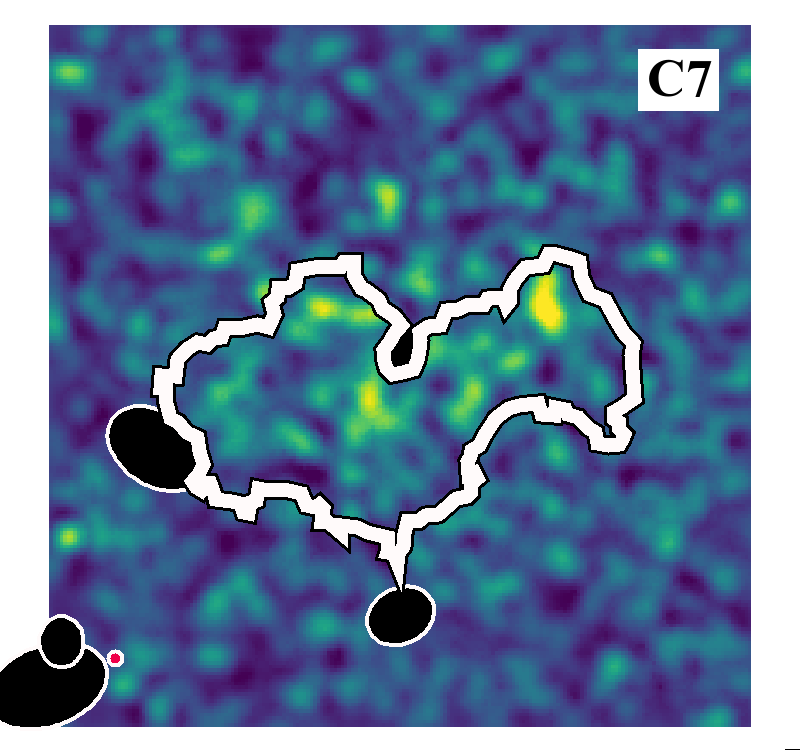}
\hspace{-.1cm}
\includegraphics[trim=2cm 2cm 2cm 1cm, clip, width=.245\textwidth]{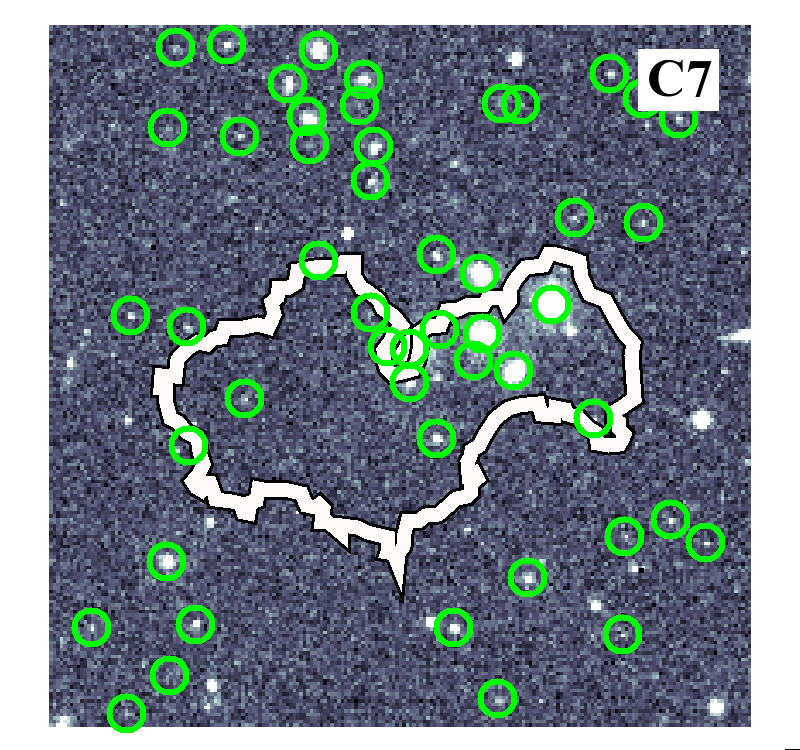}
\hspace{.05cm}
\includegraphics[trim=2cm 2cm 2cm 1cm, clip, width=.245\textwidth]{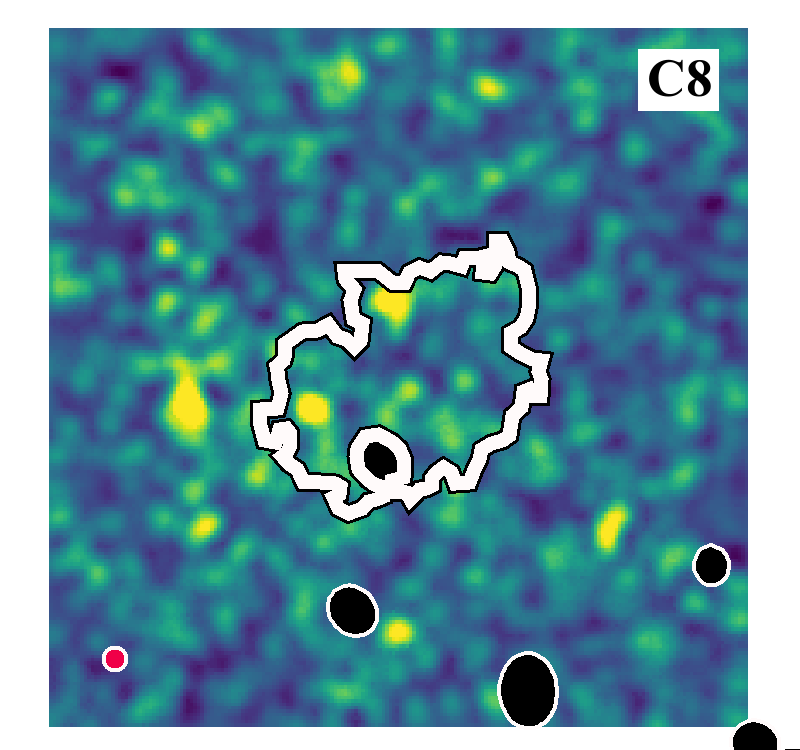} 
\hspace{-.1cm}
\includegraphics[trim=2cm 2cm 2cm 1cm, clip, width=.245\textwidth]{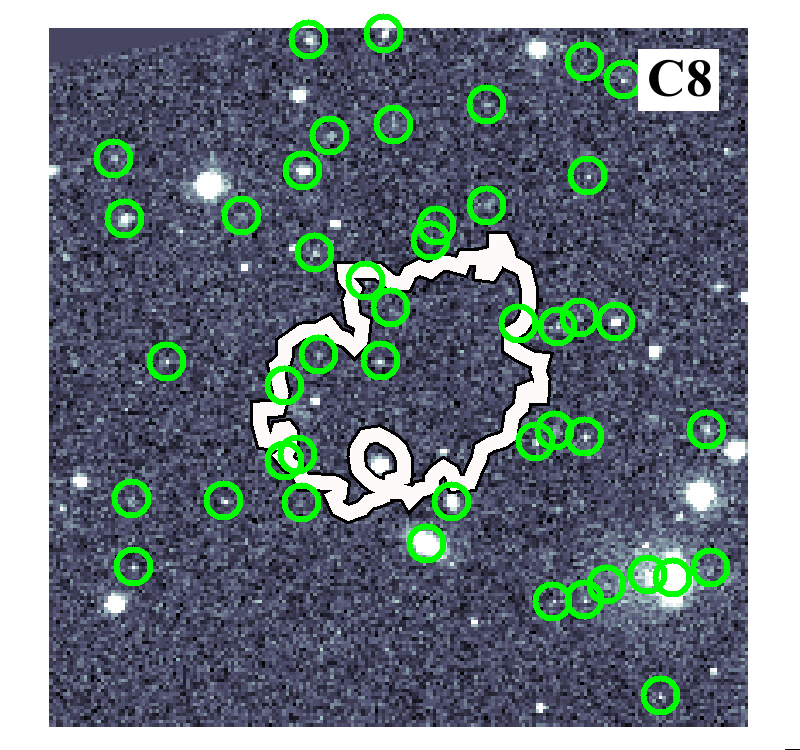} \\
\vspace{.02cm}
\includegraphics[trim=2cm 2cm 2cm 1cm, clip, width=.245\textwidth]{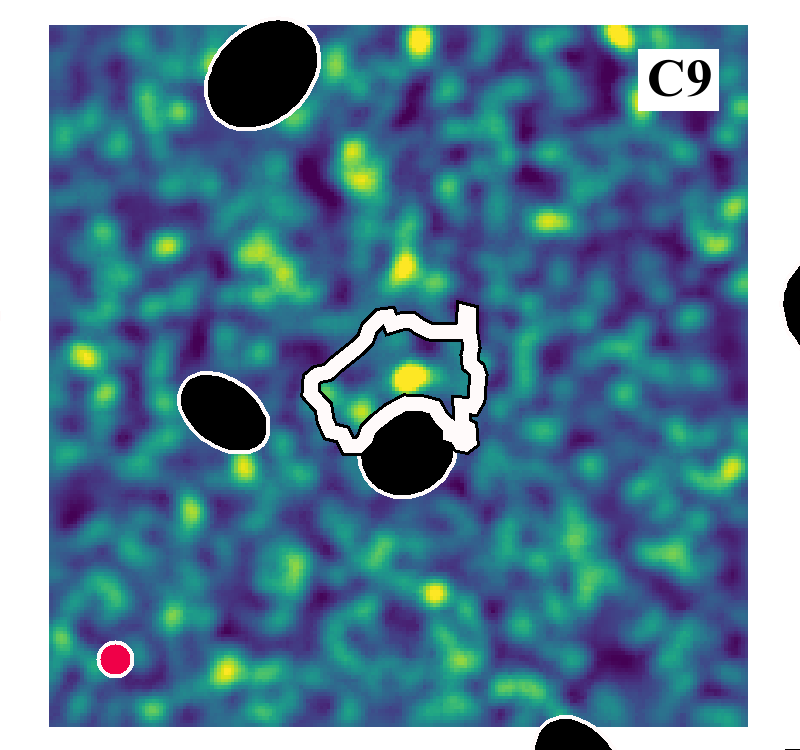}
\hspace{-.1cm}
\includegraphics[trim=2cm 2cm 2cm 1cm, clip, width=.245\textwidth]{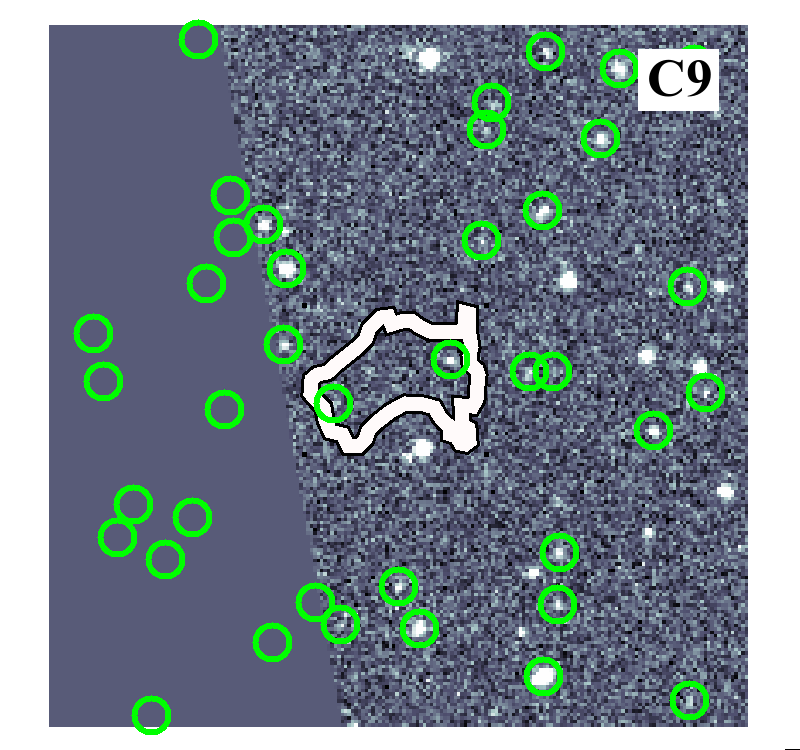}
\hspace{.05cm}
\includegraphics[trim=2cm 2cm 2cm 1cm, clip, width=.245\textwidth]{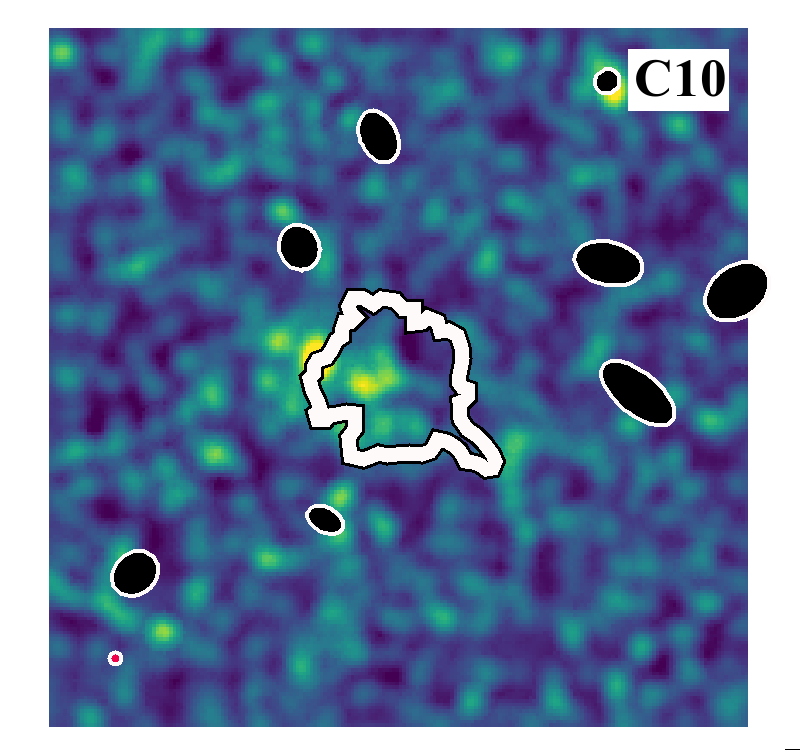}
\hspace{-.1cm}
\includegraphics[trim=2cm 2cm 2cm 1cm, clip, width=.245\textwidth]{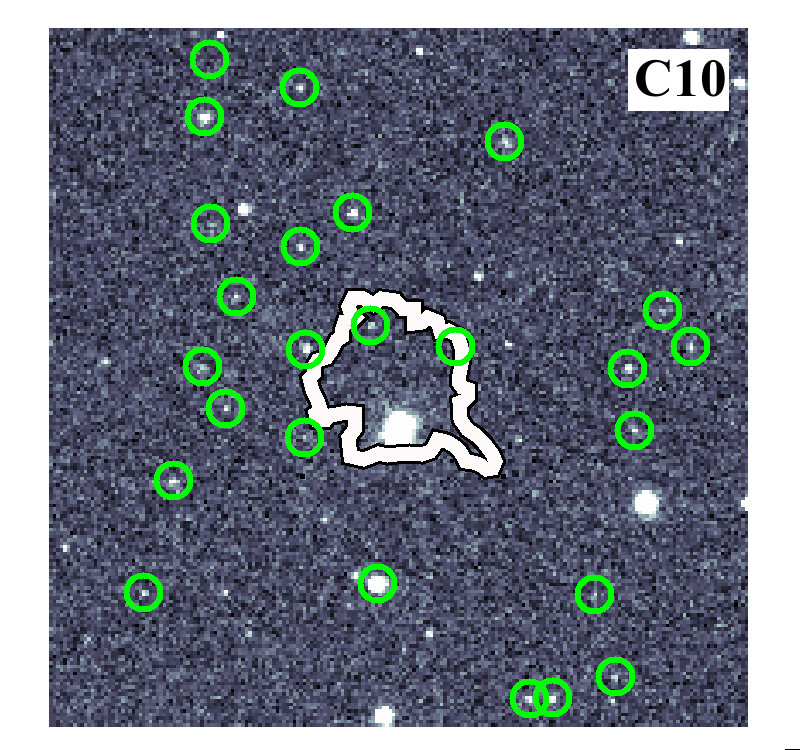}
\vspace{-.5cm}
\caption{\textit{Chandra} $0.7-2$\,keV (left panel, color images) and SDSS $r-$band cutout images (right panel, greyscale images) showing the individual clump candidates (white, irregular-shaped regions) and their environment within a $200 \arcsec \times 200 \arcsec$ area. The \textit{Chandra} images are smoothed with a Gaussian of radius $6$ to accentuate the X-ray excess. Filled black ellipses represent excluded \texttt{wavdetect} sources, while the filled red circle at the bottom left of each \textit{Chandra} image marks the local minimum size of PSF. As most of the clump candidates overlap with multiple SDSS galaxies (green circles), possible physical connections must be addressed to isolate genuine clumps (Section\,\ref{sec:opticalcounterparts} and \ref{sec:individualclumps}).}
\label{fig:stamps}
\end{figure*}
\renewcommand{\thefigure}{4 (continued)}

\begin{figure*}[h!]
\includegraphics[trim=2cm 2cm 2cm 1cm, clip, width=.245\textwidth]{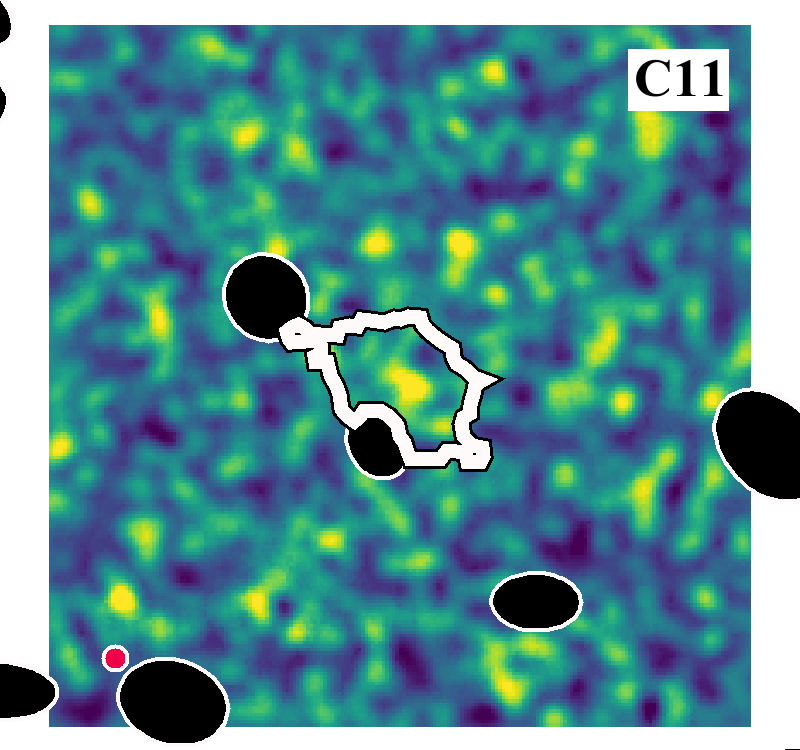}
\hspace{-.1cm}
\includegraphics[trim=2cm 2cm 2cm 1cm, clip, width=.245\textwidth]{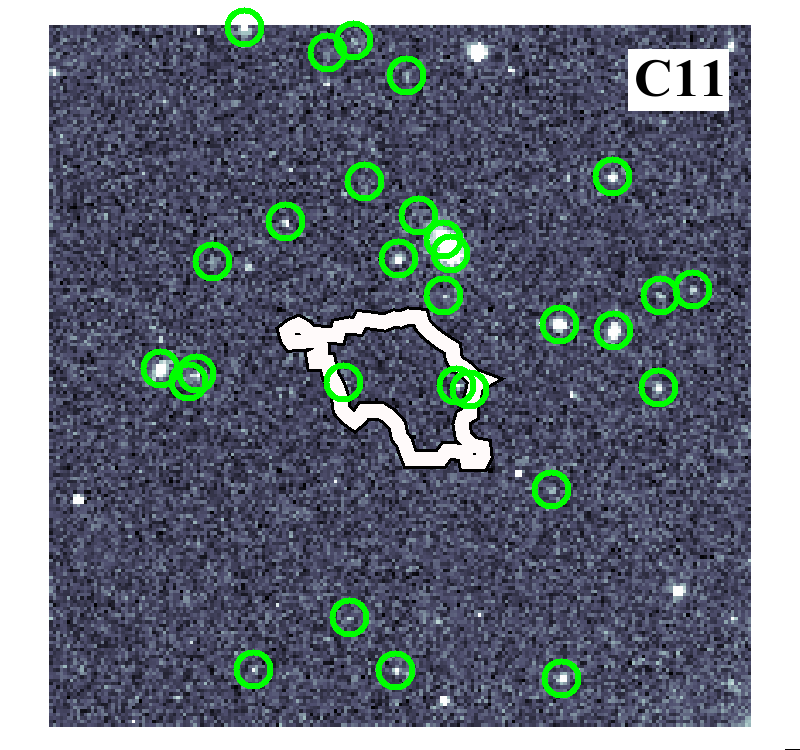}
\hspace{.05cm}
\includegraphics[trim=2cm 2cm 2cm 1cm, clip, width=.245\textwidth]{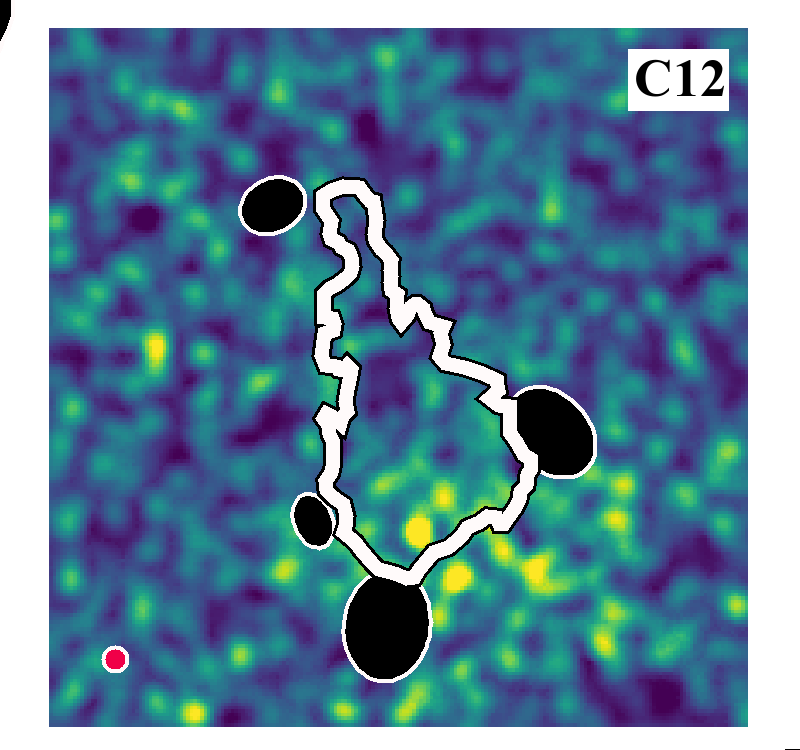}
\hspace{-.1cm}
\includegraphics[trim=2cm 2cm 2cm 1cm, clip, width=.245\textwidth]{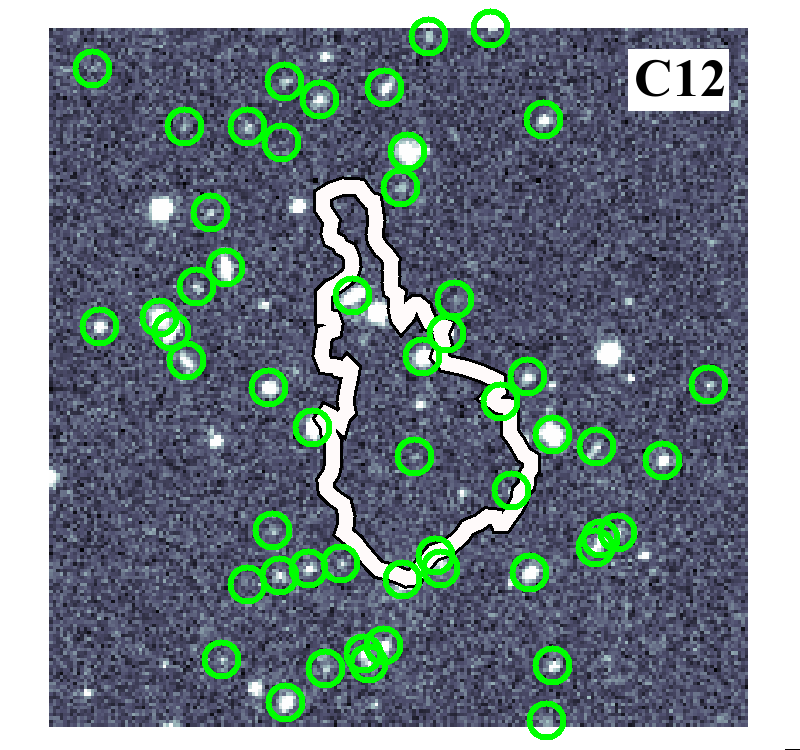} \\
\vspace{.02cm}
\includegraphics[trim=2cm 2cm 2cm 1cm, clip, width=.245\textwidth]{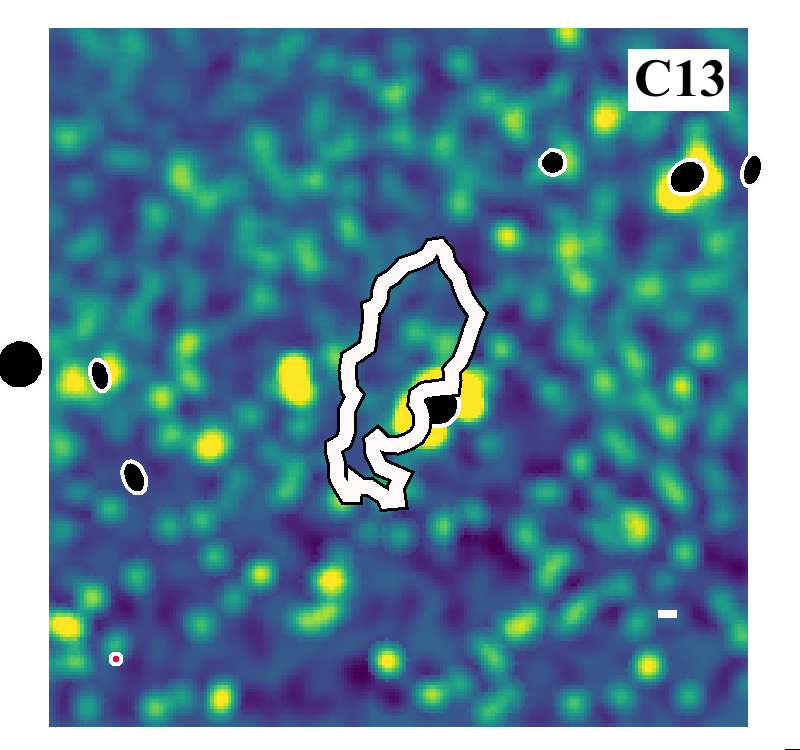}
\hspace{-.1cm}
\includegraphics[trim=2cm 2cm 2cm 1cm, clip, width=.245\textwidth]{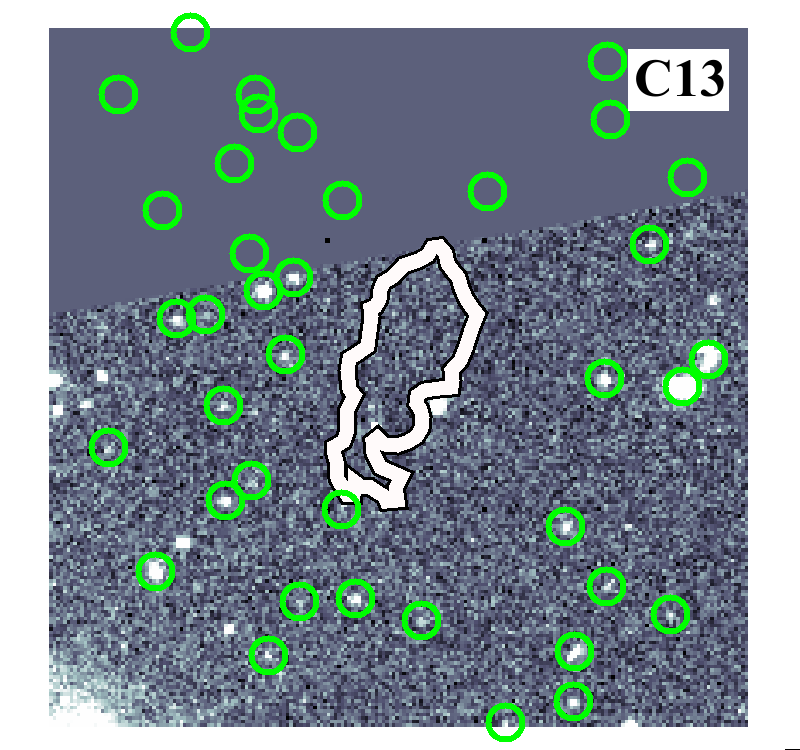}
\hspace{.05cm}
\includegraphics[trim=2cm 2cm 2cm 1cm, clip, width=.245\textwidth]{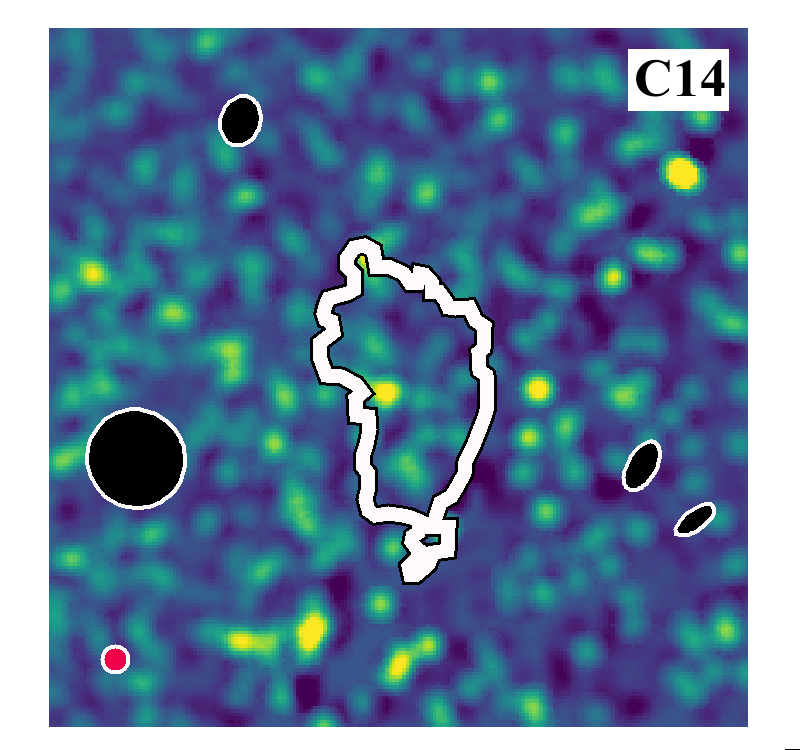}
\hspace{-.1cm}
\includegraphics[trim=2cm 2cm 2cm 1cm, clip, width=.245\textwidth]{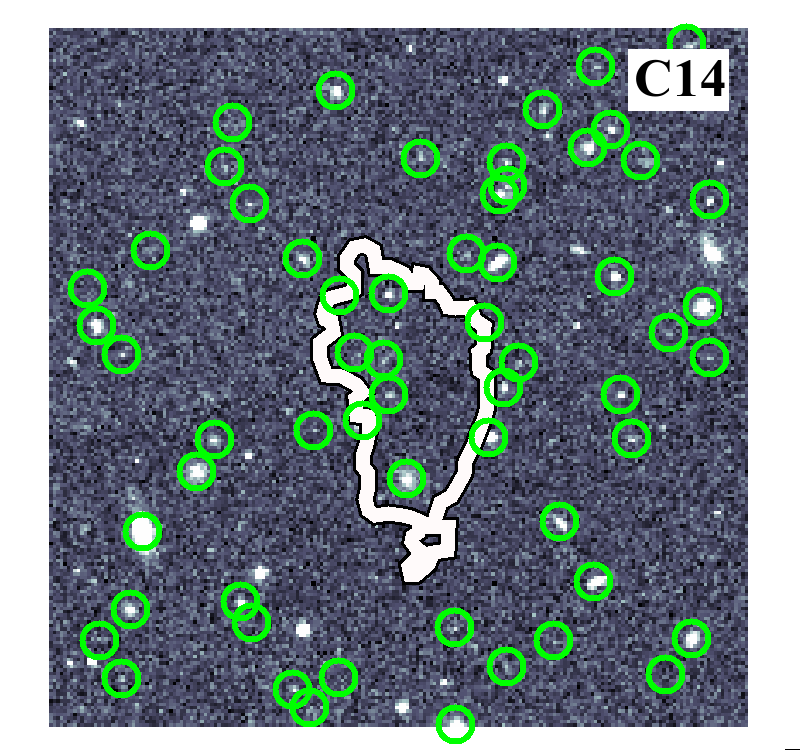} \\
\vspace{.02cm}
\includegraphics[trim=2cm 2cm 2cm 1cm, clip, width=.245\textwidth]{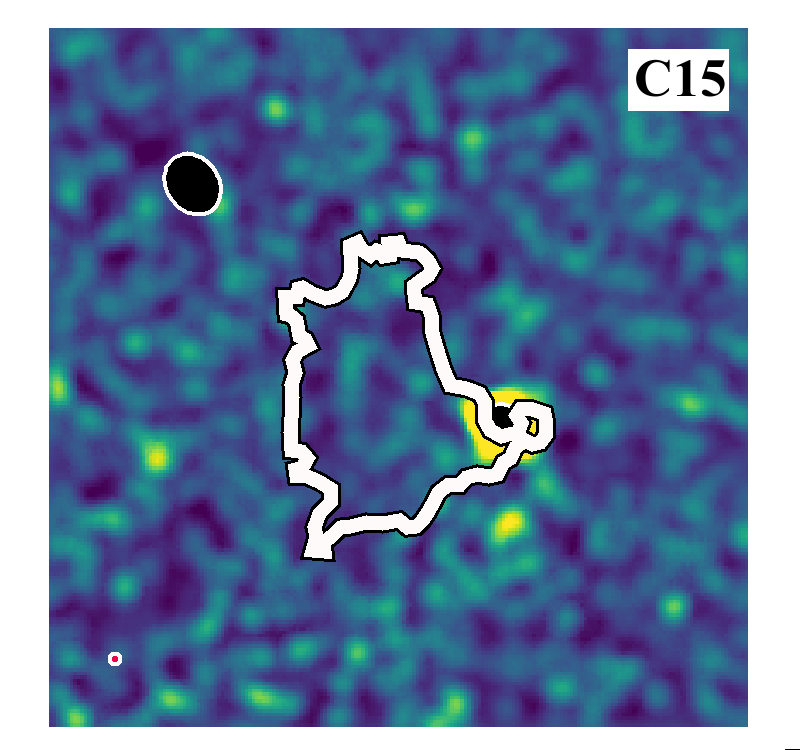}
\hspace{-.1cm}
\includegraphics[trim=2cm 2cm 2cm 1cm, clip, width=.245\textwidth]{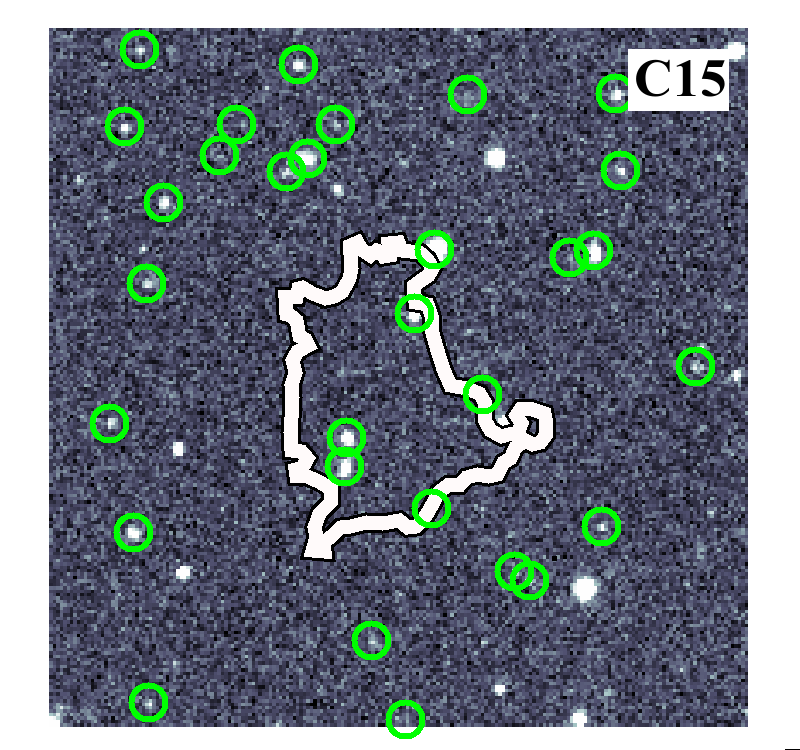}
\hspace{.05cm}
\includegraphics[trim=2cm 2cm 2cm 1cm, clip, width=.245\textwidth]{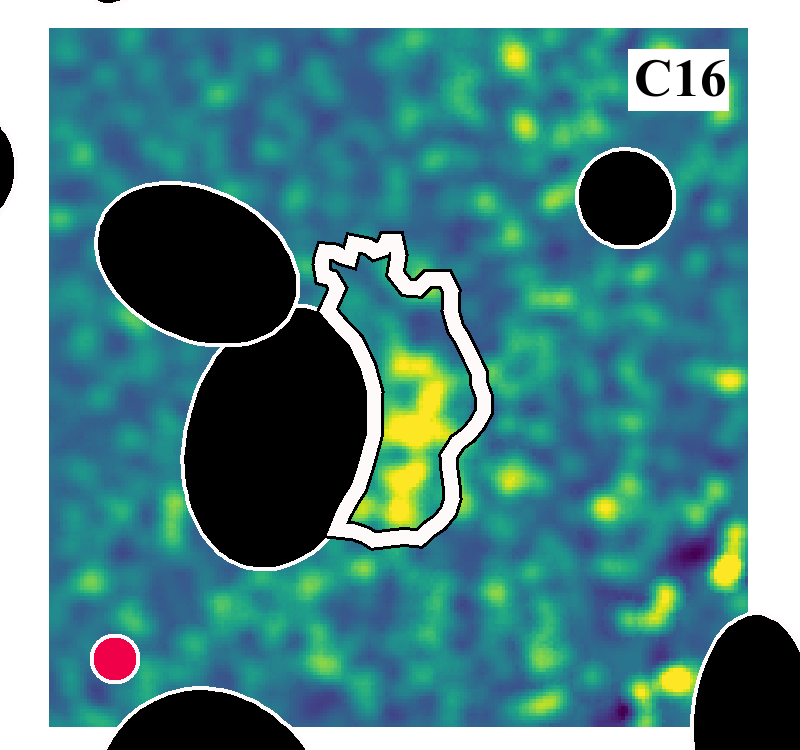}
\hspace{-.1cm}
\includegraphics[trim=2cm 2cm 2cm 1cm, clip, width=.245\textwidth]{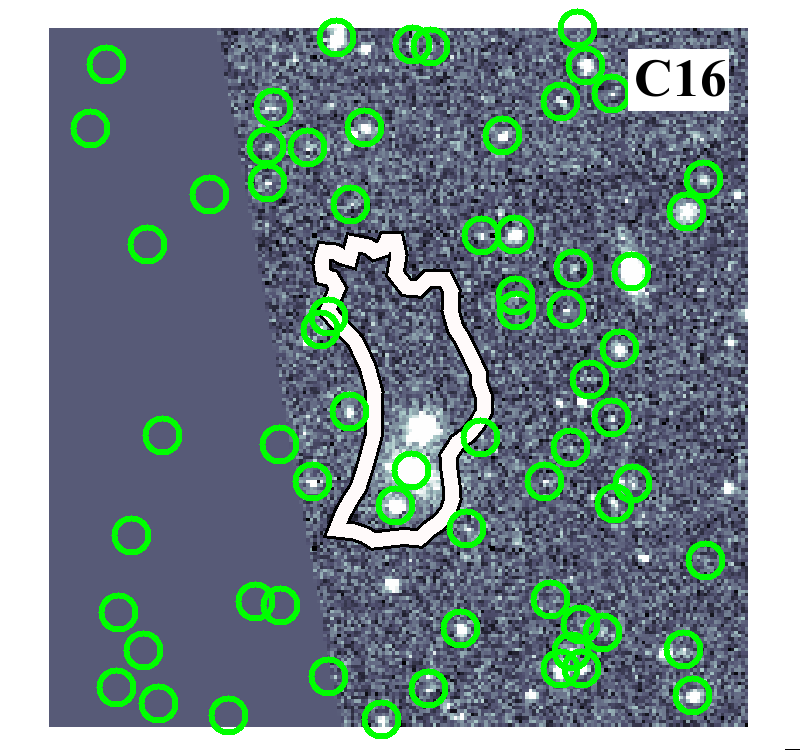} \\
\vspace{.02cm}
\includegraphics[trim=2cm 2cm 2cm 1cm, clip, width=.245\textwidth]{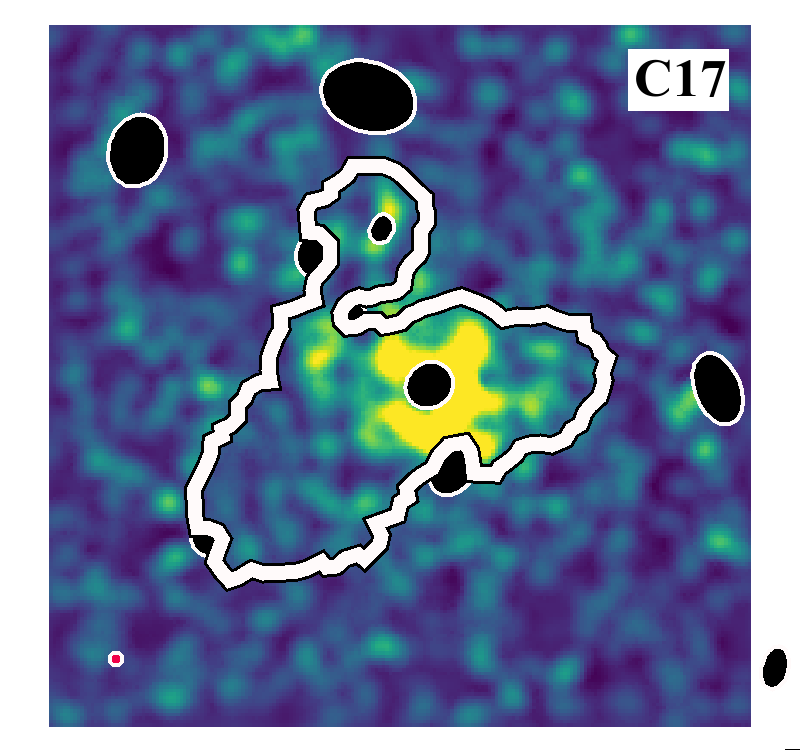}
\hspace{-.1cm}
\includegraphics[trim=2cm 2cm 2cm 1cm, clip, width=.245\textwidth]{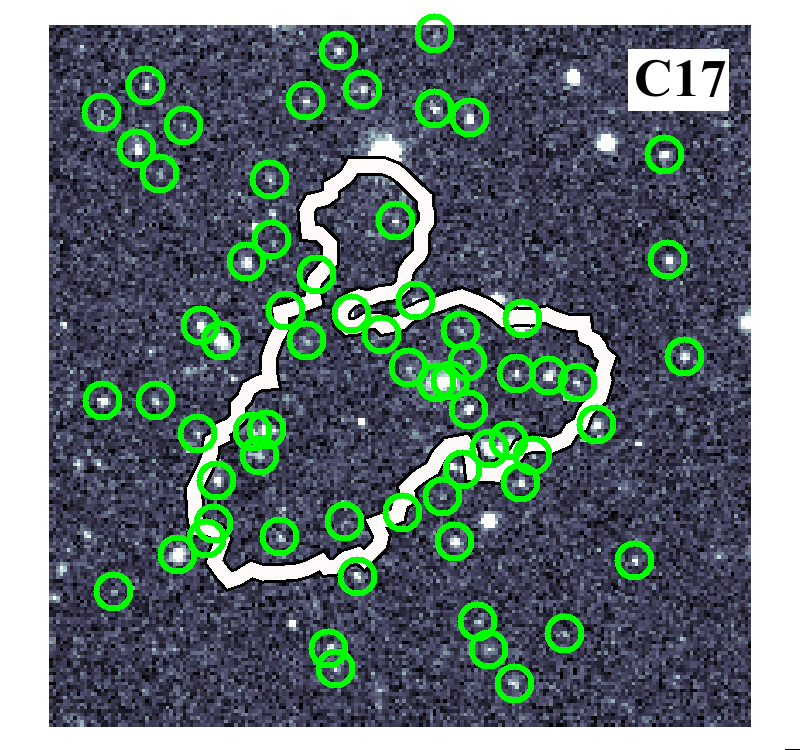}
\hspace{.05cm}
\includegraphics[trim=2cm 2cm 2cm 1cm, clip, width=.245\textwidth]{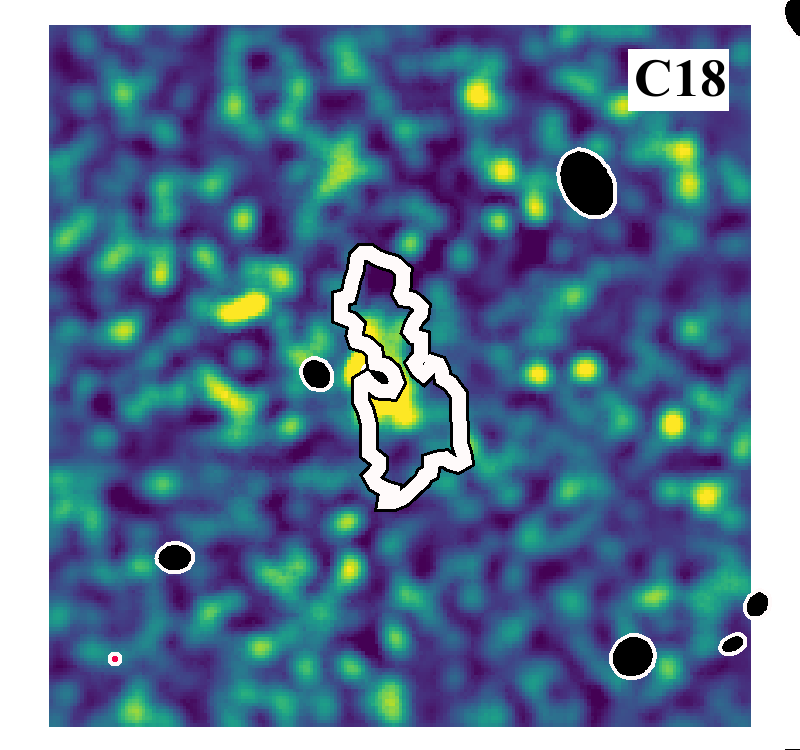}
\hspace{-.1cm}
\includegraphics[trim=2cm 2cm 2cm 1cm, clip, width=.245\textwidth]{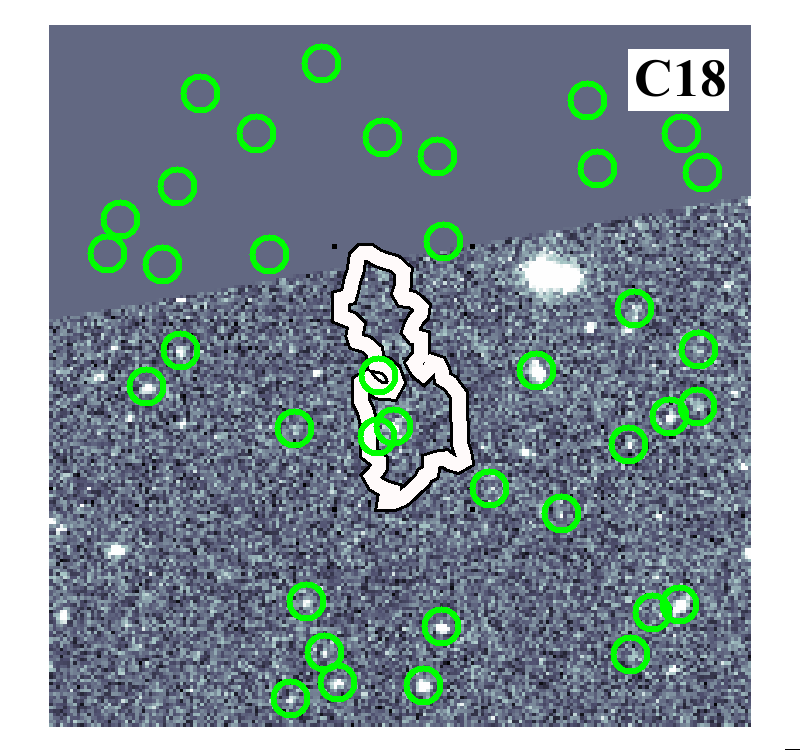} \\
\vspace{.02cm}
\includegraphics[trim=2cm 2cm 2cm 1cm, clip, width=.245\textwidth]{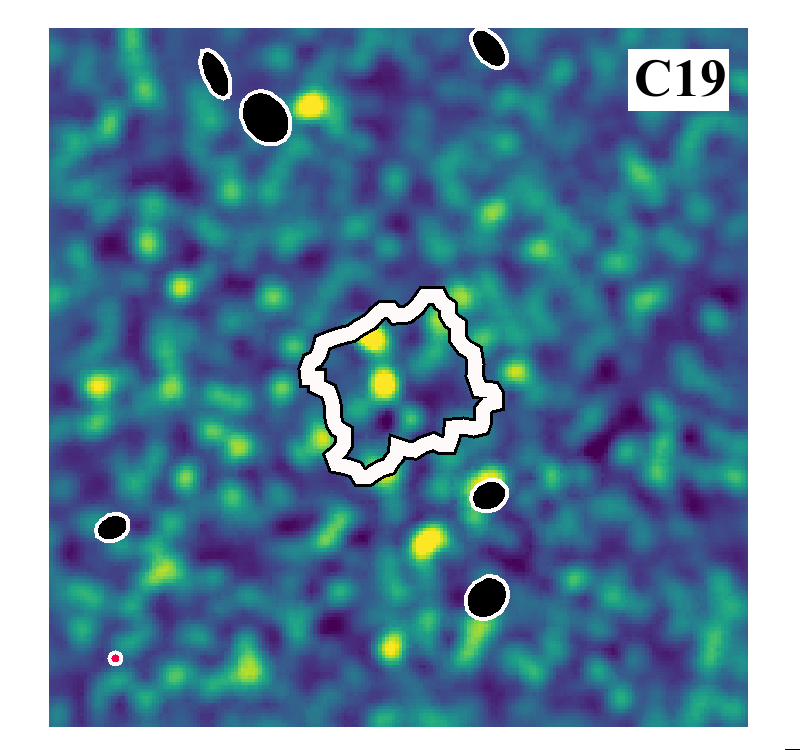}
\hspace{-.1cm}
\includegraphics[trim=2cm 2cm 2cm 1cm, clip, width=.245\textwidth]{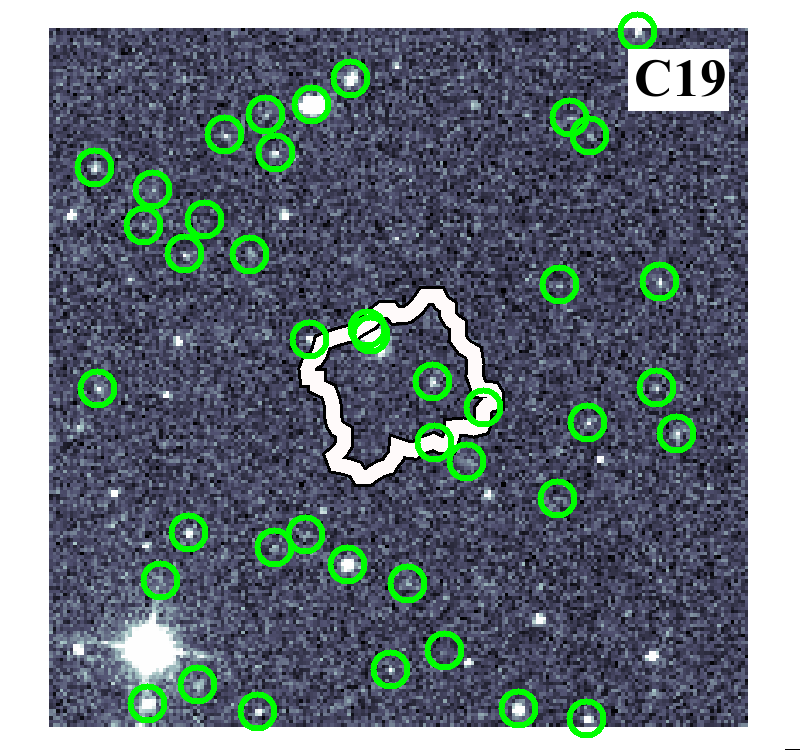}
\hspace{.05cm}
\includegraphics[trim=2cm 2cm 2cm 1cm, clip, width=.245\textwidth]{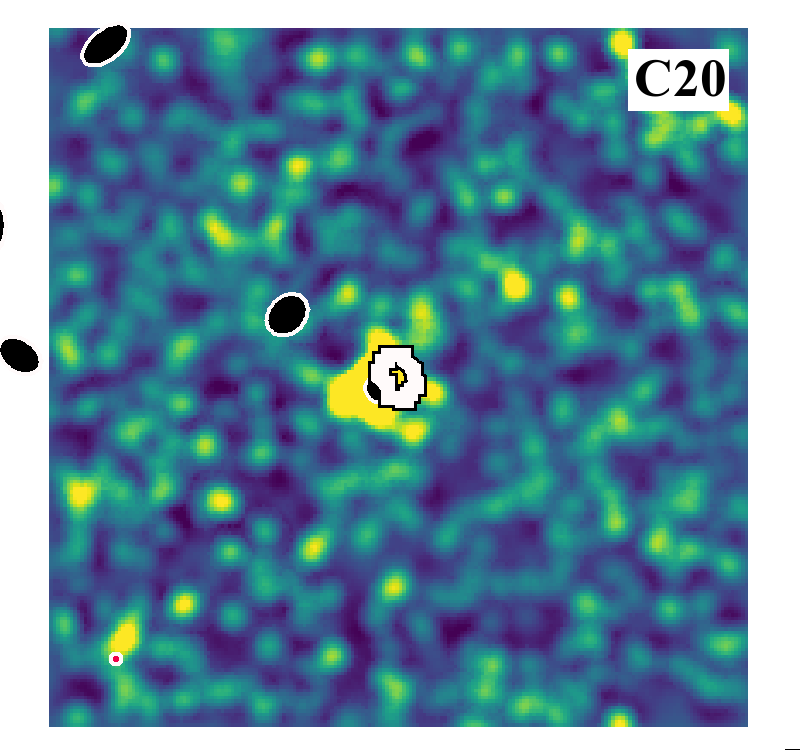}
\hspace{-.1cm}
\includegraphics[trim=2cm 2cm 2cm 1cm, clip, width=.245\textwidth]{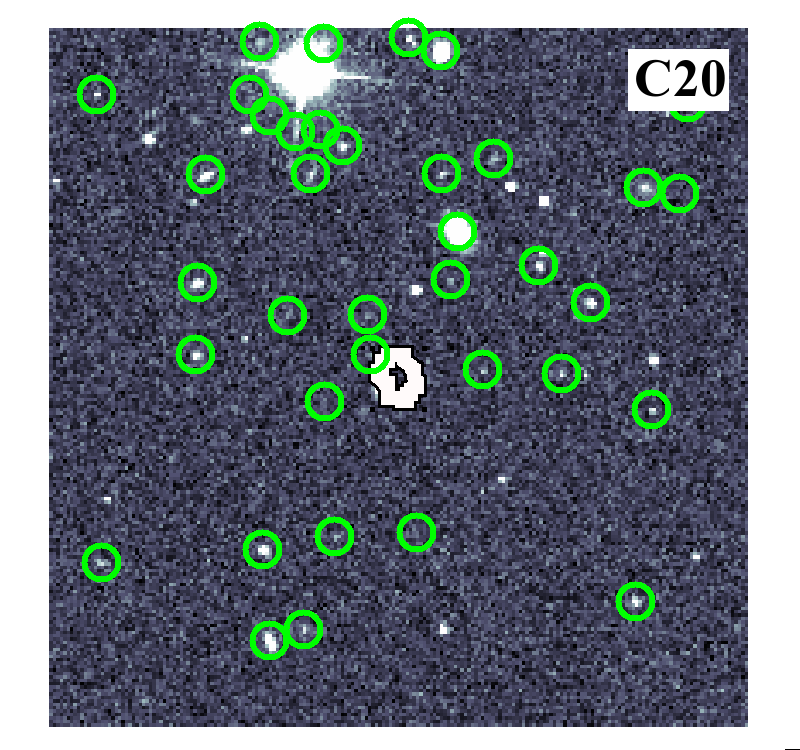}
\caption{}
\end{figure*}

\begin{figure*}[h]
\includegraphics[trim=2cm 2cm 2cm 1cm, clip, width=.245\textwidth]{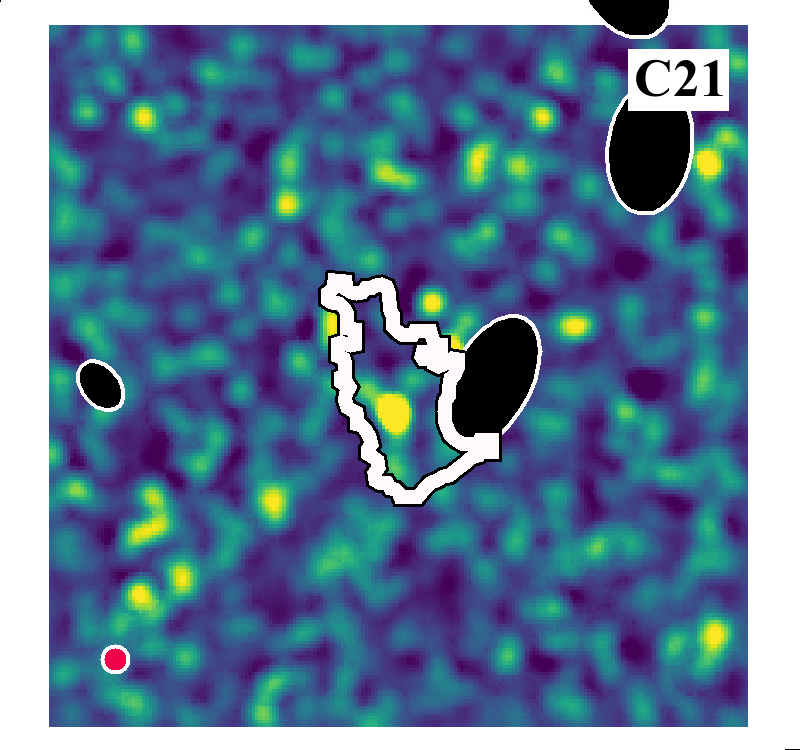}
\hspace{-.1cm}
\includegraphics[trim=2cm 2cm 2cm 1cm, clip, width=.245\textwidth]{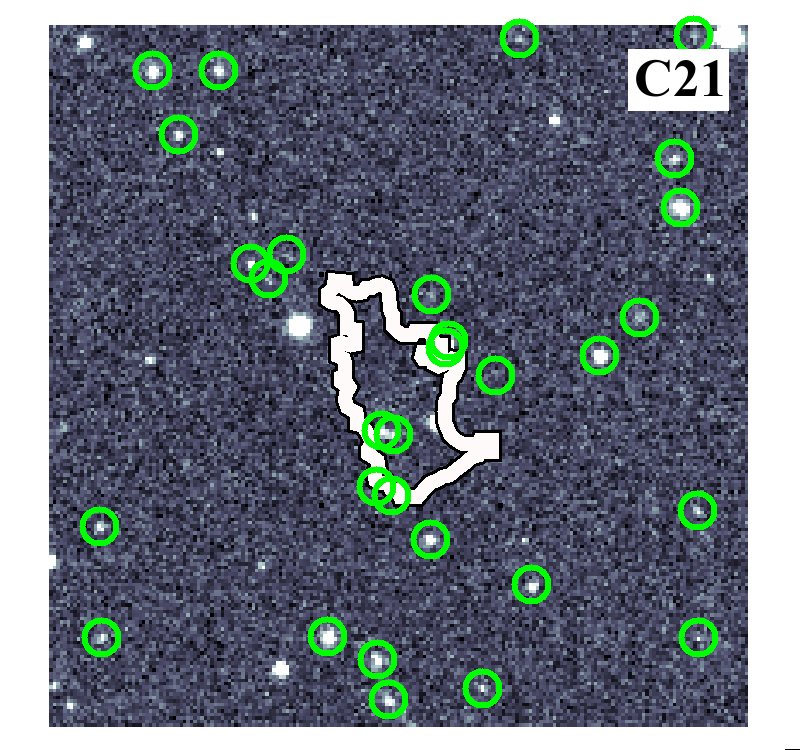}
\hspace{.05cm}
\includegraphics[trim=2cm 2cm 2cm 1cm, clip, width=.245\textwidth]{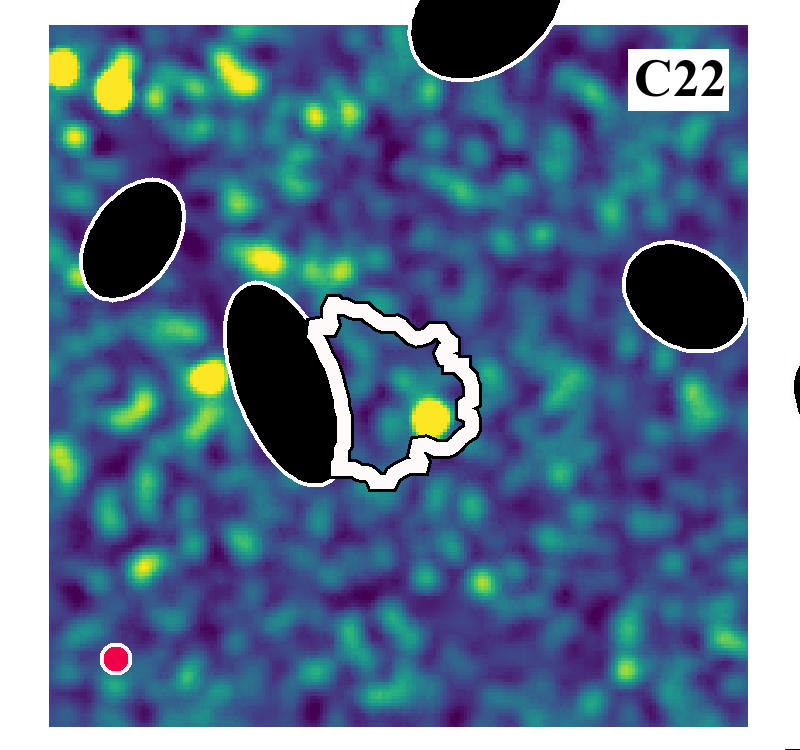}
\hspace{-.1cm}
\includegraphics[trim=2cm 2cm 2cm 1cm, clip, width=.245\textwidth]{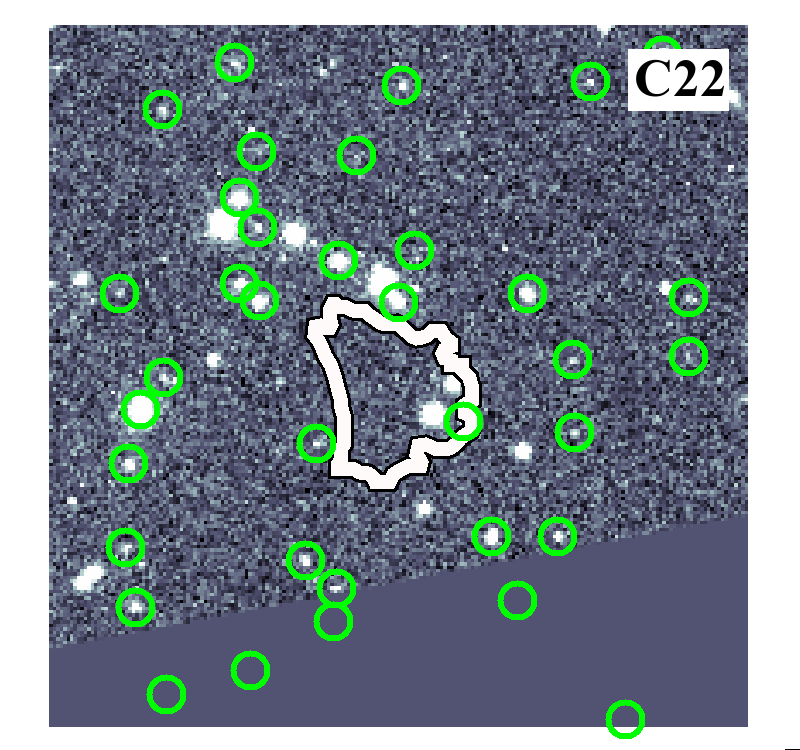}\\
\vspace{.02cm}
\hspace{-.16cm}
\includegraphics[trim=2cm 2cm 2cm 1cm, clip, width=.245\textwidth]{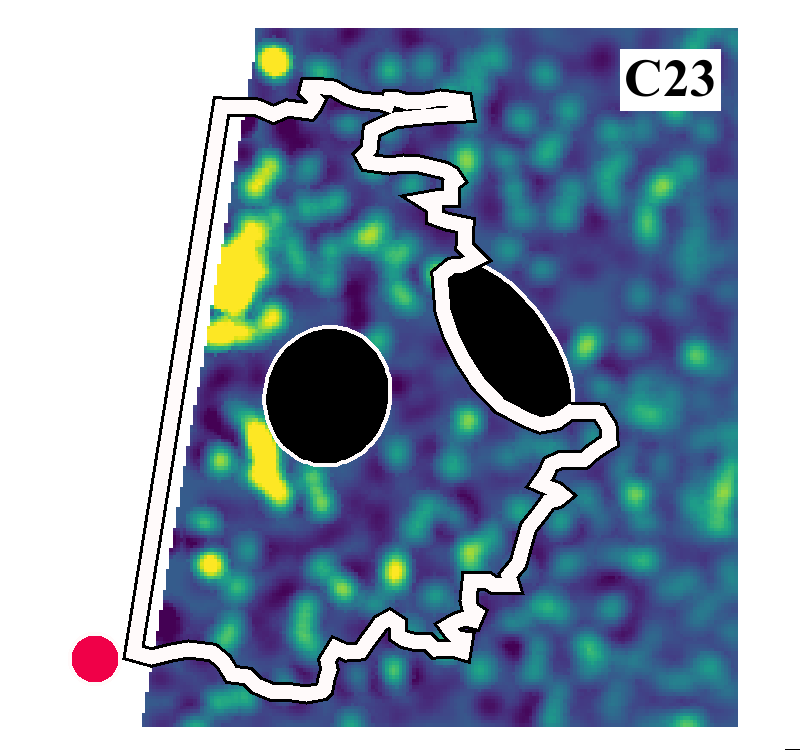}
\hspace{-.1cm}
\includegraphics[trim=2cm 2cm 2cm 1cm, clip, width=.245\textwidth]{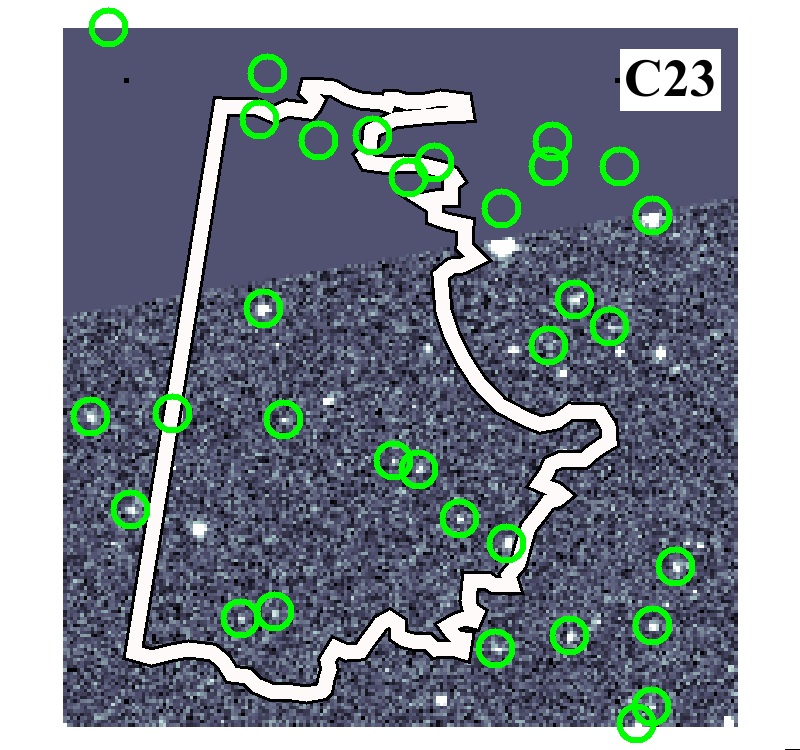}
\hspace{.05cm}
\includegraphics[trim=2cm 2cm 2cm 1cm, clip, width=.245\textwidth]{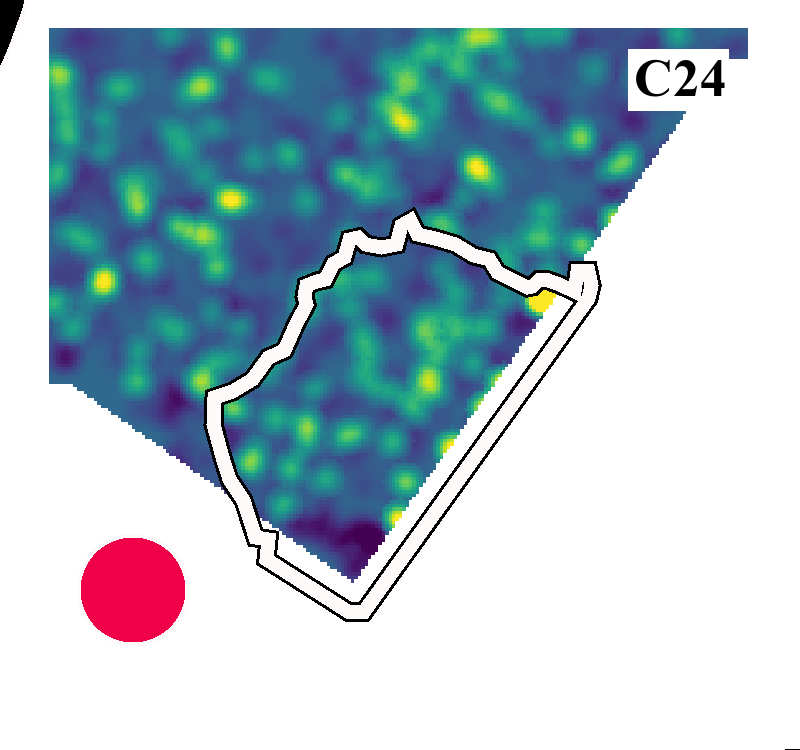}
\hspace{-.1cm}
\includegraphics[trim=2cm 2cm 2cm 1cm, clip, width=.245\textwidth]{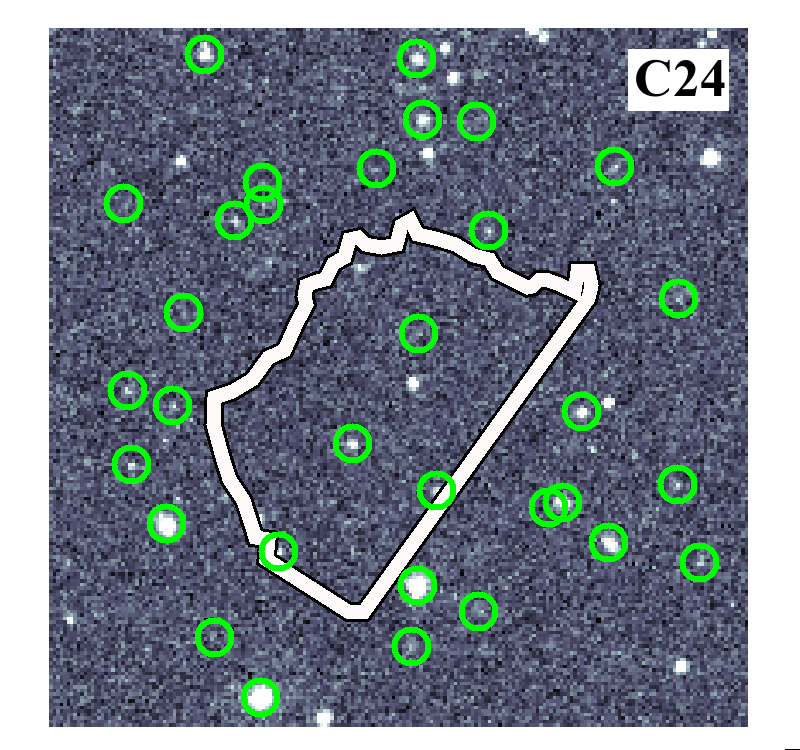}
\caption{}
\end{figure*}

\setcounter{figure}{4}
\renewcommand{\thefigure}{\arabic{figure}}

\subsection{Identifying ICM inhomogeneities} \label{sec:clumping}
A robust method for identifying ICM inhomogeneities was presented by \cite{2013MNRAS.428.3274Z}.
Based on this, the gas density of a perfectly homogeneous ICM should follow a log-normal distribution measured within a narrow spherical shell.
In the presence of gas inhomogeneities, however, a `high-density tail' emerges in the distribution, distorting its log-normal shape.
As a consequence, the mean and median of the distribution separate, with the median staying around the peak while the mean shifts to higher densities.
Suggested by numerical simulations \citep[e.g.][]{2008ApJ...687..936K,2013MNRAS.428.3274Z,2013MNRAS.431..954K,2014ApJ...791...96R}, the same applies to the surface brightness or the flux (note that the ICM's X-ray emission scales with the square of gas density), which are also log-normally distributed.
The insensitivity of the {surface brightness} median to ICM inhomogeneities was also demonstrated by \cite{2015MNRAS.447.2198E}.
Instead of relying on the gas density, which cannot be measured directly, we searched for ICM inhomogeneities in the {surface brightness} distribution of the cluster.

To measure the {surface brightness} distribution in a radial range, we first binned the total counts mosaic image using the \texttt{contbin} algorithm \citep{2006MNRAS.371..829S}.
The algorithm is based on an adaptive binning technique and divides the input image into bins having approximately the same signal-to-noise ratio (SNR).
For the SNR calculation, we supplied the stowed background data for \texttt{contbin}, and set the SNR threshold to $>$\,$3$.
Point sources were masked both from the input and background mosaic.
Since ICM inhomogeneities are expected to concentrate on the outskirts, the cluster core was also masked to reduce computational time.
With these settings, the cluster core-excluded mosaic was divided into $1132$ irregular-shaped and unequal-sized bins by \texttt{contbin}.
The smallest bins have a scale of $\lesssim 1\arcsec \times 1 \arcsec$ which corresponds to $1.2\times1.2 \,\rm kpc$ at the distance of Abell\,1795.
The resulting binmap was then applied to the background- and exposure-corrected image, which was then divided by concentric annuli (Figure\,\ref{fig:contbin}), and the {surface brightness} in each bin was measured.
The annuli were defined to have a logarithmically increasing radius so that the number of enclosed bins is sufficiently large to build the {surface brightness} distribution, {even at the outermost annulus with the lowest {surface brightness} and thus, the lowest number of bins.}

In Figure\,\ref{fig:histograms}, we present the {surface brightness} distributions between $\sim$\,$13\arcmin$ ($0.5\,r_{200}$) and $\sim$\,$42.5\arcmin$ ($\approx 1.6\,r_{200}$).
{At larger cluster radii, the `bulk' (i.e., log-normal) and `high-surface-brightness tail' components of the histograms are clearly distinguishable.}
{Accordingly}, while the median surface brightness coincides well with the peak, the mean is shifted towards high {surface brightness}, implying more weight in the right tail of the distribution.
To quantify this asymmetry, we calculated the Fisher-Pearson coefficient of skewness of the distributions.
The mean, median, and skewness, {along with annulus properties}, are listed in Table\,\ref{tab:histograms}.
The error bars represent $1$$\sigma$ bootstrap confidence intervals with $10^5$ resamplings.
We found that the skewness is positive in each annulus and shows an increasing tendency with radius.

{We fitted each annulus' distribution with a log-normal function, then, using the $\sigma$ standard deviation of the best-fit function, we selected $>\!+2\sigma$ outliers from the fit centroid (listed in Table\,\ref{tab:histograms}), and localized them on the image. }
After merging adjacent outlier bins, we identified {$24$} clump candidates.
Their properties are reported in Table\,\ref{tab:clumpcandidates}, and their spatial distribution is displayed in Figure\,\ref{fig:contbin}.
{The significance of multi-bin clump candidates was calculated from the overall surface brightness measured within the merged bins.}
{An increasing trend with radius is apparent in the frequency of the sources, with a higher projected concentration at the western hemisphere of the cluster.}
{Both the spatial distribution of individual sources and the skewness} suggest that the effect of ICM inhomogeneities on the measured {surface brightness} is more significant at larger radii{, in accordance with simulation predictions.}

\subsubsection{Optical counterparts}\label{sec:opticalcounterparts}
Figure\,\ref{fig:stamps} shows cutout images centered around the clump candidates in the $0.7-2 \, \rm keV$ band.
{Most candidates are connected with one or multiple \texttt{wavdetect} sources, which are surrounded by or adjacent to the clump candidate region}, suggesting that their emission may originate from the same X-ray source.
It might also imply that the candidates are not genuine clumps.
To uncover the nature of clump candidates and explore their surroundings, we searched for optical counterparts in the Sloan Digital Sky Survey (SDSS).
We collected images from SDSS Data Release\,16 for each candidate and the catalog of {galaxies and their photometric properties} in the surroundings.
Figure\,\ref{fig:stamps} shows SDSS {$r-$band images presenting the same FOV as the \textit{Chandra} cutout images.}
{We found that $21$ of $24$ candidates coincide with multiple, while $1$ candidate (C$22$) with one SDSS source. Clump candidates C$13$ and C$20$ remained with no co-spatial SDSS galaxies.}

To test whether the X-ray emission of clump candidates originates from an overlapping SDSS galaxy, we executed an X-ray flux comparison between the sources.
Based on visual inspection of X-ray and optical images, we selected only those galaxies for this test that are associated with excess X-ray emission within {(or in certain cases around)} the corresponding clump candidate region.
We set this criterion to keep our results reasonable in cases where the number of overlapping galaxies is too high {(e.g., C$17$)}, or where galaxies are obviously not X-ray emitters {(e.g., C$15$)}.
{Candidates with co-spatial relation of a background galaxy cluster/group or near the detector edge have not been subjected to this analysis.}
{Thus, we selected a total of $26$ galaxies from SDSS, which can be associated with clump candidates C$1$, C$3-8$, C$10$, C$14$, C$16$, C$18-19$, and C$21-22$.}

When estimating the expected X-ray flux of galaxies, we assessed the contribution of the unresolved population of low-mass X-ray binaries (LMXBs) and, for galaxies with a stellar mass of $M_{*} > 7 \times 10^{10} \, \rm M_{\sun}$, the diffuse emission of the hot interstellar medium (ISM).
Below this particular mass threshold, we do not expect the galaxies to host luminous X-ray halos according to local scaling relations \citep[e.g.][]{2011MNRAS.418.1901B,2013ApJ...776..116K,2019MNRAS.488.1072K}.
We did not consider the X-ray emission from AGN and the high-mass X-ray binary population. 
The latter is only significant for galaxies with star formation rate (SFR) $\gtrsim$\,$1$\,\citep{2003MNRAS.339..793G}, and, although the SFR of the selected galaxies is unknown, their average {$u$$-$$r = 2.14$ color index} indicates that the galaxy sample consists mainly of early-type galaxies \citep{2001AJ....122.1861S}, which typically exhibit low SFR.

The stellar mass of the galaxies was derived from SDSS broadband photometric data ($r$ magnitudes and $u$$-$$r$ color indices) and photometric redshift using the stellar mass--to--light ratio \citep{2003ApJS..149..289B}.
We found that only {about} one-third of the galaxies are massive enough to host an X-ray luminous halo.
Based on the derived stellar masses, the LMXB X-ray luminosity contribution was estimated using the $L_{X}^{\rm{LMXB}} - M_{*}$ scaling relation \citep{2004MNRAS.349..146G}, and the diffuse X-ray emission luminosity was estimated from the $L_{X}^{\rm{gas}} - M_{*}$ scaling relation \citep{2019MNRAS.488.1072K}.
The resulting X-ray luminosities were converted into $0.7-2 \, \rm keV$ fluxes and added up within each clump candidate.
For the conversion of LMXB luminosities, we assumed a power-law model with a slope of $\Gamma = 1.56$ \citep{2003ApJ...587..356I}, and for the ISM, an \textit{apec} model with $kT=0.5 \, \rm keV$ \citep[e.g.][]{2011MNRAS.418.1901B,2016ApJ...826..167G}.

Similarly, the measured {stowed background subtracted} X-ray flux was added up within each clump candidate region, {corrected for the sky background}, then converted from photon flux to energy flux units.
For the conversion, we assumed an absorbed \textit{apec} with $kT = 1 \, \rm keV$, describing the diffuse emission of cluster outskirts beyond $r_{200}$.
Note that the source of X-ray emission in clump candidates is uncertain, but assuming a power-law model, for instance, would lead to a $\sim$\,$19\%$ increment in the conversion factor.

{Our results are reported in Section\,\ref{sec:individualclumps} only for those sources where flux comparison was conclusive.}

\begin{figure}
\centering
\includegraphics[width=.5\textwidth]{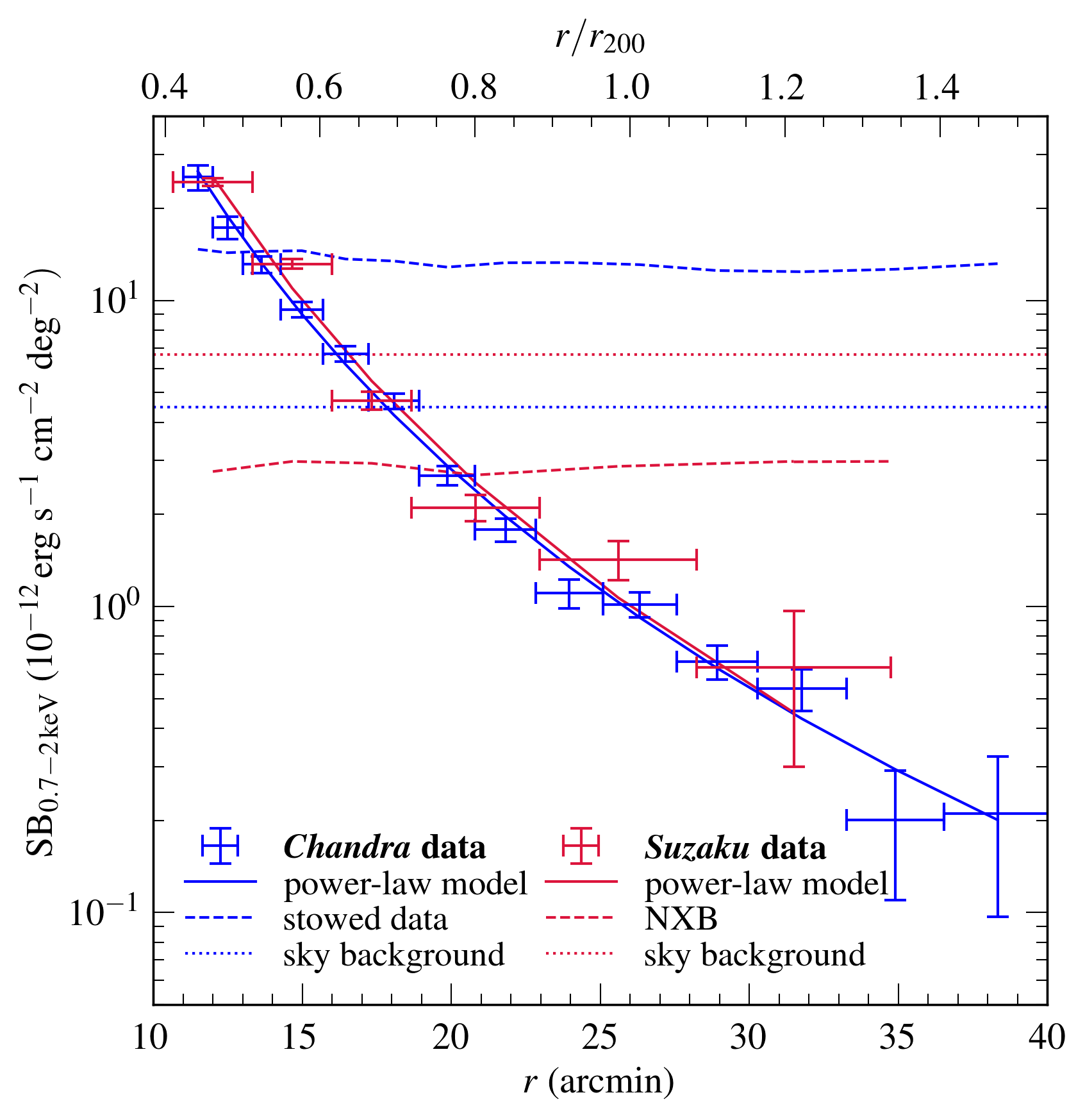}
\caption{
$0.7-2$\,keV background-subtracted azimuthally averaged surface brightness profiles of Abell\,1795 outskirts measured with \textit{Chandra} and \textit{Suzaku} with the corresponding best-fit power-law models overplotted. The background components are also shown. Point sources were removed down to a photon flux of $2\times 10^{-6} \, \rm photons \, cm^{-2} \, s^{-1}$ from both datasets, and the elevated \textit{Suzaku} sky background level reflects the excess emission from point sources whose PSF is broader than the corresponding source region.} 
\label{fig:sb1}
\end{figure}

\subsubsection{Hardness ratio}\label{sec:hardnessratio}

To further assess the source of X-ray emission in clump candidates, we gathered their rough spectral properties by measuring their hardness ratios.
The hardness ratio is defined as $H\!R=\frac{A-B}{A+B}$, where $A$ and $B$ are the exposure corrected counts measured in the $2-7 \,\rm keV$ and $0.5-1.5 \,\rm keV$ bands, respectively.
A hardness ratio of $H\!R \approx0$ is consistent with a power-law spectrum with a slope of $\Gamma = 1.4-1.7$, while $H\!R \approx -0.5$ and $H\!R \approx -0.8$ characterize an \textit{apec} spectra with $kT = 2 \,\rm keV$ and $kT = 1 \,\rm keV$, respectively.
Thus, hardness ratios provide the means to distinguish between the emission of a background AGN and thermal emission from diffuse gas at or below the temperature of cluster outskirts.
Note that clumps are predicted to have $kT_{\rm clump} < kT_{\rm ICM}$ \citep{2013MNRAS.429..799V}, and $kT_{\rm ICM} \simeq 2 \,\rm keV $ between $r_{500}$ and $r_{200}$ for Abell\,1795.

Due to the Poissonian distribution of counts, we calculated the hardness ratios and the associated confidence limits using the Bayesian Estimation of Hardness Ratios (\texttt{BEHR}) code \citep{2006ApJ...652..610P}.
We defined a local background around each clump candidate as an annulus with a width of $0.5\arcmin$ and the inner radius defined as the radius of the smallest fitting circle around the irregular-shaped source.
Source and background counts were extracted from exposure-corrected images with \texttt{wavdetect} sources excluded, then fed to \texttt{BEHR}.
The obtained hardness ratios are listed in Table\,\ref{tab:clumpcandidates}; {they could not be constrained for $6$ out of $24$ clump candidates.}

\begin{figure}
\centering
\includegraphics[width=.5\textwidth]{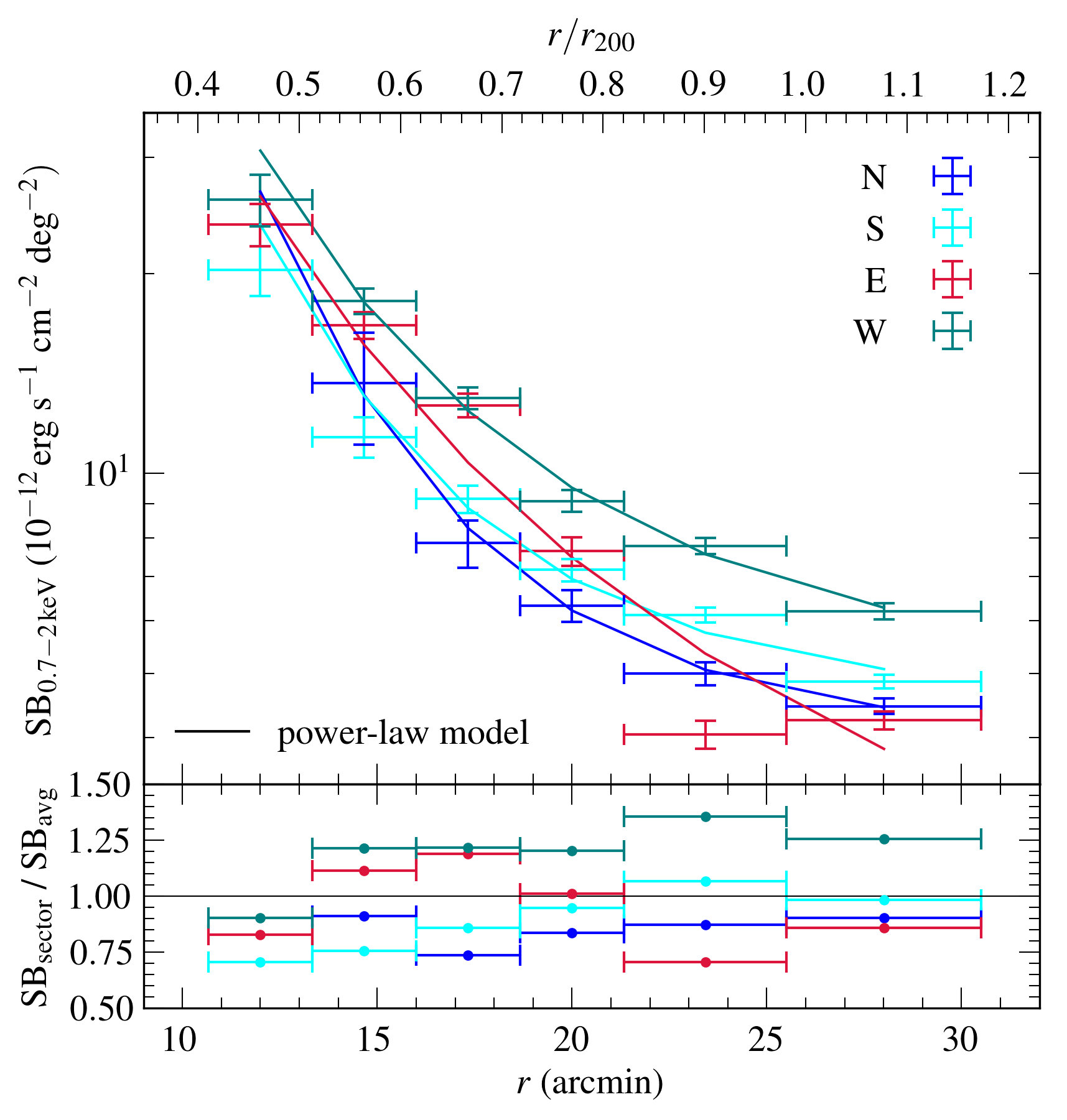}
\caption{$0.7-2\,\rm keV$ surface brightness profile of Abell\,1795 outskirts extracted from \textit{Chandra} data in four different directions. Azimuthal variations relative to the average profile are displayed in the bottom panel.}
\label{fig:sb-wedges}
\end{figure}

\subsubsection{The origin of individual clump candidates} \label{sec:individualclumps}

\begin{figure}
\centering
\includegraphics[width=.49\textwidth]{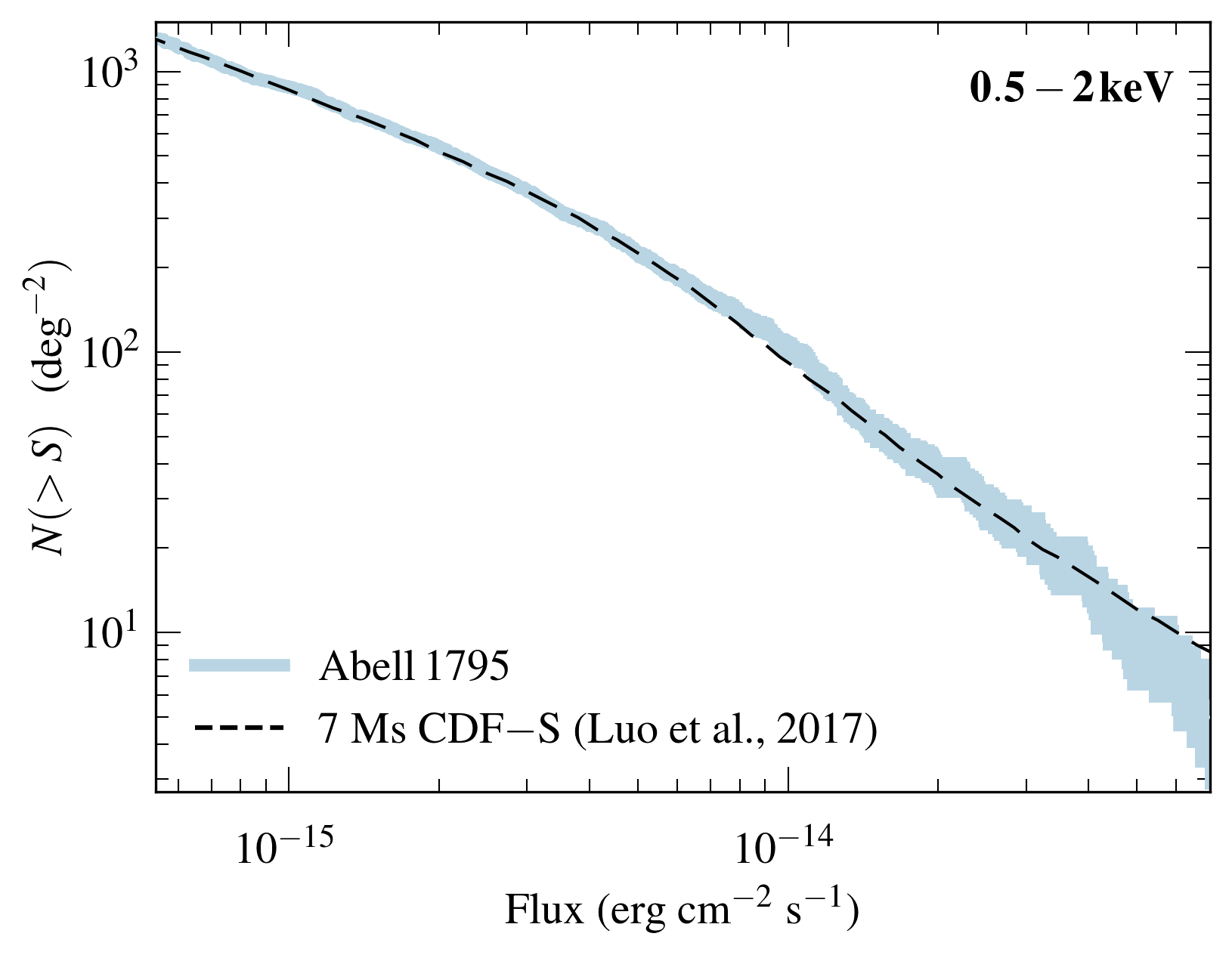}
\caption{Log$N-$log$S$ plot for the sources detected in the Abell\,1795 field (shaded curve with $1\!\,\sigma$ uncertainty) and in the $7$\,Ms Chandra Deep Field South\,(\cite{2017ApJS..228....2L}). Despite the excess in the number of Abell\,1795 field sources seen at $\sim\!10^{-14} \rm \, erg \, cm^{-2} \, s^{-1}$, which can be attributed to the cosmic variance, the distributions are in good agreement at fluxes below $\sim\!6\times10^{-14} \rm \, erg \, cm^{-2} \, s^{-1}$. No excess can be explicitly attributed to unidentified clumps among \texttt{wavdetect} sources.}
\label{fig:lognlogs}
\end{figure}

{We based the classification of clump candidates on 
the results from  Sections\,\ref{sec:opticalcounterparts} and \ref{sec:hardnessratio}.}

{\noindent\textbf{C1} is located near the chip edge and therefore has a large PSF, which complicates source characterization. The positive hardness ratio, however, is not indicative of diffuse thermal emission expected from gas clumps.}

{\noindent\textbf{C2} does not exhibit a definite X-ray peak, although the local PSF is relatively large. It overlaps with $3$ background galaxies. However, based on the estimated X-ray flux of these galaxies, they do not account for the detected X-ray emission. The positive hardness ratio suggests a possible AGN origin.}

{\noindent\textbf{C3}'s well-defined excess emission peaks around a previously detected and removed \texttt{wavdetect} source and coincides with $4$ SDSS-galaxies at relatively high redshifts ($z = 0.41 - 0.59$). As the estimated X-ray flux from the overlapping galaxies cannot explain the measured flux, it is likely that the main contributors are the AGN in these background galaxies, smeared out in the X-ray image.}

{\noindent\textbf{C4} is associated with a previously detected X-ray source's extended emission. This X-ray source coincides with $3$ background galaxies, of which none is expected to display a detectable amount of X-ray emission. With an offset of $0.012\arcmin$, the NASA/IPAC Extragalactic Database (NED) lists a galaxy cluster (X-CLASS 1684) with an unknown redshift. In addition, with a larger offset of $\sim\! 0.12\arcmin$, a rich galaxy cluster identified in the DESI Legacy Imaging Surveys\,\citep{2021ApJS..253...56Z} lies at $z \approx 0.80$, encompassing the source of interest.
We speculate that the background clusters, the X-ray source, and the clump candidate are the same sources.
The temperature $kT = 3.3 \pm 0.6 \, \rm  keV$ from the spectral fit of the X-ray source is consistent with this explanation.
}

{\noindent\textbf{C5}'s peak emission encloses two previously excluded \texttt{wavdetect} sources, one of them exhibiting extended emission, which was missed by \texttt{wavdetect}, then identified as a clump candidate. For this \texttt{wavdetect} source, an Abell\,1795 member galaxy counterpart was found in the SDSS data at $z_{\rm spec}=0.061$ with a stellar mass of $M_* \approx 3 \times 10^{11} \, \rm M_{\sun}$. The expected X-ray flux of the galaxy, $F_{0.7-2\,\rm keV} \approx 1.0 \times 10^{-14} \,\rm erg\,cm^{-2}\,s^{-1}$, is comparable with the flux measured for the clump candidate ($F_{0.7-2\,\rm keV} \approx 1.2 \times 10^{-14} \,\rm erg\,cm^{-2}\,s^{-1}$). The galaxy's spectrum, extracted from \textit{Chandra} data, can be primarily described with an \textit{apec} emission with $kT = (0.82 \pm 0.10) \, \rm keV$, which, along with the hardness ratio, indicates that the galaxy's diffuse emission is what was misidentified as a clump.}

{\noindent\textbf{C6} coincides with an SDSS galaxy at $z\approx0.18$ with an estimated stellar mass of $M_* \approx 3.6 \times 10^{11} \, \rm M_{\sun}$ and an estimated X-ray emission ($F_{0.7-2\,\rm keV} \approx 1.8 \times 10^{-15} \,\rm erg\,cm^{-2}\,s^{-1}$) comparable to C6's measured flux ($F_{0.7-2\,\rm keV} \approx 2.6 \times 10^{-15} \,\rm erg\,cm^{-2}\,s^{-1}$). The match between C6 and the co-spatial galaxy is also reflected in the candidate's $HR \approx 0.5$ hardness ratio.}

{\noindent\textbf{C7}, having a total flux of $F_{0.7-2\,\rm keV} \approx 7.7 \times 10^{-15} \,\rm erg\,cm^{-2}\,s^{-1}$, might be physically connected with a $z\approx0.19$ background galaxy, which is the most massive one in the sample with $M_* \approx 4.5 \times 10^{11} \, \rm M_{\sun}$ and $F_{0.7-2\,\rm keV} \approx 3.0 \times 10^{-15} \,\rm erg\,cm^{-2}\,s^{-1}$. In addition, centered at a $\sim\!0.3\arcmin$ offset to the east from this galaxy, a galaxy group (SDSSCGB 11608) is located at $z=0.173$ \citep{2009MNRAS.395..255M} within the C7 region.}

{\noindent\textbf{C8} shows two X-ray peaks, one overlapping with a low-mass SDSS galaxy expected to display no significant X-ray emission and another point-like optical source not listed among the SDSS galaxies but registered in NED as a red galaxy (WISEA J134810.31+260947.3). As the X-ray peaks are positioned at these objects and the estimated hardness ratio, although exhibiting large error bars, is close to zero, we suspect a physical connection between the clump and the AGN within these galaxies.}

{\noindent\textbf{C9}'s peak cannot be associated with an SDSS galaxy, however, an infrared object classified as an irregular source is detected by WISE (WISEA J134655.49+262236.9) at the peak's position, indicating a distant or low-luminosity AGN identified as clump candidate.}

{\noindent\textbf{C10} shows diffuse emission extending outside the candidate's region, where a massive ($M_* \approx 3.7 \times 10^{11} \, \rm M_{\sun}$), but relatively distant ($z\approx0.55$) SDSS galaxy is located. Its expected X-ray emission is approximately an order of magnitude lower than that measured within the C10 region. The extension of the diffuse X-ray emission within C10, similarly to C9, can be associated with an irregular source identified with WISE (WISEA J134844.78+270520.8).}

{\noindent\textbf{C11}'s positive hardness ratio does not indicate diffuse gas emission, however, no objects in SDSS are reported at the position of the X-ray excess, nor at the position of the closest \texttt{wavdetect} source making a possible AGN origin also questionable.}

{\noindent\textbf{C12} can be associated with a background galaxy cluster, WHL J134645.8+264625 at $z=0.4271$\,\citep{2015ApJ...807..178W} located at and extending beyond the southern region of C12, where an X-ray excess is indeed prominent.}

{\noindent\textbf{C13} is visibly linked with a background AGN (WISEA J134636.14+262715.7) located at $z=0.305$\,\citep{2009ApJS..180...67R}. This connection is supported by the positive hardness ratio measured within the candidate's region.}

{\noindent\textbf{C14}'s peak emission coincides with a low-mass background galaxy, which, however, cannot explain the detected X-ray excess. The overall positive hardness ratio might suggest the presence of an AGN hosted by this galaxy but note that the northern part of the clump candidate, including the peak, possibly overlaps with the outskirts of a $z\approx0.74$ galaxy cluster identified in DESI.}

{\noindent\textbf{C15}'s classification is identical to that of C13, with the candidate's emission described as excess power-law radiation (note the close to zero hardness ratio) from a background AGN (WISEA J135022.71+261003.8) at $z=1.343806$.}

{\noindent\textbf{C16} is associated with two background galaxies. The closest ($z_{\rm spec} = 0.178$) is the more massive one with $M_* \approx 2.7 \times 10^{11} \, \rm M_{\sun}$, and with an estimated X-ray flux of $F_{0.7-2\,\rm keV} \approx 8.9 \times 10^{-16} \,\rm erg\,cm^{-2}\,s^{-1}$, a factor of $\sim$\,$6.6$ fainter compared to the total flux measured within C16. Comparison with a smaller region at the galaxy's position within C16, however, yields good agreement between the measured and the expected X-ray fluxes. North of the massive galaxy lies a possible AGN detected as an irregular IR source by WISE (WISEA J134817.26+260338.8). Although the negative hardness ratio measured throughout C16 indicates diffuse emission, the relatively large PSF, along with the small projected distance between these objects, complicates the source classification.}

{\noindent\textbf{C17} overlaps with a background galaxy cluster (WHL\,J\,$134950.2+270623$) lying at $z\approx0.42$ \citep{2015MNRAS.453...38R}, and the X-ray emission detected as a clump candidate may originate from its ICM. It is confirmed by the negative hardness ratio and its small error bars.}

\begin{figure}
\flushleft
\includegraphics[width=.49\textwidth]{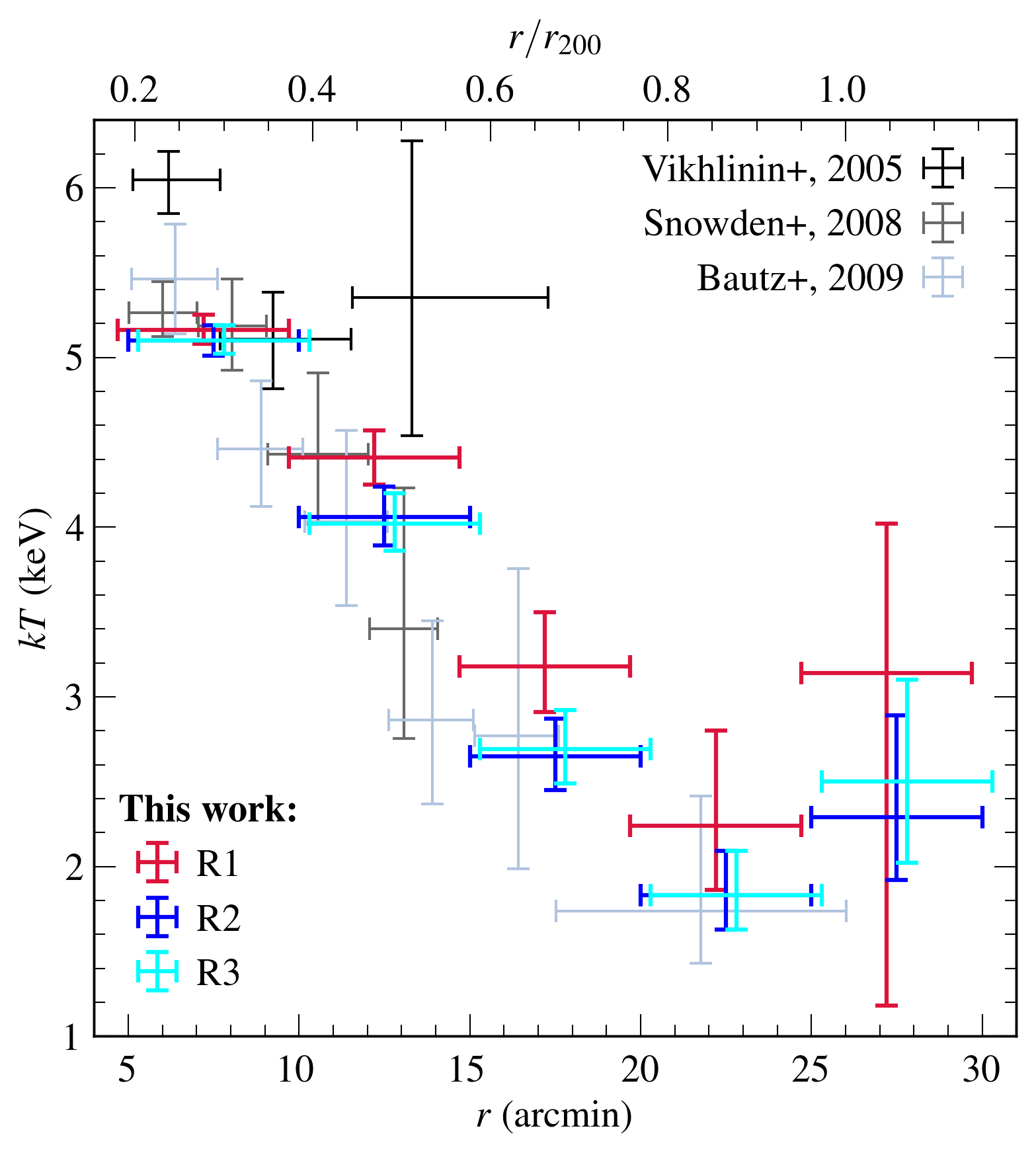}
\caption{Projected temperature profiles of the Abell\,1795 cluster outskirts from three rounds (R1$-$3) of spectral extraction. In the R1 data, only the \textit{Suzaku} resolved point sources, while from the R2 data, all point sources down to a flux of $5.8\times 10^{-15} \,\rm erg\, cm^{-2} s^{-1}$ were removed. The R3 profile, on top of that, accounts for the effect of the resolved clump candidates, which is insignificant. For comparison, we included three additional profiles from previous works.
}
\label{fig:T}
\end{figure}

{\noindent\textbf{C18} is arranged around a \texttt{wavdetect} source, registered in SDSS as a background galaxy ($z\approx0.6$), and encompasses two low-mass background galaxies ($z\approx0.4$). Based on the analysis of SDSS data, none of these can explain the measured X-ray flux, but the co-spatial alignment of the X-ray and optical sources, as well as the hardness ratio, point towards an AGN origin for C18's emission.}

{\noindent\textbf{C19}'s weak excess consists of two equally bright peaks, of which one is co-aligned with two SDSS galaxies, one lying in the background, and the other one possibly gravitationally bound to A1795 with a photometric redshift of $\sim\!0.095\pm0.027$. The second X-ray peak has no optical counterpart. Although C19 is a weak source and its hardness ratio is inconclusive, it cannot be ruled out that it includes a genuine clump.}

{\noindent\textbf{C20} may be physically connected with a background quasar (WISEA\,J\,$134643.44+265417.6$) lying at $z\approx1.75$
\citep{2009ApJS..180...67R}. The estimated hardness ratio of $HR \approx 0.1$ also supports this connection.}

{\noindent\textbf{C21} overlaps with two low-mass SDSS galaxies in the background ($z\approx0.4-0.5$), none of which is expected to account for the measured X-ray flux. The galaxy to which C21 is attached and was detected by \texttt{wavdetect} shares similar properties, but it is cataloged in WISE as an irregular source, thus, it is likely an AGN. Although the large error bars on the hardness ratio and the large local PSF in this field complicate source characterization, it cannot be ruled out that C21's emission can be associated with a background AGN.}

{\noindent\textbf{C22}'s observed emission remains elusive. The single SDSS galaxy coinciding with the candidate is a low-mass background system. Within $0.1\arcmin$ of the X-ray peak, three objects are registered in NED, two point sources and an ultraviolet source. In addition, within $\sim\!0.65\arcmin$ to the northeastern direction lies a distant galaxy cluster (GMBCG J206.84924+27.13581) at $z=0.402$\,\citep{2010ApJS..191..254H} (note that the smallest fitting circle of C22 is $\sim\!0.45\arcmin$).}

{\noindent\textbf{C23 and C24}, located at the chip edge, are smeared by the large local PSF and cannot be classified.}

\subsection{Surface brightness profile} \label{sec:sb}

\subsubsection{Azimuthally averaged surface brightness profiles} \label{sec:sbazimuthallyavg}

Extraction and fitting of the \textit{Chandra} and \textit{Suzaku} surface brightness profiles of Abell\,1795 were carried out using the \texttt{PROFFIT} software package \citep{2011A&A...526A..79E}.
The azimuthally averaged profiles were constructed considering only the outskirts of Abell\,1795 beyond a radius of $\sim\!11.25\arcmin$ ($\sim\!0.8\,\rm Mpc$).

To generate the surface brightness profiles of the diffuse emission, we excluded bright point sources from the extraction regions. 
Because the \textit{Chandra} and \textit{Suzaku} observations of Abell\,1795 have different source detection sensitivities, it is important to define a specific limiting flux above which point sources are removed from both data sets.
We used the broad-band source list identified by \texttt{wavdetect} (Section\,\ref{sec:sourcedetection}); for \textit{Suzaku}, we tailored it to its larger PSF as described in Section\,\ref{sec:suzaku-cxb}, and applied a uniform photon flux threshold of $2\times10^{-6} \rm \,photon\,cm^{-2}\,s^{-1}$ {, which corresponds to an energy flux of $\sim$$5.8\times10^{-15} \, \rm erg \,cm^{-2} \, s^{-1} $ for both telescopes.
We found that $68\,\%$ of the \textit{Chandra} detected sources lie above this threshold.}

When building the profile, we binned the $0.7-2$\,keV photon counts within $>\!60\arcsec$ (for \textit{Suzaku} within $>\!160\arcsec$) width concentric annuli centered on the cluster's X-ray peak (J2000(RA, Dec) = ($207.2194$, $26.5920$) in FK5 coordinates), applying logarithmic binning.
The instrumental background (stowed background for \textit{Chandra}, and NXB for \textit{Suzaku}) and exposure maps were also taken into account.

We fitted the obtained \textit{Chandra} and \textit{Suzaku} profiles separately with a power-law model of $S_x \propto r^{-\alpha} + B$, where $B$ represents the sky background constant.
We note that the power-law model provides a better description for the outskirts compared to the standard $\beta$-model due to the strong coupling between the $\beta$ parameter and the core radius\,\citep{1999ApJ...517..627M}, which falls outside the fitting region, therefore could not be constrained.

In Figure\,\ref{fig:sb1}, we display the resulting background subtracted surface brightness profiles and the best-fit models in units of $\rm erg$ $\rm s^{-1}$ $\rm cm^{-2}$ $\rm deg^{-2}$ to allow a direct comparison between the two datasets.
For the photon flux to energy flux conversion, we assumed an 
\textit{apec} model with $kT=2\,\rm keV$ and with $0.2$ solar abundance. 
We also display the instrumental and sky background profiles to demonstrate their significant contribution to the outskirts. Note the factor of $\sim\!4$ difference between the level of \textit{Chandra}'s stowed background and \textit{Suzaku}'s NXB, and the dominant contribution of the stowed background to the surface brightness beyond $\sim\!0.5\,r_{200}$.

The two profiles are in agreement with a best-fit power-law exponent of $4.05 \pm 0.14 $ and $4.18 \pm 0.19$ for \textit{Chandra} and \textit{Suzaku}, respectively.
These relatively high exponents represent a steep fall in the cluster surface brightness beyond $0.4\, r_{200}$. 
We emphasize that this degree of agreement between the \textit{Chandra} and \textit{Suzaku} extracted profiles is the consequence of the coordinated source removal procedure, which we explain further in Section\,\ref{sec:suzaku_sb_discussion}.

\subsubsection{Azimuthal surface brightness profiles} \label{sec:sb_sectors}

Azimuthal variations in the structure of the ICM -- and so in the surface brightness profile -- may be present even in dynamically relaxed systems \citep[e.g.][]{2011MNRAS.414.2101U,2012A&A...541A..57E,2014MNRAS.437.3939U}.
In Abell\,1795, \cite{2009PASJ...61.1117B} reported excess emission in the northern direction based on $0.5-2\,\rm keV$ \textit{Suzaku} data, with an apparent peak near $23\,\arcmin$.
To probe asymmetries in the cluster's diffuse emission with \textit{Chandra}, we built the $0.7-2\,\rm keV$ surface brightness profiles from circular wedges toward the north, south, east and west directions with opening angles of $90^{\circ}$.

Source removal in this case was carried out without the limiting flux threshold applied in Section\,\ref{sec:sbazimuthallyavg}, i.e.,\ considering the full list of broad-band \texttt{wavdetect} sources.
To achieve signal-to-noise ratios comparable to those obtained for the azimuthally averaged profile, we increased the bin width to $>\!160\arcsec$.
{Because of the uneven azimuthal coverage} (Figure\,\ref{fig:mosaic}), we introduced a uniform fitting range of $7\arcmin-31\arcmin$ and we fitted the profiles within this range in a way identical to that applied for the azimuthally averaged profile (Section\,\ref{sec:sbazimuthallyavg}).
{Note that the fitted sky background constant has no effect on the obtained power-law exponents, therefore, we did not subtract its contribution for plotting purposes.}
The obtained sky background-included profiles are displayed in Figure\,\ref{fig:sb-wedges} with the best-fit power-law models.
The corresponding power-law exponents are $4.50 \pm 0.20$, $3.99 \pm 0.17$, $2.78 \pm 0.12$, and $3.29 \pm 0.15$ for north, south, east, and west, respectively.

We found that the excess brightness from the north between $r_{500}$ and $r_{200}$ recorded by \cite{2009PASJ...61.1117B} in the \textit{Suzaku} data is not present in the \textit{Chandra}-extracted profile.
This fluctuation likely comes from a point source located at $207.26^{\circ}$, $26.98^{\circ}$ in the RA/Dec plane, which exhibits an extremely broad PSF resulting in an apparent source radius of $\sim\!3\arcmin$ in the \textit{Suzaku} observations.
Although this point source was successfully identified by \texttt{wavdetect}, the size of the source region required manual adjustment when removing it from the \textit{Suzaku} data for the azimuthally averaged surface brightness profile.
We emphasize that the \textit{Suzaku} profile indeed displayed an excess at $\sim\!25\arcmin$ without this adjustment.

In the bottom panel of Figure\,\ref{fig:sb-wedges}, we present the azimuthal variations in the level of the surface brightness in the four extraction directions.
Although the profiles exhibit a significant scatter, an obvious enhancement towards the western direction is seen beyond $\sim\!18\arcmin$ with an average significance of $4.2\,\sigma$.
A possible explanation for this deviation is given in Section\,\ref{sec:sb_excess_west}.

\subsection{$\log{N}-\log{S}$ plot}\label{sec:logn-logsplot}

The $\log{N}-\log{S}$ plot accounts for the cumulative number of sources ($N$) as a function of source flux ($S$) per square degree within a selected field and includes mostly point sources, i.e., AGN.
Galaxies become significant contributors only in the soft X-ray band at $\lesssim\!10^{-17} \, \rm erg \, cm^{-2} \,s^{-1}$ \,\citep{2017ApJS..228....2L}, which is below the detection limit of the present analysis.

We built the $\log{N}-\log{S}$ plot incorporating all sources detected on the $0.5\!-\!2\,\rm keV$ Abell\,1795 mosaic, inclusive of the central pointing, with a detection significance of $\geq\!3\sigma$ ($\sim\!94\%$ of the \texttt{wavdetect} identified sources).
Source fluxes were calculated using the \texttt{srcflux} CIAO tool assuming an absorbed power-law spectrum with a photon index of $\Gamma = 1.6$.
Local background annuli were defined with \texttt{roi} in consideration of overlapping regions and with the inner background radius of twice the source radius.
To accurately account for the high variability of the FWHM across the FOV, the PSF fraction in each source and background region was computed with \texttt{marx} simulations.
Finally, for the $\log{N}-\log{S}$ plot, we adopted the sources' intrinsic flux, i.e., the unabsorbed model flux estimates based on the power-law assumption.

For the area normalization, i.e., to specify the solid angle surveyed as a function of limiting flux, we constructed a sky coverage histogram in the soft band.
The input limiting sensitivity map for this was created {for the merged data} with the \texttt{lim\_sens} CIAO tool assuming a $\Gamma=1.6$ power law model.
This sensitivity map determines the $3\,\!\sigma$ limiting flux variations across the FOV due to the spatial variations of the PSF, effective exposure and background level with the inclusion of the sky, instrumental, and the ICM background components.

We present the $0.5-2\,\rm keV$ $\mathrm{log}N\!-\!\mathrm{log}S$ plot for the Abell\,1795 field in Figure\,\ref{fig:lognlogs} compared to the results from the $7$\,Ms Chandra Deep Field South (CDF-S) data \citep{2017ApJS..228....2L}. 
A minimum flux threshold of $5.4 \times 10^{-16} \, \rm erg \, cm^{-2} \, s^{-1}$, adjusted to a minimum survey area of $5\%$ of the FOV, was applied to reflect the high uncertainties at low sky coverage, while a maximum flux threshold of $7.0 \times 10^{-14} \, \rm erg \, cm^{-2} \, s^{-1}$ was also introduced to correct for the lowest exposure survey areas.
Within $1\,\sigma$ uncertainty, the source number counts in the field of Abell\,1795 agree with those of the CDF$-$S with only minor discrepancies at the high-flux-tail (source deficit) and at $\sim\!10^{-14} \, \rm erg \, cm^{-2} \, s^{-1}$ (source excess).

Besides the CDF-S field, we compared the present results with other surveys to test the significance of the excess, which is a potential contribution from ICM clumps.
Specifically, the cumulative source number counts at $S>10^{-14} \, \rm erg \, cm^{-2} \, s^{-1}$ exhibit a scatter of $\sim\!45$ source counts in a sky area of one square degree based on the the Lockman Hole surveyed by \textit{XMM-Newton}\,\citep{2001A&A...365L..45H}, a combination of six different surveys performed with \textit{ROSAT}, \textit{Chandra}, and \textit{XMM-Newton}\,\citep{2003ApJ...588..696M}, and the XMM$-$COSMOS field\,\citep{2007ApJS..172..341C}.
As this scatter surpasses the excess seen in the Abell\,1795 data, we attribute this source excess to cosmic variance.

\begin{figure*}[h!]
\centering
\includegraphics[trim={-0.6cm 0 0 0cm},clip, width=.47\textwidth]{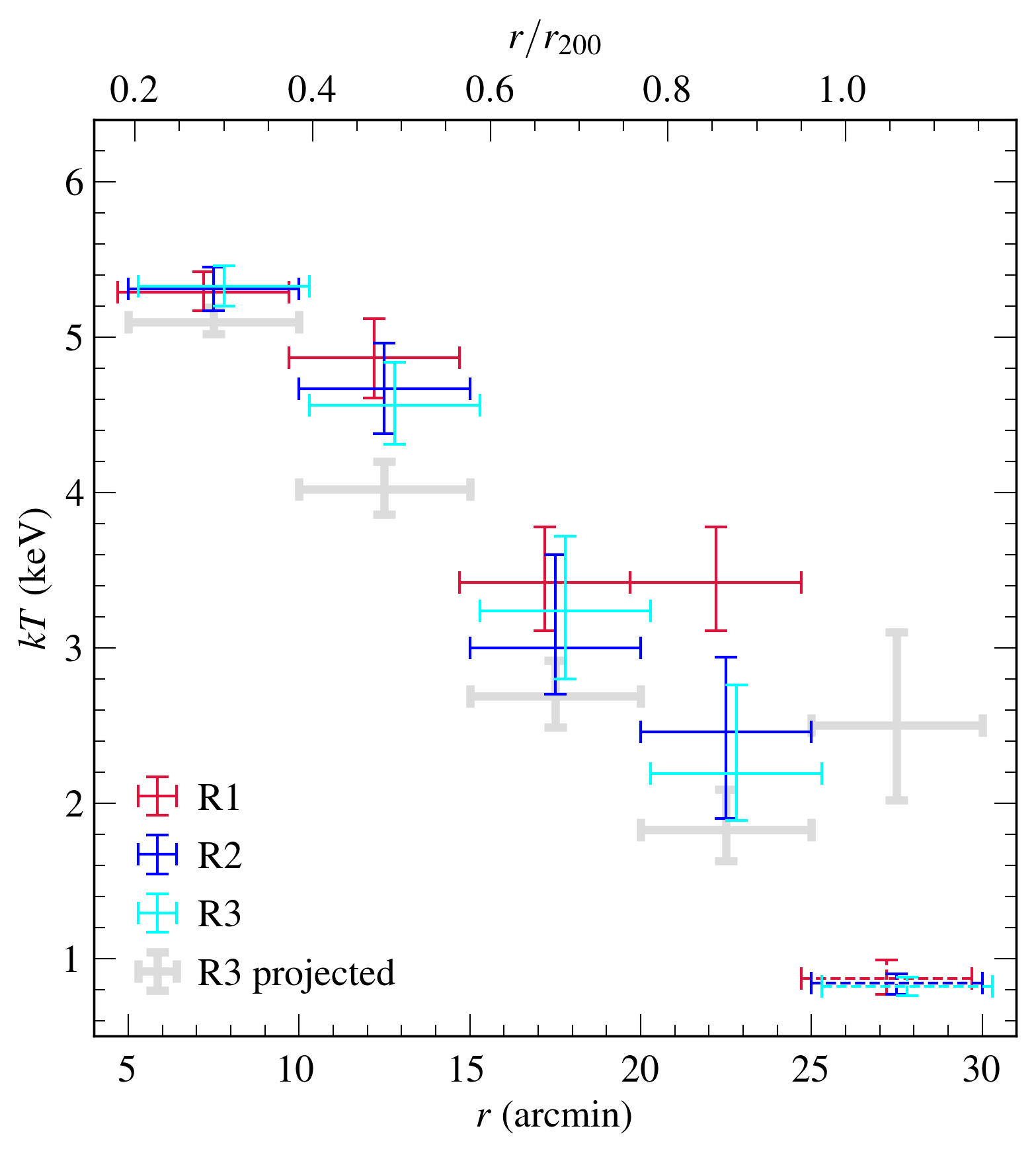}
\includegraphics[trim={0 0 0 0},clip, width=.48\textwidth]{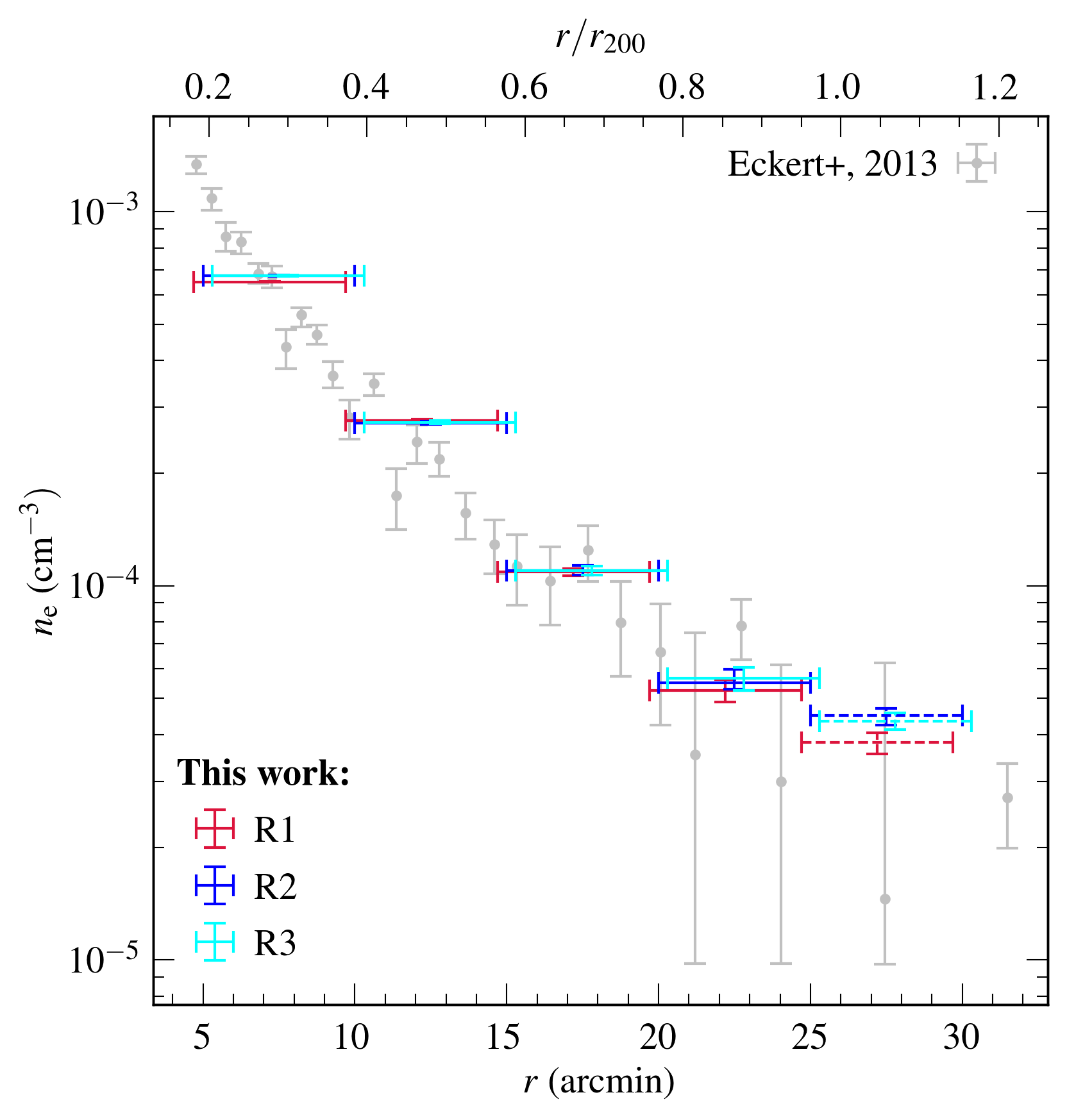} \\
\includegraphics[trim={0 0 0 0},clip, width=.47\textwidth]{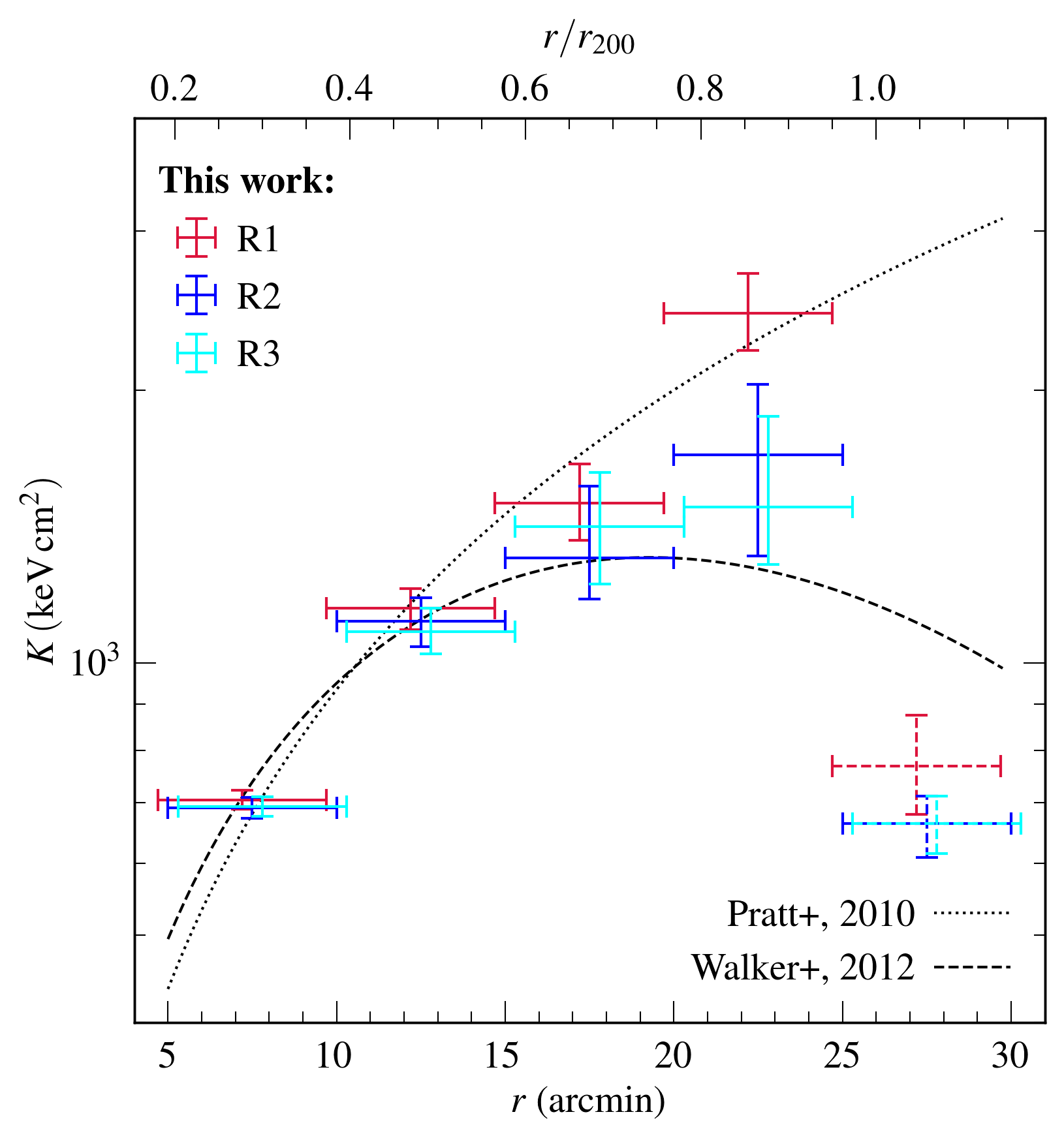}
\includegraphics[trim={0 0 0 0cm},clip, width=.48\textwidth]{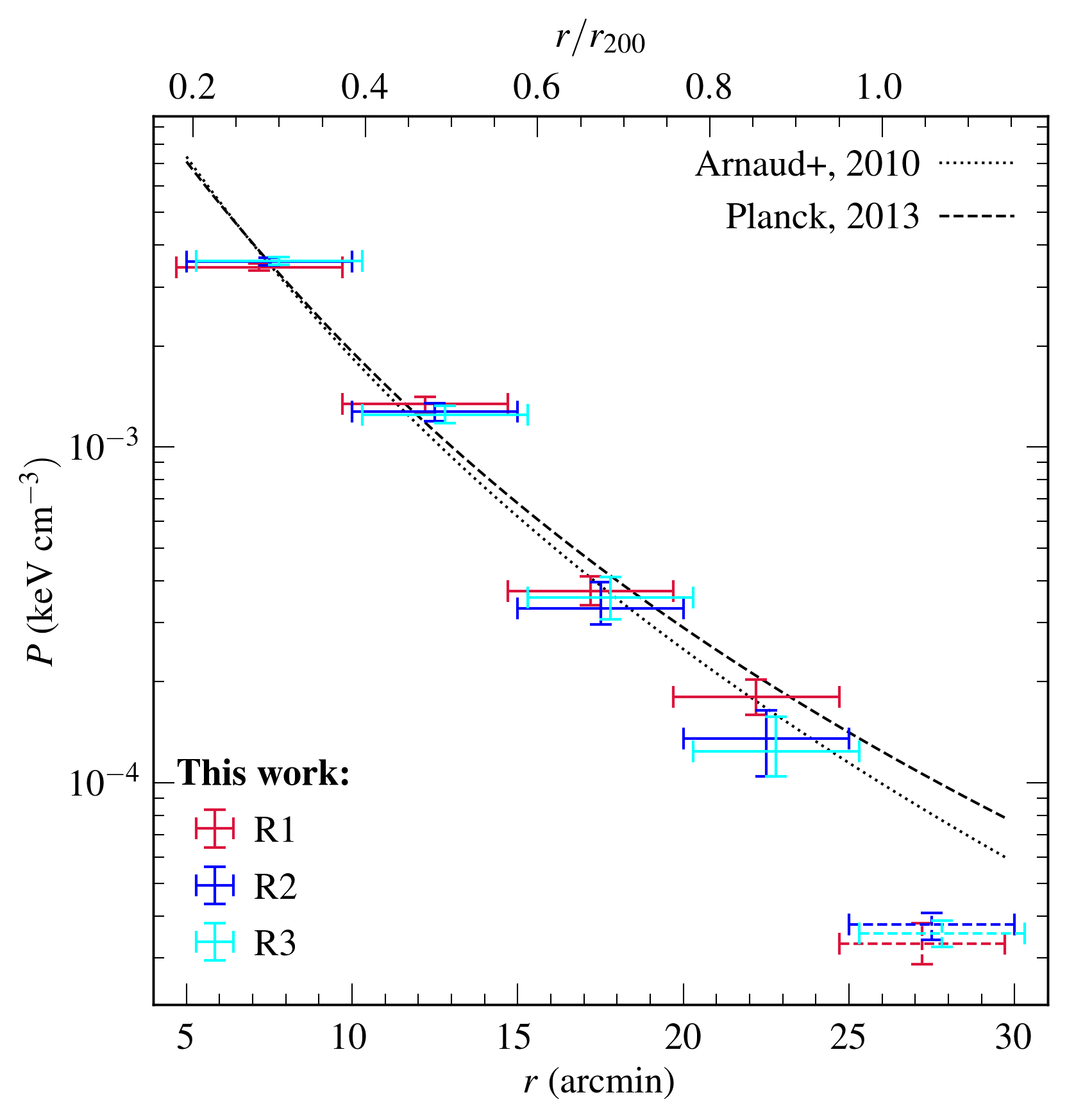}\\
\caption{
Deprojected profiles of the cluster outskirts with R1$-$3 denoting the deprojected equivalents of the profiles shown in Figure\,\ref{fig:T} (Section\,\ref{sec:suzaku-cxb}).
{The outermost data points, shown with dashed lines, exhibit large uncertainties introduced with the deprojection.} 
\textbf{\textit{Top left:}} Comparison between the deprojected and the projected temperature profiles.
\textbf{\textit{Top right:}} Electron density profiles overplotted with \textit{ROSAT} measurements of the same cluster by \cite{2013A&A...551A..22E}.
\textbf{\textit{Bottom left:}} Entropy profiles compared with the profile of \cite{2010A&A...511A..85P}, which models a galaxy cluster formed by gravitational collapse with no additional heating or cooling and with the analytic function of \cite{2012MNRAS.427L..45W} fitted on a sample of \textit{Suzaku} observed clusters.
\textbf{\textit{Bottom right:}} Pressure profile overplotted with the best-fit models of \cite{2010A&A...517A..92A} and the \cite{2013A&A...550A.131P}
.}
\label{fig:deproj}
\end{figure*}

\subsection{Thermodynamic profiles}\label{sec:thermo_prof}
The radial profiles of the ICM's thermodynamic properties were derived from spectral analysis of the \textit{Suzaku} data in a way identical to that described in Zhu et al. 2023 (submitted).
We grouped the spectral counts so that each channel contains at least $1$ count, then modeled the XIS spectra with \texttt{Xspec}.
Parameter estimation was carried out using the C-statistics due to the Poissonian distribution of the data.
We described the ICM emission as an absorbed, collisionally ionized thermal plasma with a $tbabs \times apec$ model, and the background contribution to the detected X-ray signal is considered as explained in Sections\,\ref{sec:suzaku-foreground}-\ref{sec:suzaku-particle-bg}.
Spectral fitting was carried out in three rounds (R$1-3$, see Section\,\ref{sec:suzaku-cxb}) in $5$ radial bins enclosing a $5\,\arcmin-30\,\arcmin$ annulus centered around the cluster.
Based on \cite{2013Natur.502..656W} and \cite{2017MNRAS.470.4583U} we fixed the abundance at $Z = 0.3\,Z_{\odot}$ beyond $r_{500}$, where $Z_{\odot}$ refers to the value presented in \cite{2009LanB...4B..712L}.

The ICM emission we measure this way is superposed from 3D shells along the line of sight, meaning that the cluster properties derived from the 2D spectra suffer from projection effects.
To correct for this effect, we deprojected the 2D spectra using \texttt{Xspec}'s model-dependent \texttt{PROJCT} tool assuming spherical symmetry.
The ICM emission in each shell was modeled with an absorbed \textit{apec}. For background modeling, we used a smoothed NXB generated with \texttt{fakeit} {for the correct propagation of the uncertainties in the NXB spectrum}.

We obtained the temperature profiles directly from the best-fit \textit{apec} models, and derived the electron density from the {deprojected} \textit{apec} normalization, then calculated the plasma pressure and entropy from the deprojected properties.
{The outermost data points of cluster properties derived from deprojection at $25\,\arcmin-30\,\arcmin$ might be biased by significant systematic uncertainties, likely caused by non-zero ICM emission beyond $30\,\arcmin$ or by morphological asymmetries (Section\,\ref{sec:sb_excess_west}), thus might be incorrect. In Figure \,\ref{fig:deproj}, these measurements are drawn with dashed lines but are not considered when discussing the results.}

\begin{figure}
\centering
\includegraphics[trim={0.2cm 0 0 0},clip, width=.49\textwidth]{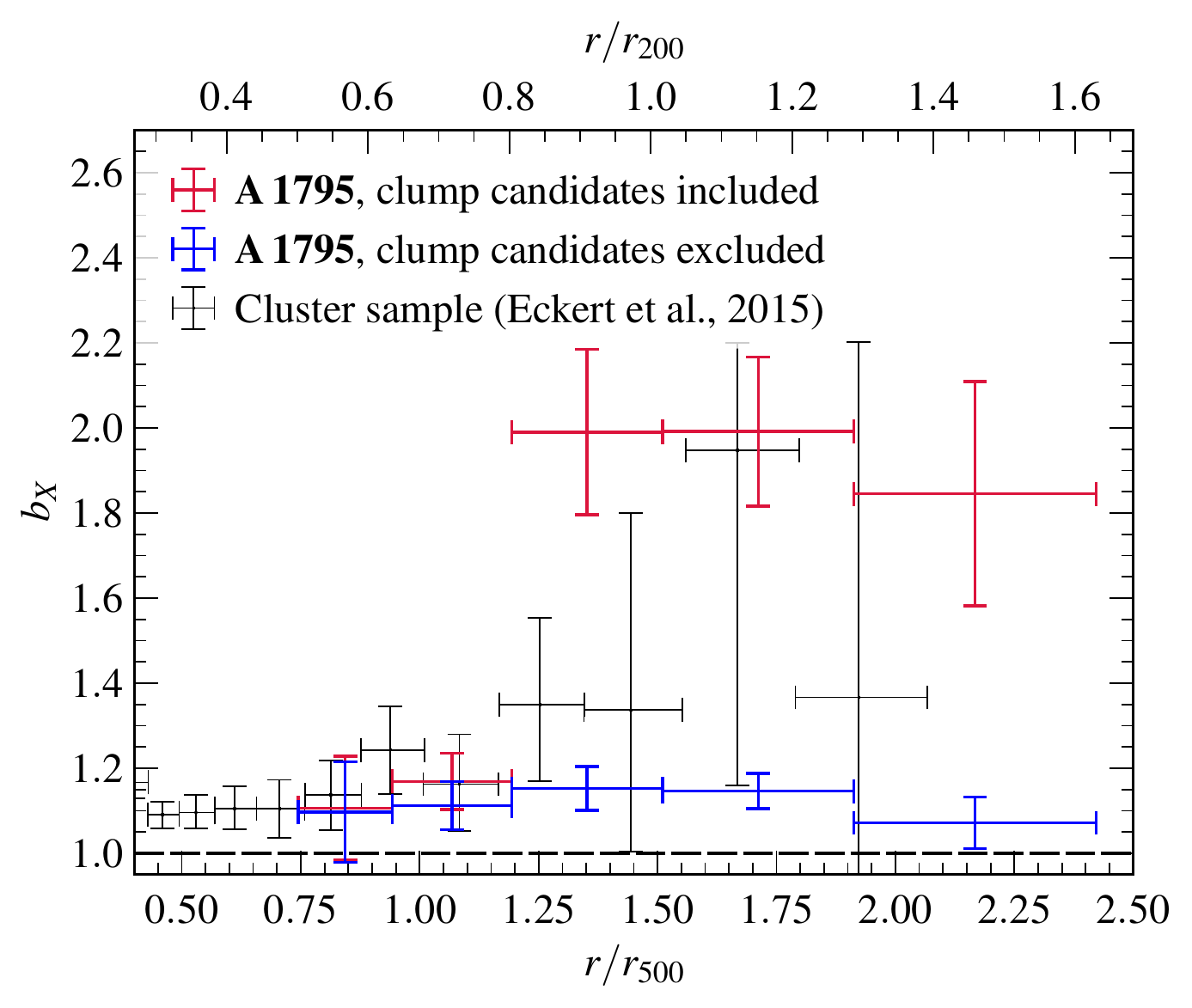}
\caption{Radial variation of the $b_X$ emissivity bias when each clump candidate is included/excluded, compared to the results from the \textit{ROSAT} sample of $31$ galaxy clusters\,\citep{2015MNRAS.447.2198E}.
{The $24$ clump candidates, although being not genuine clumps, cause an increased level of bias above $\sim\!0.8 \, r_{200}$ in Abell\,1795. When each clump candidate is excluded, the profile flattens at $b_X \geq 1$. 
Note that $b_X$ is subject to projection effects and influenced by the sky background contribution, which exhibit very different levels for the \textit{Chandra} and \textit{ROSAT} data sets.}}
\label{fig:fluxbias}
\end{figure}

\subsubsection{Temperature profile}\label{sec:kT}

Figure\,\ref{fig:T} shows projected gas temperature profiles for each run.
We emphasize that the present dataset, extending beyond $r_{200}$, probes the outskirts at their lowest-surface-brightness regime that was, so far, only poorly explored.
Beyond $r_{200}$, the R1 profile {becomes statistically unconstrained}, while the R2 and R3 profiles flatten at {$\gtrsim\!2\,\rm keV$.}
A slight tendency of increasing temperature difference with increasing radii can also be observed between the R1 and R2/R3 profiles.
We find that the R2 and R3 profiles are in agreement, suggesting no influence on the projected temperature from the resolved fraction of clump candidates at the scales of the present study.
Note however that systematic effects, although relatively low, introduce additional uncertainties, which we discuss in Appendix\,\ref{appendix:systematic}.

{In Figure\,\ref{fig:T}, we also present results obtained from \textit{Chandra} observations \citep{2005ApJ...628..655V}, from \textit{XMM-Newton} observations \citep{2008A&A...478..615S}, and from \textit{Suzaku} observations of the northern cluster region \citep{2009PASJ...61.1117B}.
Note that the \textit{Suzaku} observations of Abell\,1795 point in multiple directions, but in the work of \cite{2009PASJ...61.1117B}, only the northern temperature measurements extend beyond $r_{500}$.
Within the error bars, the present results are in agreement with those obtained from \textit{XMM-Newton} and from the former \textit{Suzaku} analysis, while the measurements from \textit{Chandra} show, on average, higher temperatures within $r_{500}$.
We also emphasize that the present \textit{Suzaku} results are not identical to those obtained by \cite{2009PASJ...61.1117B}, not only because of the different azimuthal coverage but also because the changes in \textit{Suzaku} calibration and \textit{apec} code affect the best-fit temperature determination.}

We show the deprojected temperature profiles obtained for R1$-$3 in the top left panel of Figure\,\ref{fig:deproj}.
The R$1-3$ plasma temperatures show a gradual decrease {with radius} reaching $2-3$\,keV at $<\!r_{200}$, and the effect of clump candidates 
is only marginal.

\subsubsection{Electron density profile}
Densities were derived from the emission measure ($EM = \int n_{\mathrm e} n_{\mathrm H}\, dV$) contained in the best-fit \textit{apec} model normalization:
\begin{equation}
\mathrm{Norm} = \frac{10^{-14}} {4 \pi [D_{\mathrm A} (1+z)]^2} \int n_{\mathrm e} n_{\mathrm H}\, dV
\end{equation}
where $D_{\mathrm A}$ is the angular distance to the Abell\,1795 cluster, $dV$ is the volume element, while $n_{\mathrm e}$ and $n_{\mathrm H}$ are the electron and hydrogen number densities, respectively.
We assumed a fully ionized plasma, where $n_{\mathrm e}=1.2\,n_{\mathrm H}$.
{The emitting volume was expressed as $\frac 4 3 (r_{\mathrm out}^3 - r_{\mathrm in}^3) \pi$, and the correction for FOV irregularities was handled by the ARF information.}

We compare the electron number density distribution in the top right panel of Figure\,\ref{fig:deproj} with that obtained by \cite{2013A&A...551A..22E} from \textit{ROSAT} observations of the same cluster.
{Note that the latter one follows the isothermal $\beta$-model with $\beta \approx 2/3$, which is the canonical value for large clusters \citep{1984ApJ...276...38J}.}
Both the different \textit{Suzaku} runs 
(R1-3) and the \textit{ROSAT} profiles are in agreement with each other suggesting a gradual density decrease with radius.
Although \textit{ROSAT} observations extend beyond $r_{200}$, most of the $>\!0.8\,r_{200}$ data points exhibit large errorbars, while the present data confirm $n_{\mathrm e}$ to be $\sim\!5\times 10^{-5} \,\rm cm^{-3}$ at $\sim\!(0.8-1)\,r_{200}$.

\begin{figure}
\centering
\includegraphics[trim={0 0 0 0},clip, width=.49\textwidth]{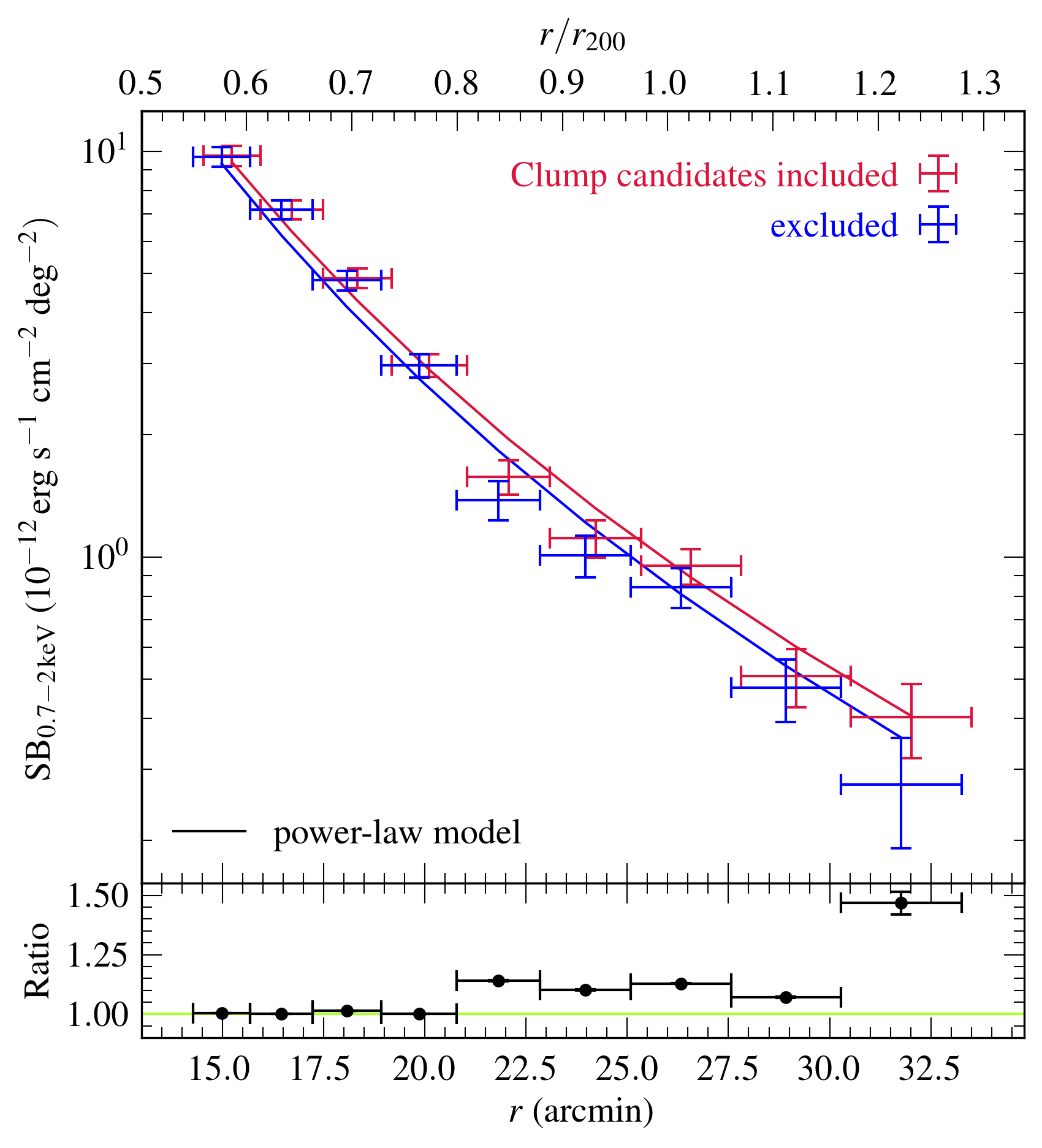}
\caption{Azimuthally averaged surface brightness profile of Abell\,1795 {from \textit{Chandra} data} with clump candidates included and excluded. The bottom panel shows the difference, where a slight decrement in the profile at the radius of the clump candidates becomes prominent when removing these sources. {The best-fit power-law also becomes moderately steeper with the $24$ clump candidates removed, with the best-fit $\alpha$ increasing from $4.18\!\pm\!0.13$ to $4.34\!\pm\!0.09$. This steepening, however, is attributed mainly to the relatively large difference between the last data points and the associated uncertainties.}}
\label{fig:sb_with_and_wo_clumps}
\end{figure}

\subsubsection{Entropy profile}\label{sec:entropy}

The radial distribution of entropy, defined as $K = kT / n_{\mathrm e}^{2/3}$, is presented in the bottom left panel of Figure\,\ref{fig:deproj}.
The observed profile is compared with that predicted assuming pure gravitational heating, where the entropy is expected to follow a power-law relation with $K \sim r^{1.1}$ \citep{2005MNRAS.364..909V}. Expressed with the dimensionless entropy, the profile follows
\begin{equation}
\frac{K(r)}{K_{500}} = 1.47 \left( \frac{r}{r_{500}} \right )^{1.1}
\end{equation}
where $K_{500} = 106 \, \mathrm{keV\,cm^{-2}} \left ( \frac{M_{500}}{10^{14}\mathrm{M}_{\odot}} \right ) ^{\frac 2 3} \left ( \frac{1}{f_{\mathrm{b}}} \right ) ^{\frac 2 3} E(z)^{-\frac 2 3}$ is the characteristic entropy \citep{2005MNRAS.364..909V,2010A&A...511A..85P}, $M_{500} = 6 \times 10^{14}\,\rm M_{\odot}$ is the cluster mass within $r_{500}${\,\citep{2006ApJ...640..691V},} the $f_{\mathrm{b}}$ baryon fraction is taken to be $0.15$, and $E(z) = \sqrt{\Omega_{\mathrm{m}} (1+z)^3 + \Omega_{\Lambda}}$ is the ratio of the Hubble constant at $z$ with its $z=0$ value.
The pure gravitational collapse of a galaxy cluster, however, may be disrupted by non-gravitational processes (e.g., AGN feedback and sloshing close to the cluster core or clumping and accretion shocks at the outskirts), which modify the above picture.
We, therefore, compare our results with the analytical function yielded for a sample of cool core galaxy cluster outskirts ($r>0.2r_{200}$) at $z\leq0.25$ (including Abell\,1795) by \cite{2012MNRAS.427L..45W}:
\begin{equation}
\frac{K(r)}{K(0.3r_{200})} = A \left( \frac{r}{r_{200}} \right ) ^{1.1} \, e^{- \left( \frac{r}{Br_{200}} \right ) ^2}
\end{equation}
with $(A,B) = (4.4^{+0.3}_{-0.1}, 1.00^{+0.03}_{-0.06})$.
This formula predicts an entropy flattening and turnover at $\gtrsim0.4\, r_{200}$.

{The R$2-3$ profiles run between the two model curves and feature no turnover points.
The positive radial entropy gradient suggests that the ICM is convectively stable. However, the measured profile is still lower than the theoretical curve calculated under the assumption of pure gravitational heating. A possible explanation for this deviation is presented in Section\,\ref{sec:electron}.}

{Interestingly, in contrast to Abell\,1795, the entropy profile of Abell\,133 shows a definite flattening following the Walker profile, even upon the removal of clump candidates (R3 profile). For details, refer to the companion paper of Zhu et al. 2023 (submitted).}

\subsubsection{Pressure profile}\label{sec:pressure}
The $P = n_{\mathrm{e}} kT$ radial pressure profile is shown in the bottom right panel of Figure\,\ref{fig:deproj}, over-plotted with the best-fit models of \cite{2010A&A...517A..92A} and the \cite{2013A&A...550A.131P}, both 
parametrizing the generalized form of the pressure profile proposed by \cite{2007ApJ...668....1N}:
\begin{equation}
\frac{P(r)}{P_{500}} = \frac {P_0} { (c_{500}x)^{\gamma} \, [1+(c_{500}x)^{\alpha}]^{(\beta-\gamma)/\alpha}}
\end{equation}
where $P_{500} = 1.65 \times 10^{-3} \, E(z)^{\frac 8 3} \left [ \frac{M_{500}}{3 \times 10^{14} h_{70}^{-1} \mathrm{M}_{\odot}}\right ]^{\frac 2 3} h_{70}^2 \, \rm keV \, cm^{-3}$ is the characteristic pressure which scales with the cluster mass purely based on gravitation, $x=r/r_{500}$, $c_{500}$ is the concentration parameter at $r_{500}$, and $\alpha$, $\beta$, and $\gamma$ are the profile slopes in the intermediate, outer and central regions, respectively.
From a sample of $z<0.2$ clusters explored within $0.6\,r_{200}$ with \textit{XMM-Newton}, \cite{2010A&A...517A..92A} obtained a best-fit parameter set of $[P_0,\,c_{500},\,\alpha,\,\beta,\,\gamma] = [8.403, 1.177, 1.0510, 5.4905, 0.3081]$, while from the sample of \textit{Planck} clusters explored at $0.02\,r_{500} < r < 3\,r_{500}$, the \cite{2013A&A...550A.131P} obtained $[P_0,\,c_{500},\,\alpha,\,\beta,\,\gamma] = [6.41, 1.81, 1.33, 4.13, 0.31]$, yielding a slightly flatter profile.
{The present measurements, overall, are consistent with the models, although the R3 data point above $\sim\!0.75\,r_{200}$ might be indicative of a steeper pressure drop in the outskirts compared to the \textit{Planck} baseline profile.}

\begin{figure*}
\flushleft
\includegraphics[trim=.26cm 0 0 0, clip, width=.325\textwidth]{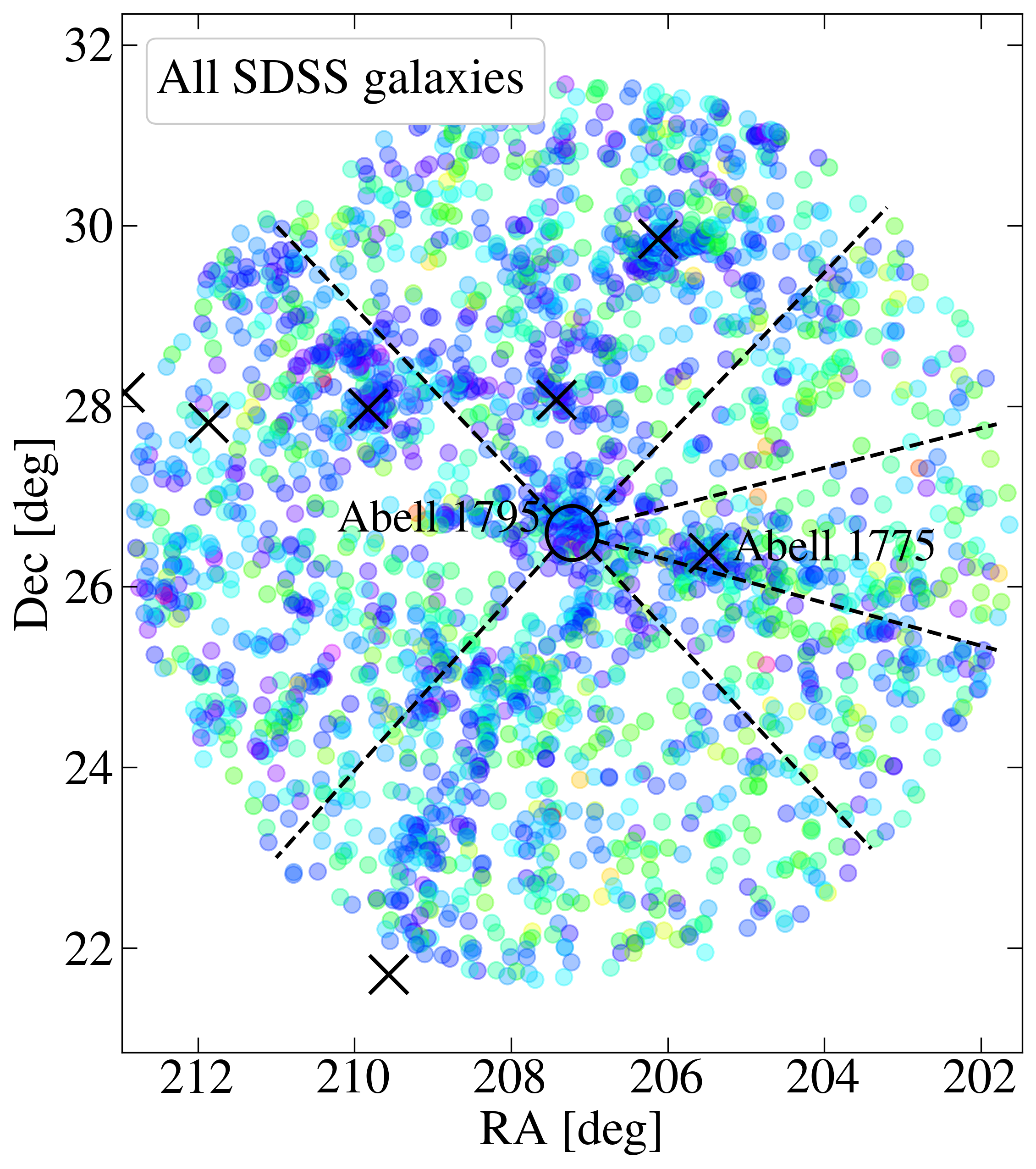}
\includegraphics[trim=.26cm 0 0 0, clip, width=.2905\textwidth]{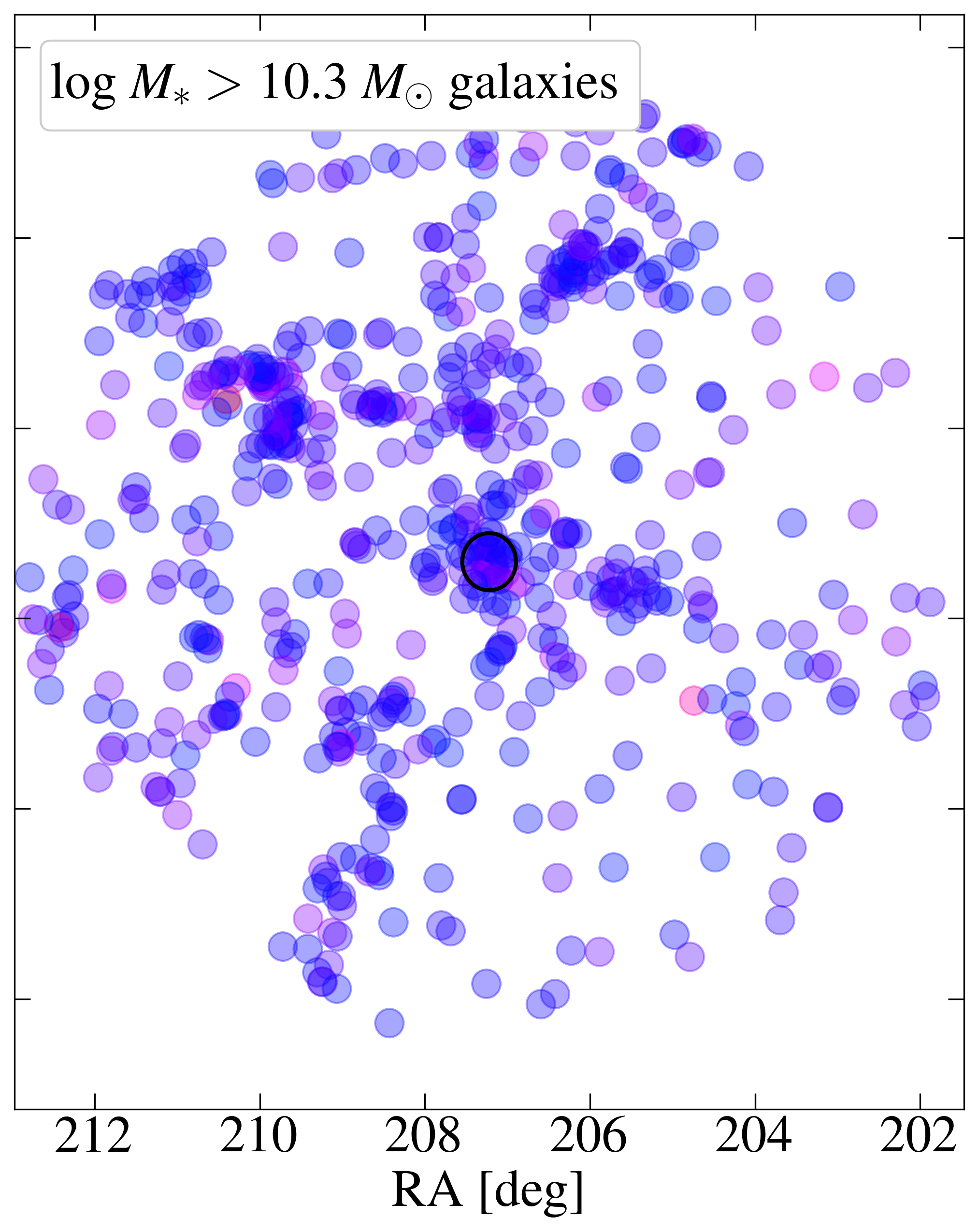}
\includegraphics[trim=.26cm 0 0 0, clip, width=.3675\textwidth]{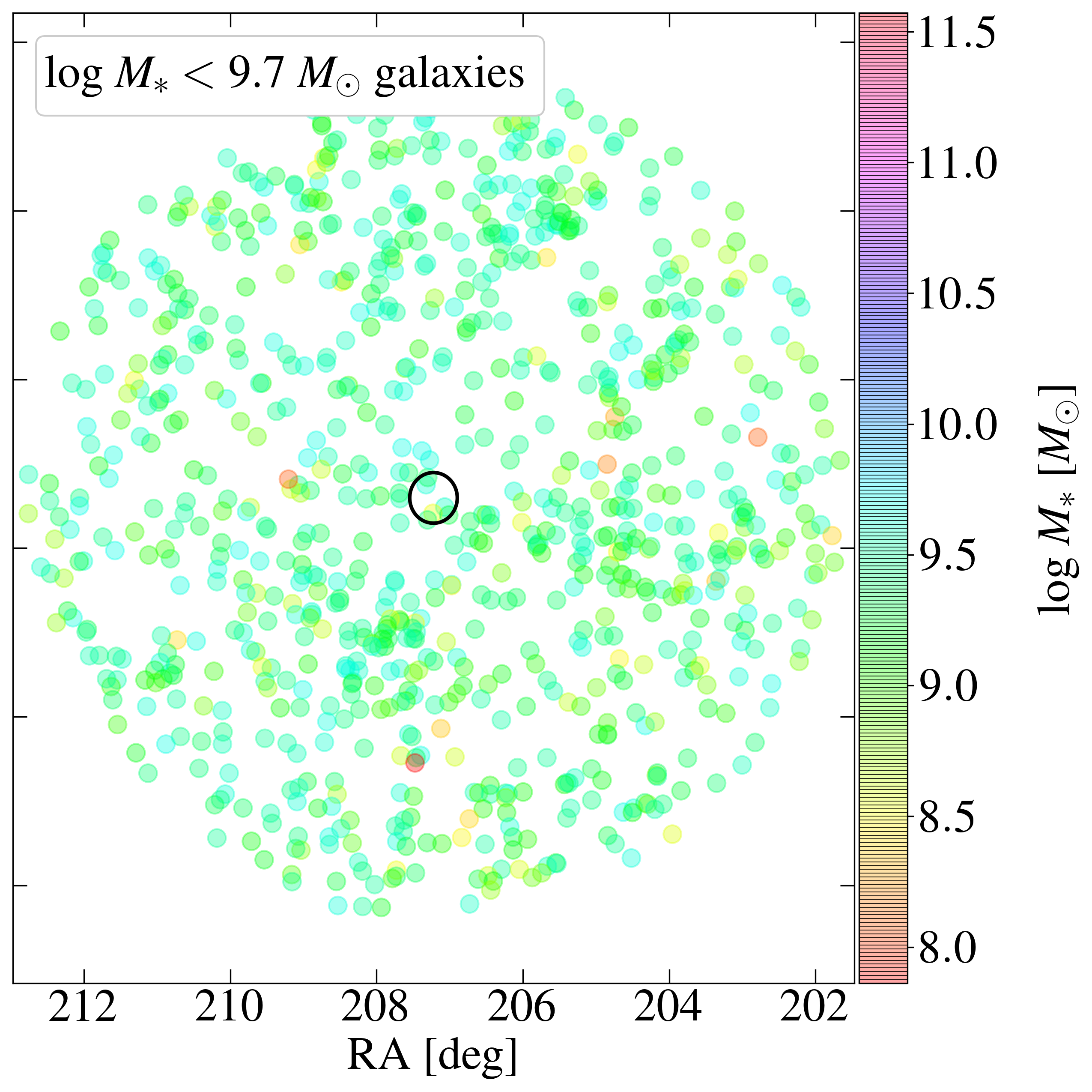}
\caption{\textbf{\textit{Left:}} Spatial distribution of galaxies in the environment of Abell\,1795. Dashed lines denote the direction of the surface brightness extraction regions, but note that the \textit{Chandra} FOV is limited to the area marked by the central black circle. Black crosses indicate the location of neighboring clusters, among which Abell\,1775 is one of the richest in galaxies, and towards which a surface brightness excess was observed in the outskirts of Abell\,1795. \textbf{\textit{Middle:}} Filamentary structures in the distribution of high-mass galaxies. \textbf{\textit{Right:}} Low-mass galaxies, populating volume-filling sheets and voids, producing a homogeneous distribution.}
\label{fig:galdistr}
\end{figure*}

\section{Discussion}\label{sec:discussion}

\subsection{Emissivity bias}\label{sec:fluxbias}
{Almost all clump candidates reported in Section\,\ref{sec:individualclumps} are associated with background objects projected at Abell\,1795 (Table\,\ref{tab:clumpcandidates}).
Although these are not genuine clumps, their effect can mimic clumps seen in simulations.}
In Section\,\ref{sec:thermo_prof}, we compare the thermodynamic profiles with clump candidates included (R2) and removed (R3), but no influence on the overall picture was found.

{More generally, the effect of clump candidates can be assessed, following \cite{2015MNRAS.447.2198E}, by calculating the ratio between the mean and the median of the surface brightness distribution (Figure\,\ref{fig:histograms}).
The $b_X = \mathrm{SB}_{\textrm{mean}} / \mathrm{SB}_{\textrm{median}}$ emissivity bias profile obtained this way is displayed in Figure\,\ref{fig:fluxbias} and listed in Table\,\ref{tab:histograms} with the error bars marking $1\!\,\sigma$ bootstrap confidence intervals from $10^5$ resamplings.
We emphasize that $b_X$ is a purely observational quantity that is strongly influenced by the contribution of the sky background and projection effects.
A perfectly uniform ICM is characterized by $b_X = 1$, when these effects are not present.}

{In Figure\,\ref{fig:fluxbias}, we compare the emissivity bias in the outskirts of Abell\,1795 with the average measurements of \cite{2015MNRAS.447.2198E} for a sample of $31$ \textit{ROSAT/PSPC} observed galaxy clusters.
The Abell\,1795 profile shows a sharp jump at $\sim\!0.8 \, r_{200}$ followed by a plateau with $b_X \sim 2$.
In contrast, the \textit{ROSAT} sample, which has large uncertainties, features a gradual increase with radius.
Upon the removal of the resolved clump candidates from the Abell\,1795 data, the emissivity bias profile flattens at $\gtrsim\!1$ (Figure\,\ref{fig:fluxbias}).
We note that the \textit{Chandra} measurements exhibit a lower level of sky background contamination due to the higher fraction of resolved CXB sources compared to \textit{ROSAT}. Therefore, we propose that the difference between the \textit{Chandra} and \textit{ROSAT} emissivity bias profiles is primarily attributed to the contribution of background sources.}

{Simulations predict that clumps populate the periphery of clusters \citep{2011ApJ...731L..10N}. As a result, the emissivity bias introduced by clumps is expected to increase with radius.
However, the emissivity bias caused by unresolved background objects, such as those identified as clump candidates in the Abell 1795 field, is also expected to follow the same trend.
It is because, at smaller cluster radii, the source detection sensitivity is significantly lower due to the bright ICM emission than in the outskirts of clusters. Therefore, it cannot be ruled out that the population of unresolved sources in the \textit{ROSAT} data, which accounts for the measured emissivity bias, is contaminated with background objects.}

In contrast to the emissivity bias, which is affected by projection effects and sky background contribution, the significance of ICM inhomogeneities in numerical simulations is usually expressed with the $C$ clumping factor measured on the 3-dimensional profile\,\citep{1999ApJ...520L..21M}.
Observational measurements of $C$ are also reported by, e.g., \cite{2015MNRAS.447.2198E} and \cite{2023MNRAS.518.2954D}.
The clumping factor can be computed in the same fashion as the emissivity bias is computed from the observed profile.
To reconstruct the 3-dimensional ICM structure, the observed profile needs to be deprojected, however, its technical implementation is hindered by the low number of counts in the \textit{Chandra} data, where we resolved the clump candidates.
Note, however, that a positive correlation between the emissivity bias and the clumping factor was shown by \cite{2015MNRAS.447.2198E}, who derived a best-fit relation of $b_X \sim \sqrt{C}$ based on a set of synthetic galaxy clusters.
It confirms that the emissivity bias provides a robust estimate for the density bias when calculating the clumping factor is not feasible.

\begin{figure*}
\centering
\includegraphics[width=.45\textwidth]{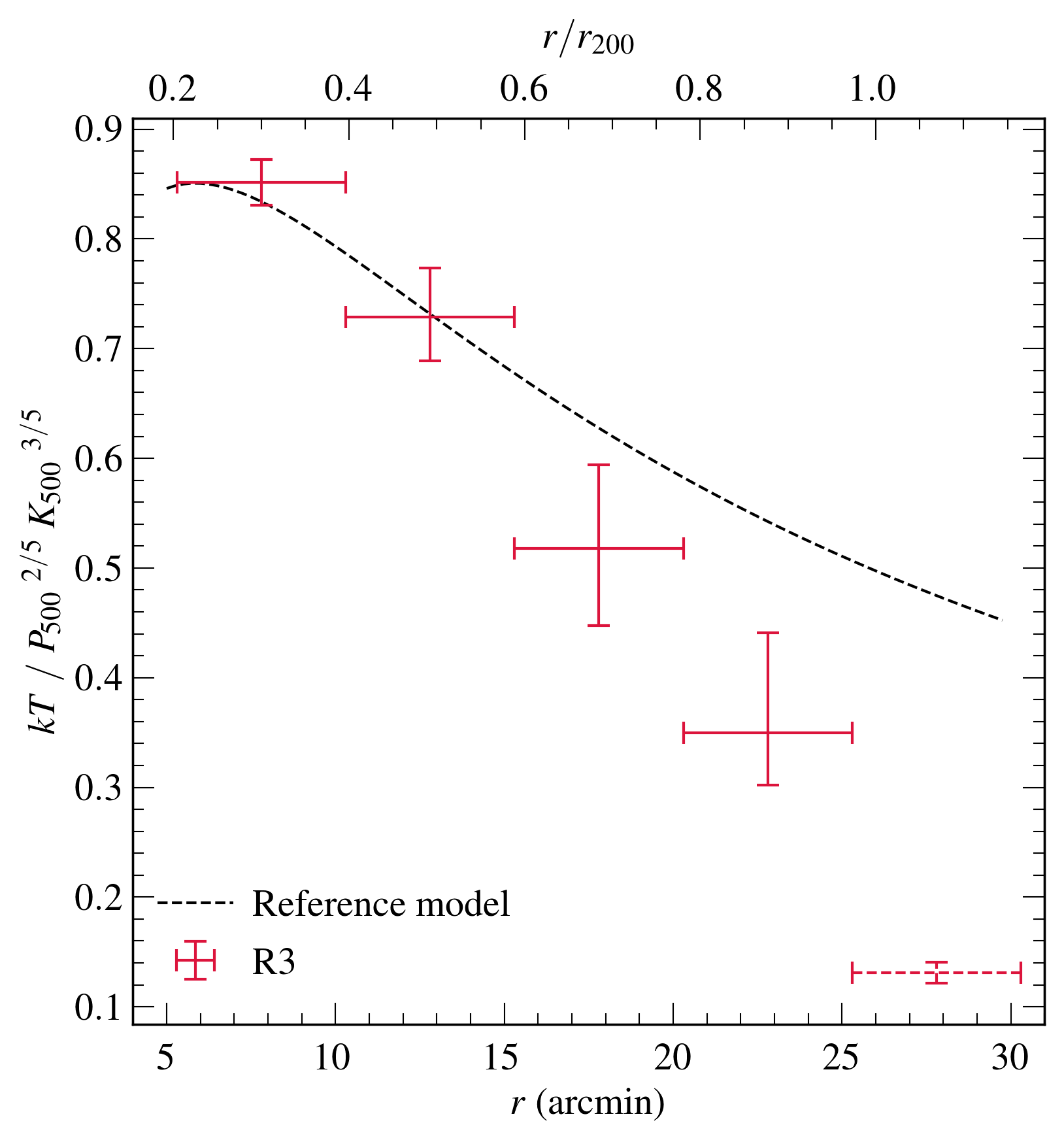}
\hspace{.5cm}
\includegraphics[width=.465\textwidth]{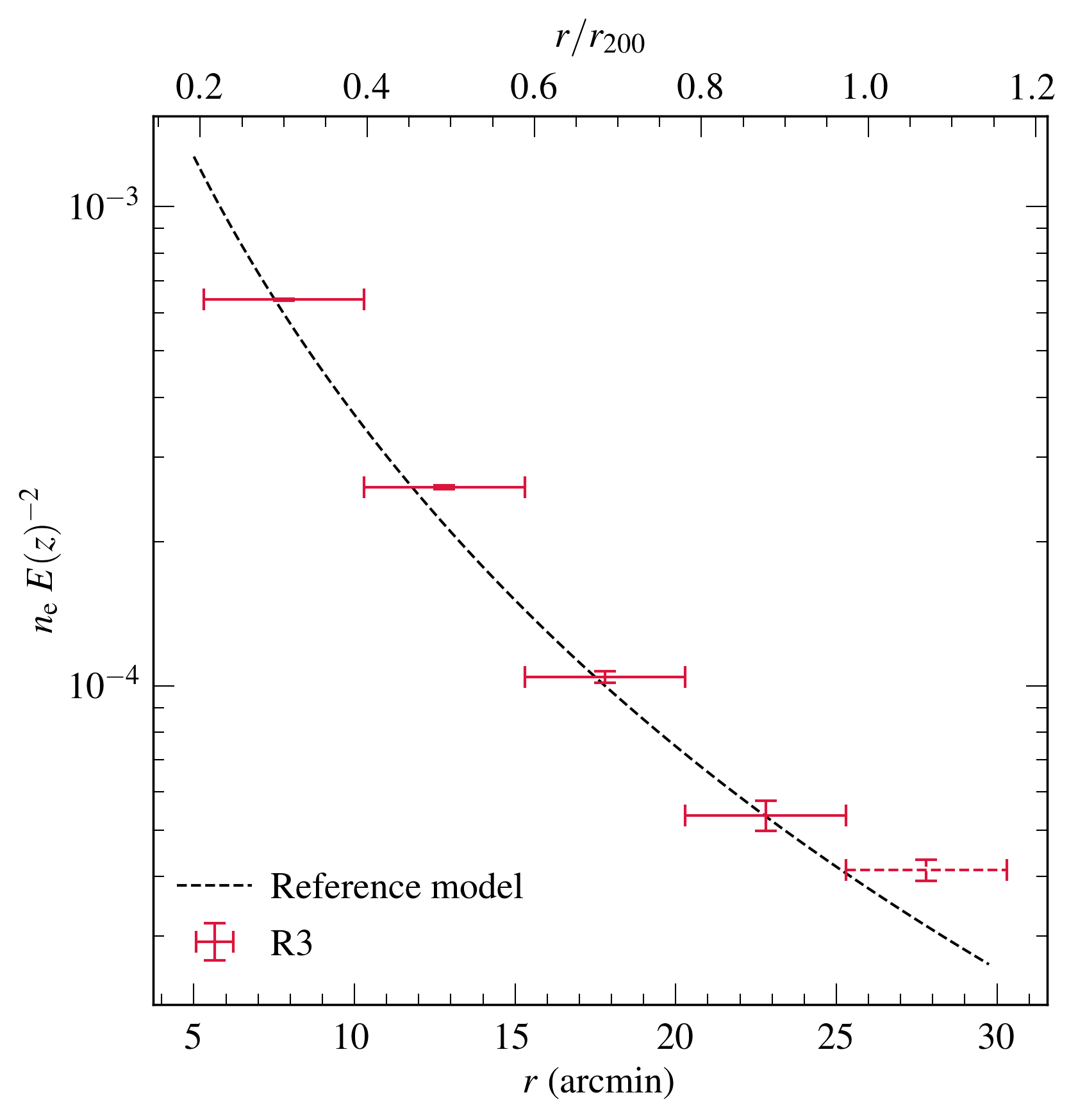}
\caption{{Self-similarly scaled profiles of measured and modeled properties of Abell\,1795. The reference models represent an ICM where the universal pressure profile from the \cite{2013A&A...550A.131P} and the entropy profile of \cite{2010A&A...511A..85P} apply. Note that the outermost measurements are not reliable due to the large uncertainties introduced during deprojection.
\textbf{\textit{Left:}} The lower-than-expected temperature in the outskirts of Abell\,1795 indicates the influence of complex dynamics in the physics of the outskirts. This may support theories suggesting electron-ion non-equilibrium and non-thermal pressure support to be at play in the cluster outskirts. \textbf{\textit{Right:}} The outskirts of Abell\,1795 show no excess in the electron number density, implying the successful resolution of point sources and the absence of resolved or unresolved clumping in the ICM gas.}}
\label{fig:self}
\end{figure*}

\subsection{The effect of {clump candidates} on the surface brightness profile}

In the presence of gas clumping, the observed ICM surface brightness profile, which is conventionally estimated assuming a homogeneous gas distribution, will be overestimated by the $C(r)$ clumping factor \citep[e.g.][]{2011ApJ...731L..10N}.
To measure the significance of this effect, we built the surface brightness profile of the cluster outskirts {from the \textit{Chandra} data} {with the $24$ clump candidates} identified in Section\,\ref{sec:individualclumps} removed, and compared it with that including the clump candidates.
The extraction of the profiles was done as described in Section\,\ref{sec:sb} with all the point sources masked.
We compare the profiles in Figure\,\ref{fig:sb_with_and_wo_clumps}.
Marginal differences between the two surface brightness distributions are seen at the radial location of the clump candidates, and they are more pronounced in the $\rm SB_{incl} / SB_{excl}$ ratio plot.
In Figure\,\ref{fig:sb_with_and_wo_clumps}, we also show the best-fit power-law models.
{The model describing the profile without the clump candidates shows a slight steepening, primarily due to the relatively large difference and uncertainty in the outermost surface brightness measurements.}

\subsection{Variations in the \textit{Suzaku} surface brightness profile caused by unresolved point sources}\label{sec:suzaku_sb_discussion}
With the high spatial resolution of \textit{Chandra}, having a PSF that is $100$ times smaller than \textit{Suzaku}'s, we resolved $1137$ point sources in the $0.5-2$\,keV energy band down to fluxes of $>\!10^{-15}\, \rm erg \, s^{-1} \, cm^{-1}$ (Section\,\ref{sec:sourcedetection} and \ref{sec:logn-logsplot}).
In contrast, the number of sources resolved by \textit{Suzaku} is 29 (Section\,\ref{sec:sbazimuthallyavg}).
Note, however, that the field coverage of the two telescopes is not identical (Figure\,\ref{fig:mosaic}).
Despite \textit{Suzaku}'s excellence in low surface brightness studies, mistreatment of source exclusion may distort measurements.

To probe the impact of \textit{Suzaku}'s low spatial resolution on imaging measurements, we extracted and fitted a surface brightness profile based on purely \textit{Suzaku} data, i.e., masking only the $29$ \textit{Suzaku}-resolved sources, which we compared with the profile presented in Section\,\ref{sec:sbazimuthallyavg}.
Within error bars, the two profiles are in agreement with only $\lesssim 8\%$ difference measured between the best-fit power-law exponents, suggesting only a slight improvement with the inclusion of \textit{Chandra} data.

\subsection{Excess ICM emission towards the west}\label{sec:sb_excess_west}

The surface brightness distribution extracted in the western direction displays a $4.2\,\sigma$ enhancement relative to the azimuthally averaged profile in the \textit{Chandra} observations of the cluster outskirts (Section\,\ref{sec:sb_sectors}, Figure\,\ref{fig:sb-wedges}).
For further analysis, we sub-divided the western extraction region into three equal-sized sectors with an opening angle of $30^{\circ}$ and measured the surface brightness profile within each sector.
We found that the surface brightness excess is the most prominent in the southernmost, then the central of the western sub-sectors.

To explore the cluster's cosmic environment, we built the galaxy distribution in the large-scale surroundings and looked for structures (i.e.,\ galaxy concentrations) that may be linked to the excess surface brightness.
We searched for galaxies in SDSS DR16 within a $5^{\circ}$ radius around Abell\,1795 and {within a redshift range of $0.0552-0.0692$ \citep{2018ApJ...867...14W}.}
With the same method described in Section\,\ref{sec:opticalcounterparts}, we also estimated the stellar mass of the galaxies so that a stellar mass filter can be applied when mapping the galaxy distribution.
The resulting maps are displayed in Figure\,\ref{fig:galdistr}.

The distribution of galaxies shows filamentary structures, which are the most prominent when plotting only the relatively large-mass galaxies ($\mathrm{log} \,M_{*} > 10.3\,M_{\odot}$) (Figure\,\ref{fig:galdistr}, middle panel), while the distribution of low-mass galaxies ($\mathrm{log} \,M_{*} < 9.7\,M_{\odot}$) is relatively homogeneous (Figure\,\ref{fig:galdistr}, right panel).
Filtering the sample based on stellar mass is reasonable because massive galaxies ($\mathrm{log}\,M_{\mathrm{halo}} \gtrsim 11.5 M_{\odot}$) almost exclusively reside within filaments, while low-mass galaxies are typically scattered around the field, populating walls and voids, thus, producing a more even distribution \citep{2007ApJ...655L...5A,2007MNRAS.375..489H,2014MNRAS.441.2923C}.
The spatial configuration of galaxy concentrations in Figure\,\ref{fig:galdistr} is in agreement with the filaments found in UV absorption studies by \cite{2004ApJ...614...31B}.
The thickness (i.e., the galaxy richness) and direction of the filaments, however, vary with the location. In addition, the scales on which the filaments are already outlined in the galaxy distribution, are not comparable to the \textit{Chandra} FOV of Abell\,1795.
These, therefore, complicate the association between the SB excess and large-scale filaments.

In the left panel of Figure\,\ref{fig:galdistr}, we also indicate the position of the neighboring galaxy clusters, which are, together with Abell\,1795, members of the Bo{\"o}tes supercluster.
The richest of the members is Abell\,1795, along with Abell\,1775 \citep{1989ApJS...70....1A}, which lies in the projected direction of the surface brightness excess present in Abell\,1795.
As no indication of excess towards Abell\,1795 was found in the surface brightness profile of Abell\,1775 extracted from \textit{XMM-Newton} data out to $\gtrsim\!r_{200}$, we speculate that the measured surface brightness excess might be related to infall along a filament.

{The surface brightness excess towards the west is also well pronounced when selecting azimuthal slices specifically based on the direction of galaxy filaments.
Using this approach, we found no surface brightness excess towards other filaments.
}

\subsection{{Electron-ion non-equilibrium and non-thermal pressure support}}\label{sec:electron}
{Clumps disrupt the homogeneity of the ICM, and bias its density profile high. It leads to the classical observational signs of ICM clumping, namely an increased pressure ($P \sim n_{\mathrm{e}}$) and, more importantly, a flattened entropy ($K \sim n_{\mathrm{e}}^{-2/3}$) compared to the profiles with pure gravitational heating. Note that cluster observations suggest that the entropy is more sensitive to dynamical history and non-gravitational processes\,\citep{2010A&A...517A..92A}.}
{In Abell\,1795, however, the entropy ($K \sim kT$) measurements are biased low in the outskirts, and the pressure ($P \sim kT$) is consistent with the empirical model of \cite{2010A&A...517A..92A}, while slightly steeper in the outskirts compared to the baseline of \cite{2013A&A...550A.131P} (Figure\,\ref{fig:deproj}, R2-3 profiles), which, in turn, is indicative of a temperature bias.}

{The presence of a temperature bias and absence of a density bias in the outskirts of Abell\,1795 are also confirmed by the self-similarly scaled profiles (Fig. \ref{fig:self}) built following \cite{2013MNRAS.432..554W}. 
The reference models adopt the universal pressure profile from \cite{2013A&A...550A.131P} and the baseline entropy profile proposed by \cite{2010A&A...511A..85P} to describe the ICM.
The mass within an overdensity of 500 ($M_{500} = 5.33 \times 10^{14} \,M_{\odot}$) and the corresponding radius ($r_{500} = 16.82 \arcmin$) of Abell\,1795 were adopted from \cite{2017AN....338..293E}.
Both the models and measurements are scaled by the $K_{500}$ self-similar entropy (Section\,\ref{sec:entropy}) and $P_{500}$ self-similar pressure (Section\,\ref{sec:pressure}) at $r_{500}$, enabling a comparison of shape and normalization between the baseline profiles and measurements.}

{Low temperature values in the outskirts could be attributed to non-equilibrium effects \citep{2010PASJ...62..371H,2012PASJ...64...49A,2016ApJ...833..227A}. When a plasma of electrons and ions travels through a shock, the heavier ions tend to absorb most of the kinetic energy and heat to $T_{\rm i}\!>\!>\!T_{\rm e}$. Thus, while the system eventually reaches equilibrium, the electron temperature may be lower than the ion temperature.
As CCD spectroscopy is only able to measure the electron temperature, in the presence of non-equilibrium conditions, the mean gas temperature may be underestimated.
In the absence of physical processes able to couple the temperature of particles (e.g., interactions driven by magnetic fields), the equilibration timescale for a fully ionized ICM is given by
\begin{equation}
t_{\mathrm{ei}} \approx 6.3 \times 10^8 \mathrm{yr} \left( \frac{T_\mathrm{e}}{10^7 \mathrm{K}}\right)^{\frac 3 2} \left(\frac{n_\mathrm{i}}{10^{-5}\mathrm{cm}^{-3}}\right)^{-1} \left(\frac{\mathrm{ln}\,\Lambda}{40}\right)^{-1}
\end{equation}
\citep{1962pfig.book.....S,2009ApJ...701L..16R}, where $\mathrm{ln}\,\Lambda$ is the Coulomb logarithm. We substituted $T_{\rm e}$ and $n_{\rm i}$ from the R3 measurements of the outermost annulus and obtained $t_{\mathrm{ei}} \approx 1.7 \times 10^8 \rm yr$ for Abell\,1795, which is much smaller than the cluster age. 
Therefore, if indeed present, electron-ion non-equilibrium would indicate a recent dynamical event in the outskirts.
}

{In addition, the thermodynamic properties of the ICM might also be influenced by non-thermal pressure due to turbulent and bulk gas motions in the plasma.
However, the measurements of \cite{2019A&A...621A..40E} at $r_{500}$ and $r_{200}$ in Abell 1795 indicate non-thermal pressure ratios of only $2.2\%^{+5.6\%}_{-2.2\%}$ and $6.7\%^{+6.0\%}_{-4.5\%}$ respectively.
These fractions are below the median of the rest of the sample, which comprises clusters from the \textit{XMM-Newton} cluster outskirts project, \citep{2017AN....338..293E}, and suggest that only a small fraction of the energy is attributed to kinetic motions.}

\subsection{Unidentified clumps among point sources}
During point source detection with \texttt{wavdetect}, we applied a reduced signal threshold of $10^{-8}$ so that the weakest sources, including potential ICM clumps, remain undetected at this phase of the analysis (Section\,\ref{sec:sourcedetection}).
{Later in the analysis, no confirmed clumps were found among the list of clump candidates.
In fact, these candidates turned out to be background AGN, galaxies, or galaxy clusters/groups (Section\,\ref{sec:individualclumps}, Table\,\ref{tab:clumpcandidates}).}
Despite the reduced signal threshold, it should be verified that no clumps were detected in the \texttt{wavdetect} output source list.
For this, we picked out the most extended and least luminous sources from the list, i.e., those having the same characteristics as clumps.
As we found a slight positive trend between the PSF ratio ($\sqrt{\mathrm{ax_{major}}\times\mathrm{ax_{minor}}}/\mathrm{PSF}$) and the net count rate of the sources, we only applied a limit to the PSF ratio of $>\!1.25$, to which $35$ sources correspond.

To get an impression of the type of  selected sources, we built their combined spectrum.
This way, we ensured that the number of source counts was sufficient for spectral extraction.
Background treatment was carried out using local background annuli for each source within a radius of twice the source major axis.
We found that the combined source spectrum follows a power-law model with a best-fit photon index of $\Gamma = 1.39 \pm 0.03$.
It suggests that the most extended sources in the \texttt{wavdetect} output are most likely AGN and not ICM clumps.
This conclusion is in agreement with Section\,\ref{sec:logn-logsplot}, where we found no significant excess in the $\mathrm{log}N-\mathrm{log}S$ distribution of the \texttt{wavdetect} identified sources compared to those constructed for CDFS point sources.
Note, however, that cosmic variance, as such, may blur the effect of ICM clumps, which may hinder bulk clump detection using the $\mathrm{log}N\!-\!\mathrm{log}S$ plot.

\section{Conclusions}

We present a joint analysis of high-angular-resolution \textit{Chandra} ACIS-I and high-sensitivity \textit{Suzaku} XIS data of the Abell\,1795 cluster outskirts with the aim of isolating ICM inhomogeneities and building the radial profiles of main ICM properties out to very large radii.
The imaging analysis and the projected profiles probe the cluster outskirts beyond $r_{200}$, while the radial profile of deprojected thermodynamic properties extends to $r_{200}$.
Our findings are summarized below.
\begin{enumerate}
\item {We identified 24 clump candidates in the projected field of Abell\,1795.
Based on their multi-waveband properties, the candidates can be typically associated with background objects, thus, they are not genuine ICM clumps.}
\item The impact of the {$24$} clump candidates on the surface brightness and thermodynamic profiles of the ICM is marginal.
\item Due to the cosmic variance, the effect of clumping cannot be detected in the $\mathrm{log}N\!-\!\mathrm{log}S$ plot of field sources.
In fact, these sources are expected to be solely AGN.
\item The surface brightness profile extracted from \textit{Chandra} observations is in agreement with that extracted from \textit{Suzaku}. Excess ICM emission towards the west is seen in the direction of Abell\,1775, a rich neighbor of Abell\,1795.
\item {Upon removal of clump candidates from the ICM, the emissivity bias profile flattens at $\sim\!1$, indicating no or only marginal unresolved clumping in Abell\,1795.}
\item {{After resolving and removing all \textit{Chandra} sources (including both point sources and clump candidates), the entropy 
profile of the ICM still departs from what is expected from a purely gravitational infall. The shape of the profile indicates hints for complex physics governing the outskirts, including potential electron-ion non-equilibrium and non-thermal pressure support.}}

\end{enumerate}

\begin{acknowledgements}
OEK and NW are supported by the GA\v{C}R EXPRO grant No. 21-13491X ``Exploring the Hot Universe and Understanding Cosmic Feedback".
This paper benefited from a visit supported by European Union's 2020 research and innovation program under grant agreement No.871158, project AHEAD2020. 
The authors thank Dan Hu for her useful contribution to the \textit{XMM-Newton} analysis of Abell\,1775. 
AS and ZZ are supported by the Women In Science Excel (WISE) programme of the Netherlands Organisation for Scientific Research (NWO). AS acknowledges the Kavli IPMU for the continued hospitality. 
\'AB acknowledges support from the Smithsonian Institution and the Chandra X-ray Center through NASA contract NAS8-03060.
\end{acknowledgements}

\bibliographystyle{aa}
\bibliography{clumping}

\begin{appendix} 
\section{{Uncertainties in the projected temperature profile introduced by systematic effects}}\label{appendix:systematic}

\begin{figure}
\includegraphics[width=.5\textwidth]{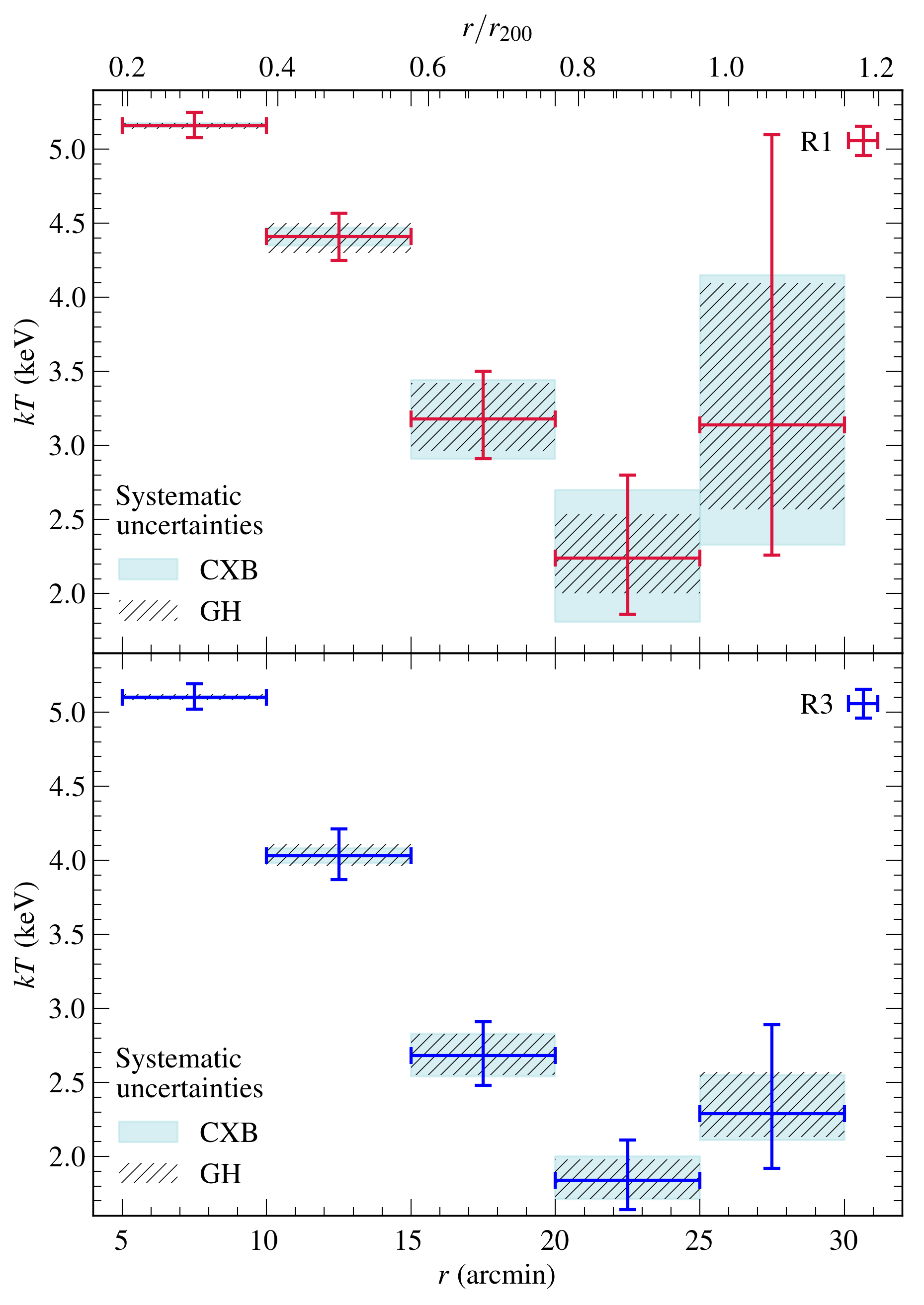}
\caption{{Demonstration of systematic effects on the projected R1 and R3 $kT$ profiles caused by spatial variations in the cosmic X-ray background and in the galactic halo emission.}}
\label{fig:systematic-uncert}
\end{figure}

{According to the estimated $25\%$ fluctuation in the level of the galactic halo (GH) emission (Section\,\ref{sec:suzaku-syst-uncert}), the R1 and R3 spectral fits were repeated with the normalization of the corresponding \textit{apec} model fixed at the $\pm 25\%$ values of the best-fit normalization. The $kT$ measurements obtained this way represent the upper and lower boundaries of the systematic error from GH fluctuations and are plotted with hatched regions in Figure\ref{fig:systematic-uncert}.
The uncertainties in $kT$ introduced by the relative cosmic variance (Section\,\ref{sec:suzaku-syst-uncert}) were computed similarly, and the resulting $kT$ intervals are illustrated in Figure\ref{fig:systematic-uncert} with shaded regions.
These results show that the systematic errors are below the statistical ones, and the present Suzaku measurements are not dominated by systematic uncertainties.
}

{}

\end{appendix}

\end{document}